\documentclass[12pt,aps]{revtex4}
\usepackage{epsfig}

\newcommand{\beeq}{\begin{equation}}
\newcommand{\eneq}{\end{equation}}
\newcommand{\be}{\begin{eqnarray}}
\newcommand{\ee}{\end{eqnarray}}
\newcommand{\bpic}{\begin{picture}}
\newcommand{\epic}{\end{picture}}
\newcommand{\bs}{\begin{scriptsize}}
\newcommand{\es}{\end{scriptsize}}

\def\dd{\partial}
\def\la{\raise.16ex\hbox{$\langle$} \, }
\def\ra{\, \raise.16ex\hbox{$\rangle$} }

\def\d{\delta}
\def\e{\epsilon}

\def\r{\rho}
\def\s{\sigma}
\def\O{\Omega}
\def\D{\Delta}

\def\l{\lambda}

\def\Box{\kern1pt\vbox{\hrule height 1.2pt\hbox{\vrule width 1.2pt\hskip 3pt
   \vbox{\vskip 6pt}\hskip 3pt\vrule width 0.6pt}\hrule height 0.6pt}\kern1pt}
\def\gtwid{\mathrel{\raise.3ex\hbox{$>$\kern-.75em\lower1ex\hbox{$\sim$}}}}
\def\ltwid{\mathrel{\raise.3ex\hbox{$<$\kern-.75em\lower1ex\hbox{$\sim$}}}}

\def\Box{\kern1pt\vbox{\hrule height 1.2pt\hbox{\vrule width 1.2pt\hskip 3pt
   \vbox{\vskip 6pt}\hskip 3pt\vrule width 0.6pt}\hrule height 0.6pt}\kern1pt}
\begin{document}
\begin{titlepage}
\begin{flushright}
astro-ph/0510414\\LPT-ORSAY-05-59
\end{flushright}
\vspace{.4cm}
\begin{center}
\textbf{Gravitational Lensing and Structural Stability\\
of\\ Dark Matter Caustic Rings}
\end{center}
\begin{center}
V. K. Onemli$^{1\;2}$
\end{center}
\begin{center}
${}^1$\textit{Department of Physics, University of Crete, GR-71003
Heraklion, Greece}
\end{center}

\begin{center}
${}^2$\textit{Laboratoire de Physique Th\'eorique, Universit\'e
Paris XI, B\^at. 210, 91405 Orsay, France}
\end{center}

\begin{center}
ABSTRACT
\end{center}
Gravitational lensing by the dual cusp ($A_{-3}$) catastrophes of
the cold dark matter (CDM) caustic rings at cosmological distances
may provide the tantalizing opportunity to detect CDM indirectly,
and discriminate between axions and weakly interacting massive
particles (WIMPs). The infall of CDM onto isolated galaxies such
as our own produces discrete number of flows and caustics in the
halo CDM distribution. Caustics are places where the CDM particles
are naturally focussed. In the CDM cosmology, caustics in the
distribution of dark matter are plentiful once density
perturbations enter the nonlinear regime. Our focus is upon the
caustic rings which are closed tubes whose cross-section is an
elliptic umbilic ($D_{-4}$) catastrophe with three dual cusps. We
call this cross-section a ``tricusp.'' Caustic rings are located
near where the particles with the most angular momentum in a given
inflow reach their distance of closest approach to the galactic
center before going back out. We reparameterize the CDM flow near
a caustic ring, and obtain the ring equations in space, as single
valued functions of an angular variable parameterizing the
tricusp. A caustic ring has a specific density profile, a specific
geometry and, therefore, precisely calculable gravitational
lensing signatures that vary on the caustic surface, depending on
the location of the line of sight. We calculate the image
magnification of point-like sources as a function of the angular
parameter of the tricusp for the line of sights that are parallel
to the galactic plane of the caustic ring and near tangent to the
surface. The magnification monotonically increases as the line of
sight approaches to the cusps where it diverges in the limit of
zero velocity dispersion. If the velocity dispersion is small but
nonzero, the divergence is cut off, because the caustic surface
gets smeared over some distance in space and the cusps are
smoothed out. The lensing effects are no longer infinite at the
cusps, but merely large.  To estimate the magnification
numerically, for the rings at cosmological distances and for the
ring closest to us (the fifth caustic ring of the Milky Way), we
choose sample points near the cusps. We find that the lensing
effects are largest at the outer cusp which lies in (or, near) the
galactic plane. Around the outer cusp, the surface is curved
toward the side with two extra flows. We call such a surface
concave. In the limit of zero velocity dispersion, at a sample
point near the outer cusp, we find $37\%$ magnification for the
CDM caustic rings at cosmological distances. This is about $250$
times stronger than the lensing effect we predicted for the
concave folds (considering the $A_2$ catastrophes enveloping the
galaxy), and $37$ times stronger than the largest lensing effect
we predicted for the CDM caustics (considering the convex folds of
the rings) before. In the presence of finite velocity dispersion,
the lower and upper bounds of the effective velocity dispersions
of the axion and WIMP flows in galactic halos may be used to
constrain the lensing effects at the smoothed cusps. For a
cosmological axion caustic ring, we find that the magnification
may range between $3\%$ and $2800\%$ at the outer cusp, and
between $2\%$ and $46\%$ at the non-planer cusps. For a
cosmological WIMP caustic ring, on the other hand, we constrain
the magnification between $3\%$ and $28\%$ at the outer cusp, and
between $2\%$ and $5\%$ at the non-planer cusps. Because the upper
bounds for the magnification at the cusps are obtained considering
the minimum primordial value of the velocity dispersions for the
CDM candidates in space where no small scale structure has formed,
and also because the observer's line of sight may not exactly be
parallel to the galactic plane of the caustic ring in general,
they should be regarded with precaution. The images of extended
sources may also show distortions that can be unambiguously
attributed to lensing by dark matter caustics. Finally, we derive
the Catastrophe Function of the triaxial caustic rings. We obtain
the flow equations as the equilibrium points of this Catastrophe
Function. The analysis of the Stability (Hessian) Matrix show that
the caustic rings are structurally stable.

\begin{flushleft} PACS
numbers: 98.80.Cq, 4.62.+v
\end{flushleft}
\end{titlepage}

\section{Introduction}
\label{sec:intro} A large amount of astronomical evidence
indicates the existence in the universe of more gravitationally
interacting matter than luminous (baryonic) matter. Most of the
matter in the universe is dark (23\% of the composition, according
to the first year data \cite{Bennett} of the Wilkinson Microwave
Anisotropy Probe, is dark matter, whereas only 4.4\% is ordinary
baryonic matter). The stuff that is responsible for holding
galaxies, and clusters of galaxies together, is a peculiar kind of
matter that we neither see nor detect by any means. The detection
dilemma of dark matter particles has been occupying physicists for
many decades by now, and the case still remains open. There are,
however, significant reasons to believe that the dark matter of
the universe is constituted, in large fraction, by collisionless
particles with very small primordial velocity dispersion. Such
particles are called cold dark matter (CDM). The attribute
``collisionless'' indicates that the particles interact so weakly
that, practically, they move purely under the influence of
gravity. Hence, unlike ordinary baryonic matter, they are
undissipative. The small primordial velocity dispersion allows CDM
to clump in galactic scales. It is believed that galaxies are
surrounded by CDM which keeps falling into the gravitational
potential wells of galaxies from all directions, and forms halos
around their baryonic disks. The leading CDM candidates are axions
and weakly interacting massive particles (WIMPs), such as
neutralinos. The information about the distribution of CDM in
galactic halos is substantial for the detection experiments, for
the understanding of galaxy formation and galactic structure. The
CDM infall onto isolated galaxies, such as the Milky Way, produces
a discrete number of flows (the minimum number of CDM flows on the
Earth is of order 100 \cite{Ipser}) and associated caustics
throughout the halos of galaxies. Caustics are places in the halos
where the CDM particles are naturally focussed and, hence, the
density is very large.

Zel'dovich \cite{Zel} emphasized the importance of caustics in
large scale structure formation, and suggested using the name
``pancakes'' for them.  The reason why galaxies tend to lie on
surfaces \cite{Huchra} is that the 3D sheet on which the dark
matter particles and baryons lie in phase space acquires folds on
very large scales, producing caustics called Zel'dovich pancakes.

\begin{figure}[ht] \centering
\includegraphics[height=7cm,width=30cm]{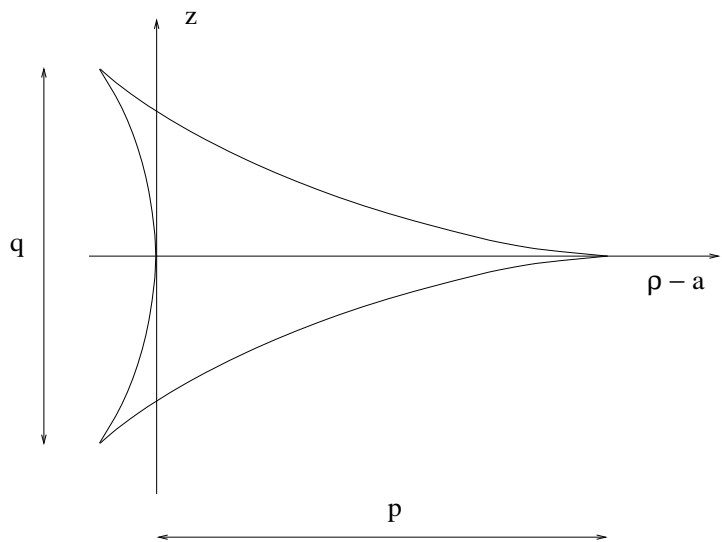}
\caption{Cross-section of a caustic ring in the case of axial and
reflection symmetry, and $p,q \ll a$.} \label{fig:fig6}
\end{figure}

Sikivie \cite{cr,sing} derived the minimal CDM caustic structure
that must occur in galactic halos. There are two types of caustics
in the halos of galaxies: inner and outer. The outer caustics are
simple fold ($A_2$) catastrophes located on topological spheres
surrounding the galaxy. Outer caustics occur where a given outflow
reaches its furthest distance from the galactic center before
falling back in. The inner caustics are rings \cite{sing,cr}. They
are located near where the particles with the most angular
momentum in a given inflow reach their distance of closest
approach to the galactic center before going back out. A caustic
ring is more precisely a closed tube with a special structure. Its
transverse cross-section is an elliptic umbilic ($D_{-4}$)
catastrophe which is a closed line with three dual cusps, one of
which points away from the galactic center; see Fig.
\ref{fig:fig6}.  We call it a ``tricusp.'' We use the terms ``cusp''
and ``dual cusp'' interchangeably. In this paper, our focus is
upon the caustic rings.

The motivations to study CDM caustics originate from their analogs
in Optics. Caustics are well known phenomena in light. They are
singularities in the propagation of light. Examples are the bright
wavy lines generated at the bottom of a swimming pool by the sun
light, or cusp-like reflection (cardioid) that is sometimes seen
in a coffee cup. Generically, caustics are boundaries between the
two regions where the light intensity decreases very sharply on
one side, and, typically two light beams superimpose on the other.
Two conditions must be satisfied \cite{Big} for caustics to occur:
(1) the propagation must be collisionless, (2) the flow must have
small velocity dispersion. Propagation of light is classically
collisionless, and the flow of light from a point source has zero
velocity dispersion. Cold dark matter flow also has both of the
properties that light has, i.e., (1) CDM particles interact only
weakly, and (2) the primordial velocity dispersion $\delta v$ of
the cold dark matter candidates is of a very small order
\cite{sing,Thesis}. For the axions, we find \beeq \delta v_a (t)
\simeq 4.6\cdot 10^{-12} \,{\rm km/s}\, \left( {10^{-5} {\rm
eV}\over m_a}\right)^{4.7/5.7}~\left({t_0\over t}\right)^{2/3}\,
,\label{delvax}\eneq where $t_0$ and $m_a$ are respectively the
present age of the universe and the axion mass. For the WIMPs, if
they decoupled before the temperature was about the electron mass
(i.e. if the particles ($\chi$s) were not affected by the
reheating of the photons created by the $e^+ e^-$ annihilations),
we find \beeq \delta v_\chi(t)\simeq 2.8\cdot 10^{-6}\,{\rm
km/s}\,\left({\frac{{\rm
GeV}}{m_\chi}}\right)^{1/2}\left({\frac{{\rm
MeV}}{T_D}}\right)^{1/2}\left(\frac{t_0}{t}\right)^{2/3}\label{delvWIMP1}\;
,\eneq where $T_D$ and $m_\chi$ are respectively the decoupling
temperature and the mass of the WIMPs, or, if they decoupled after
the temperature was about the electron mass (i.e. if the $\chi$s
were affected by the reheating), we find \beeq\delta
v_\chi(t)\simeq 3.9\cdot 10^{-6} \,{\rm km/s}\,\left(\frac{{\rm
GeV}}{m_\chi}\right)^{1/2}\left(\frac{{\rm MeV}}{T_D}\right)^{1/2}
\left(\frac{t_0}{t}\right)^{2/3}\; .\label{delvWIMP2} \eneq Hence,
as in the case of light, caustics are expected to be common in the
distribution of CDM in space. These estimates of primordial
velocity dispersions are approximate, but in the context of this
paper and of galaxy formation in general, $\delta v_a$ and $\delta
v_\chi$ are entirely negligible. Therefore, the CDM density must
be very large at a caustic. Henceforth, unless otherwise stated,
the velocity dispersion of CDM flow is set equal to zero.

Detection of the caustic structure in the halos would
revolutionize our understanding of galaxy formation, and galactic
structure. Caustics precisely predict the distribution of dark
matter in the caustic neighborhood, and hence their gravitational
lensing effects \cite{Hogan,lensing,Thesis,probing}. Near the
sharp edges of caustics, we may expect to have special signatures
in image magnification and in image structure. Point-like
background sources and extended sources, such as radio jets might
thus become probes of elusive CDM particles. Detection of the
caustic ring structure by gravitational lensing would mean
indirect detection the CDM in galactic halos. It may even be
possible to determine the velocity dispersion of dark matter from
lensing observations \cite{lensing}, and hence distinguish between
axions and neutralinos. Because WIMP annihilation rate per unit
volume is proportional to the WIMP number density squared, the
presence of caustics increase the total annihilation rate and
change the spatial distribution of emission \cite{Micro} in
galactic halos. If photons from WIMP annihilation are observed,
the caustic locations may be unveiled as sharp lines and hot spots
on the sky.

In Ref. \cite{lensing}, we studied the density profiles and
gravitational lensing effects of outer and inner (ring) caustics,
in the limit of zero velocity dispersion. For the caustic rings,
the curvature radii and density depend on the location on the
surface of the caustic ring. These quantities were expressed as
functions of position on the tricusp and estimated using the self
similar infall model. We presented a formalism which simplified
the relevant gravitational lensing calculations and applied our
formalism to the dark matter caustics in four specific cases. In
the first three, the line of sight was tangent to a surface where
a simple fold catastrophe was located. The three cases were
distinguished by the curvature of the caustic surface at the
tangent point in the direction of the line of sight: (i) the
surface curves toward the side with two extra flows; (ii) the
surface curves away from the side with two extra flows; and (iii)
the surface has zero curvature (a large fraction of the surface of
a caustic ring is saddle-shaped and therefore has tangent
directions along which the curvature of the surface vanishes). In
the fourth case (iv) studied, the line of sight was at a specific
location close to a dual cusp and parallel to the galactic plane
of the caustic ring. We estimated the lensing effects in all these
four cases at special points on the caustics. For case (i), which
we called ``concave,'' only the outer caustics were considered.
Cases (ii)-(iv) are valid only for the caustic rings. We found
that each case has characteristic lensing signatures. In three of
the cases (ii-iv) there are multiple images and infinite
magnification of point sources when their images merge (in case
(i), there are no multiple images). The effects estimated were all
small. It was found that the typical magnifications and image
distortions are of order one \% to a few \%. The most promising
possibility seemed monitoring the sky for distortions caused by
the dark matter caustics in the images of the extended sources,
such as radio jets.

In this paper, we point out that the lensing effects are enhanced
significantly near the cusps, in particular, near the outer cusp.
We first reparameterize the flow equations near a caustic ring.
The reparametrization is useful in studying the characteristics,
stability and gravitational lensing effects of the caustic rings.
We express the caustic rings generated by the CDM flow as single
valued functions $z(\psi)$ and $\rho(\psi)$, where $z$ and $\rho$
are transverse space coordinates and $\psi$ is an angular
variable, parameterizing the tricusp cross-section. These
equations show that the reason for the existence of the outer cusp
which lies in the galactic plane, is the same as the other two
cusps: both $\frac{\dd\rho}{\dd\psi}$ and $\frac{\dd z}{\dd\psi}$
vanish simultaneously at the cusps. The parametrization given in
\cite{sing} is inefficient to reveal this fact, because, the plot
of the caustic equations do not traverse through the outer cusp.
They plot half of the tricusp; the other half is obtained by the
reflection symmetry, due to the double valuedness of $z(\psi)$, in
that parametrization. Generically, caustics are surfaces
separating regions with differing number of flows. One side of a
caustic surface has two more flows than the other. The density
profile near a caustic surface is
$d=\frac{A}{\sqrt{\sigma}}\Theta(\sigma)$, where $A$ is called the
fold coefficient, and $\sigma$ is the distance to the surface, on
the side with two extra flows. In the limit of zero velocity
dispersion, the density diverges as $\frac{1}{\sqrt{\sigma}}$.
Near the surface of a caustic ring we express the fold coefficient
as a function of $\psi$. We find that $A(\psi)$ is minimum at the
middle locations (where $\psi=\frac{\pi}{3}, \pi, \frac{5\pi}{3}$)
between the cusps. It increases monotonically as one approaches to
cusps where it diverges. We also study the differential geometry
of the caustic ring surface, using the Cartan's Structure
equations. We obtain Gaussian, mean and principal curvatures as
functions of $\psi$ at any location other than the cusps. As is
shown in the Appendix, the principal curvatures are the inverses
of the principal radii of the surface, and are needed for
gravitational lensing applications. Using the equations for the
fold coefficient and the curvature radius that are derived here,
in the expressions for the magnification and image shift given in
Ref. \cite{lensing}, we obtain the lensing formulae for the line
of sights parallel to the galactic plane of the caustic ring and
near tangent to the surface at a given $\psi$. The magnification
monotonically increases as the line of sight approaches to the
cusps where it diverges in the limit of zero velocity dispersion.
We consider the concave folds of the caustic rings to estimate the
magnification and the shift numerically, for the first time in
this paper. We find that, for the axion caustic rings at
cosmological distances, when the observer's line of sight is
tangent to the surface at a sample point near the outer cusps, the
magnification is about 37$\%$. This effect is about 250 times
greater than the effect obtained for the concave folds considering
the outer caustics, and about $37$ times greater than the largest
effect predicted in Ref. \cite{lensing} considering the convex
folds of the caustic rings. We also estimate the magnifications at
sample locations near the non-planer cusps for cases (i), (ii),
(iii), and constrain the expected magnifications at the cusps in
the presence of finite velocity dispersion. When the velocity
dispersion is small but nonzero, the divergence at the cusps is
cut off, because the caustic surface gets smeared over some
distance in space and the cusps are smoothed out. The lensing
effects are no longer infinite at the cusps, but merely large. The
lower and upper bounds of the effective velocity dispersions of
the axion and WIMP flows in galactic halos may be used to
constrain the lensing effects at the cusps. For a cosmological
axion caustic ring, we find that the magnification may range
between $3\%$ and $2800\%$ at the outer cusp, and between $2\%$
and $46\%$ at the non-planer cusps. For a cosmological WIMP
caustic ring, on the other hand, we constrain the magnification
between $3\%$ and $28\%$ at the outer cusp, and between $2\%$ and
$5\%$ at the non-planer cusps. Because the upper bounds for the
magnification at the cusps are obtained considering the minimum
primordial value of the velocity dispersions for the CDM
candidates in space where no small scale structure has formed, and
also because the observer's line of sight may not exactly be
parallel to the galactic plane of the caustic ring in general,
they should be regarded with precaution. Moreover, a triangular
feature in the Infrared Astronomy Satellite (IRAS) map of the
Milky Way plane was interpreted \cite{IRAS} as the imprint on
baryonic matter of the caustic ring of dark matter nearest to us.
The nearby caustic ring, the fifth ring of the Milky Way, is 1 kpc
away (in the direction of observation) from us. We also estimate
the magnifications for the line of sights close to the cusps of
this ring. Unfortunately, even near the outer cusp, the
gravitational lensing due to a caustic ring only a kpc away from
us, is too weak to be observed with present instruments. Finally,
we study the caustic rings in the Catastrophe Theory point of
view. We derive the Catastrophe Function of the triaxial caustic
rings and obtain the flow equations as the equilibrium points of
this Catastrophe Function. The analysis of the Stability (Hessian)
Matrix show that the caustic rings are structurally stable.

To see how caustics of CDM form in galactic halos, it is
instructive to discuss the phase space structure of CDM halos.
Since the particles are in three-dimensional (3D) space, the phase
space is six-dimensional. The small velocity dispersion $\delta v$
means that CDM particles lie on a very thin 3D sheet in 6D phase
space. The thickness of the sheet is $\d v$ of the particles. The
number of CDM particles is enormous in terms of astronomical
length scales. Hence, the sheet on which the particles lie in
phase space is continuous. The flows can be described in terms of
the evolution of this sheet. When a large overdensity enters the
nonlinear regime, the particles in the vicinity of the overdensity
fall back onto it. This implies \cite{sing} that the phase space
sheet ``winds up'' in clockwise fashion wherever an overdensity
grows in the nonlinear regime. Before density perturbations enter
the nonlinear regime, there is only one value of velocity (i.e., a
single flow) at a typical location in physical space, because, the
phase space sheet covers physical space only once.  On the other
hand, inside an overdensity in the nonlinear regime, the sheet
wraps up in phase space, turning clockwise in any two dimensional
cut $(r, \dot{r})$ --- where $r$ denotes the radial coordinate and
$\dot{r}$ is the associated velocity --- of that space. The
outcome of this process is an odd number of flows at any point in
a galactic halo. One flow is associated with particles falling
through the galaxy for the first time ($n=1$); another is
associated with particles falling through the galaxy for the
second time ($n=2$), and so on. At the boundary between two
regions, one of which has $n$ flows and the other $n + 2$ flows,
the physical space density is very large, because, the phase space
sheet has a fold there. At the fold, the phase space sheet is
tangent to velocity space, and hence, in the limit of zero
velocity dispersion $(\delta v = 0)$, the physical space density
diverges, since it is the integral of the phase space density over
velocity space.  The structure associated with such a phase space
fold is called a ``caustic.'' Because the caustic surfaces occur
wherever the number of flows changes, they are topologically
stable. As mentioned earlier, in the limit of zero velocity
dispersion, the density diverges as $d \sim {1\over\sqrt{\sigma}}$
when the caustic is approached from the side with $n+2$ flows,
where $\sigma$ is the distance to the caustic.  If the velocity
dispersion is nonzero, this divergence is cut off, because, the
location of the caustic surface gets smeared over some distance
$\delta x$.  For the dark matter caustics in galactic halos,
$\delta x$ and $\delta v$ are related \cite{cr} by
\begin{equation}
\delta x \sim {R~\delta v \over v} \label{sme}\ ,
\end{equation}
where $v$ is the order of magnitude of the velocity of the
particles in the flow and $R$ is the distance scale over which
that flow turns around (i.e., changes its direction).  For a
galaxy like our own, $v = 500$ km/s and $R = 200$ kpc are typical
orders of magnitude. Using the estimates of the primordial
velocity dispersions of the CDM candidates (Eqs.
\ref{delvax}-\ref{delvWIMP2}), one finds \cite{lensing,Thesis}
that axion caustics in galactic halos are typically smeared over
\begin{equation}
\delta x_a \sim 6\cdot 10^{4}\, {\rm km} \left({10^{-5} {\rm eV}
\over m_a} \right)^{4.7/5.7} \; , \label{dxa}
\end{equation}
as a result of their primordial velocity dispersion; whereas WIMP
caustics are smeared over
\begin{equation}
\delta x_\chi
 \sim 3\cdot 10^{10}\, {\rm km} \left({{\rm GeV}
\over m_\chi}\right)^{1/2} \; , \label{dxW1}
\end{equation}
if the particles were not affected by the reheating, or
\begin{equation}
\delta x_\chi
 \sim 5\cdot 10^{10}\, {\rm km} \left({{\rm GeV}
\over m_\chi}\right)^{1/2} \; , \label{dxW2}
\end{equation}
if the particles were affected by the reheating. It should be kept
in mind, however, that a CDM flow may have an effective velocity
dispersion that is larger than its primordial velocity dispersion.
Effective velocity dispersion occurs when the sheet on which the
dark matter particles lie in phase space is wrapped up on scales
that are small compared to the galaxy as a whole.  It is
associated with the clumpiness of the dark matter falling onto the
galaxy.  The effective velocity dispersion of a flow may vary from
point to point, taking larger values where more small scale
structure has formed; and taking the minimum primordial value
where no small scale structure has formed.  For a coarse-grained
observer, the dark matter caustic is smeared over $\delta x$ given
by Eq. \ref{sme} where $\delta v$ is the effective velocity
dispersion of the flow. Little is known about the size of the
effective velocity dispersion of dark matter flows in galactic
halos. However, the sharpness of the triangular feature's edges
(in the IRAS map of the Milky Way plane) implies \cite{IRAS} an
upper limit of 15-20 pc on the distance $\delta x$ over which that
caustic is smeared (and hence an upper limit of order 50 m/s on
the effective velocity dispersion of the corresponding flow). We
will use this upper limit and the lower limits Eqs. \ref{dxa} and
\ref{dxW2} for the smearing out of caustics to constrain the
density and lensing estimates at the cusps of the caustic rings in
Sects. \ref{Density} and \ref{sec:Lensing}.

Evidences for the caustic rings were found \cite{KS,IRAS} in the
distribution of bumps in the rotation curves of spiral galaxies,
including the Milky Way. The self-similar infall model of galactic
halo formation predicts that the caustic ring radii \cite{cr}
\beeq \{a_n: n=1, 2,.\; .\; .\}\simeq(39,~19.5,~13,~10,~8,.\; .\;
.){\rm kpc}\cdot\!
\left(\frac{j_{\rm{max}}}{0.27}\right)\!\left(\frac{0.7}{h}\right)\!
\left(\frac{v_{\rm{rot}}}{220\,{\rm km/s}}\right)\; , \label{a_n}
\eneq where $j_{\rm {max}}$ is a parameter, with a specific value
for each halo, which is proportional to the amount of angular
momentum that the dark matter particles have \cite{STW1,STW2}, $h$
is the present Hubble constant in units of 100 km/(s Mpc), and
$v_{\rm rot}$ is the rotation velocity of the galaxy. The infall
is called {\it self-similar}, if it is time independent, after all
distances are rescaled by a time-dependent scale $R(t)$ at time
$t$, and all masses are rescaled by the mass $M(t)$ interior to
$R(t)$. In the case of zero angular momentum and spherical
symmetry, the infall is self-similar if the initial overdensity
profile has the form $\frac{\delta
M_i}{M_i}=\left(\frac{M_0}{M_i}\right)^\epsilon$, where $M_0$ and
$\epsilon$ are parameters \cite{FG}. In CDM theories of large
scale structure formation, $\epsilon$ is expected to be in the
range $0.2$ to $0.35$ \cite{STW1,STW2}. In that range, the
galactic rotation curves predicted by the self-similar infall
model are flat \cite{FG}. Because the caustic rings lie close to
the galactic plane, they cause bumps in the rotation curve, at the
locations of the rings. In a study \cite{KS} of 32 extended and
well-measured external galactic rotation curves, evidence was
found for the law given in Eq. \ref{a_n}. In the case of the Milky
Way \cite{IRAS}, the locations of eight sharp rises in the
rotation curve fit the prediction of the self-similar model at the
3$\%$ level. Thus, on theoretical and observational grounds, the
caustic ring radii $a_n (n=1,2,3..)$ obey the approximate law
$a_n\sim 1/n$. Evidence for the first outer caustic (the caustic
of second turnaround) has been recently found \cite{Mahdavi} in
the NGC 5846 group of galaxies. The plot of the surface number
density of galaxies was inferred to mark the first outer caustic
of radius $R_1$ by an abrupt density drop (to a roughly constant
value) and a transition from a large velocity dispersion interior
to $R_1$, to a very small velocity dispersion exterior to $R_1$,
as predicted by the CDM infall models \cite{FG}.

The outline of this paper is as follows. In Section
\ref{sec:FlowEq}, we reparameterize the flow equations that were
introduced in Ref. \cite{sing}. Using this reparametrization, we
obtain the caustic ring equations associated with the CDM flow, as
single valued functions of an angular parameter $\psi$. In Secs.
\ref{subsec:foldcoef}-B, following the procedure given in Ref.
\cite{lensing},  we derive the density on the caustic ring.
Section \ref{difgeo} is devoted to the differential geometry of
the caustic ring surface. In Sect. \ref{sec:Lensing}, using the
equations we obtained in Ref. \cite{lensing} and in Sect.
\ref{difgeo} of this paper, we derive gravitational lensing
formulae for the line of sights tangent to the caustic surface at
any point on the surface. We estimate the magnification near the
cusps where the effect is largest, for the caustic rings at
cosmological distances (and also for the nearby fifth ring of our
own galaxy), in the limit of zero velocity dispersion. We
calculate the upper and lower bounds for the lensing effects at
the cusps in the presence of finite velocity dispersion. The
reparameterized flow equations are also useful in studying the CDM
caustics in the Catastrophe Theory point of view. In Sec.
\ref{Stability}, we present the correspondence of our formulation
with the Catastrophe Theory. We derive the Catastrophe Function of
the triaxial caustic rings and obtain the flow equations as the
equilibrium points of this Catastrophe Function. The analysis of
the Stability (Hessian) Matrix show that the caustic rings are
structurally stable. In Sec. \ref{Conc}, we summarize our
conclusions.

\section{Flow Equations at a Caustic Ring}
\label{sec:FlowEq}

The caustic rings are closed tubes whose cross-section is a
$D_{-4}$ catastrophe \cite{cr,sing}. They are located near where
the particles with the most angular momentum in a given inflow are
at their distance of closest approach to the galactic center. For
simplicity, we study caustic rings which are axially symmetric
about the $z$-direction, as well as reflection symmetric with
respect to the $z=0$ plane. The dark matter flow is then
effectively $2$-dimensional. Although the flow is in 3D space, in
general, the dimension along the tube is irrelevant as far as the
caustic properties are concerned. Particles actually move in this
direction, however, the motion is a simple rotation along the
azimuthal direction. To discover the essential properties of the
caustic tubes, it is sufficient to consider the cross-section
perpendicular to the trivial direction. Thus, the flow is {\it in
effect} two dimensional, even in the absence of axial symmetry.
The essential properties of caustics are invariant under
continuous deformations. To proceed further, we assume axial
symmetry of the flow in this paper. We throw away the irrelevant
$\varphi$ coordinate and choose $\rho$ and $z$ as the cylindrical
coordinates of the particles. In galactocentric cylindrical
coordinates, the flow at such a caustic ring is described
\cite{sing} by: \be
\rho&=&a+\frac{u}{2}(\tau-\tau_0)^2-\frac{s}{2}\alpha^2\\
z&=&b\tau\alpha\; , \ee to lowest order in an expansion in powers
of $\tau$ and $\alpha$; see Ref. \cite{Thesis} for a detailed
derivation. The parameters $\tau$ and $\alpha$ label the particles
in the flow. $\rho$ is distance to the $z$-axis. $\tau$ is the
time a particle crosses the $z=0$ plane. $\alpha$ is the
declination of the particle, relative to the $z=0$ plane, when it
was at last turnaround. The constants $a, b, s, u$ and $\tau_0$
are characteristic of the caustic ring.

In this paper, we first make the following reparametrization:
\beeq {\chi_1}=\sqrt{\frac{u}{2}}\,(\tau_0-\tau)\;,\;\;
{\chi_2}=\sqrt{\frac{s}{2}}\,\alpha\; ,\label{repar} \eneq so that
\beeq \rho=a+{\chi_1}^2-{\chi_2}^2\label{flownewr}\eneq \beeq
z=2\zeta\,(\sqrt{p}-{\chi_1}){\chi_2} \; ,\label{flownewz} \eneq
where $\zeta\equiv\frac{b}{\sqrt{us}}$. These reparameterized flow
equations allow us to obtain caustic equations in terms of single
valued functions $\rho(\psi)$ and $z(\psi)$ which are useful in
studying the Differential Geometry (Sect. \ref{difgeo}),
Gravitational Lensing (Sect. \ref{sec:Lensing}) and Catastrophe
Theory (Sect. \ref{Stability}) of the caustic rings.

Let us obtain the number density $d(\rho, z, t)$ of the CDM flow,
near a caustic ring at time $t$. The total number of
particles\beeq N=\int\rho d\rho dz 2\pi d(\rho, z, t)\label{d}\;
,\eneq can also be written in parameter space as\beeq N=\int
d{\chi_1} d{\chi_2} d\phi \frac{d^3N}{d{\chi_1} d{\chi_2}
d\phi}({\chi_1}, {\chi_2}, \phi)=\int\!\! d{\chi_1} d{\chi_2}
\frac{d^2N}{d{\chi_1} d{\chi_2}}({\chi_1}, {\chi_2})\;\;\;
.\label{2.8}\eneq In the last equation, changing variables
$({\chi_1}, {\chi_2})\rightarrow(\rho, z)$, and using the
transformation identity  \beeq\frac{d^2N}{d{\chi_1}
d{\chi_2}}(\chi_1, \chi_2)\rightarrow \sum_{j=1}^{n(\rho,\; z,\;
t)}\left[\frac{d^2N}{d{\chi_1}
d{\chi_2}}(\chi_1,\chi_2)\left|\det\left(\frac{\dd(\chi_1,\chi_2)}{\dd(\rho,
z)}\right)\right|\,\right]\Bigg|_{({\chi_1},\;
 {\chi_2})=({\chi_1},\;{\chi_2})_j(\rho,\; z,\; t)}\; ,\eneq we obtain
\beeq N=\!\int\! d\rho dz\!\sum_{j=1}^{n(\rho, z,
t)}\!\frac{d^2N}{d{\chi_1} d{\chi_2}}({\chi_1}(\rho, z,
t),\;{\chi_2}(\rho, z, t))_j\frac{1}{|D_2({\chi_1},
{\chi_2})|}\Bigg|_{({\chi_1},\;
 {\chi_2})=({\chi_1},\;{\chi_2})_j(\rho,\; z,\; t)}\; .\label{2.8a}\eneq
Here, we used the facts that the determinant of the Jacobian of
the transformation $({\chi_1}, {\chi_2})\rightarrow(\rho, z)$ is
equal to the reciprocal of the Jacobian of the inverse
transformation \beeq D_2({\chi_1},
{\chi_2})=\det\left(\frac{\partial(\rho, z)}{\partial({\chi_1},
{\chi_2})}\right)\label{2.9}\; ,\eneq and $({\chi_1},
{\chi_2})_j$, with $j = 1.\; .\; .n$, are the solutions of $\rho =
\rho ({\chi_1}, {\chi_2}; t)$ and $z= z({\chi_1}, {\chi_2}; t)$.
The number of distinct solutions (flows) $n$ is a function of the
location in space $\rho$, $z$, and time $t$. Comparing Eqs.
\ref{d} and \ref{2.8a}, we find the number density of particles in
physical space as
\begin{equation}
d(\rho,z,t) = {1\over 2\pi\rho}~\sum_{j=1}^{n(\rho,\; z,\;
t)}~{d^2N\over d{\chi_1} d{\chi_2}}~(({\chi_1},
{\chi_2})_j(\rho,\; z,\; t))~{1\over\mid D_2({\chi_1},
{\chi_2})\mid}\Biggl|_{({\chi_1}, {\chi_2}) = ({\chi_1},
{\chi_2})_j(\rho,\; z,\; t)}\; , \label{dens}
\end{equation}
where the determinant $D_2({\chi_1},{\chi_2})$ of the Jacobian
matrix \be
{\cal{D}}({\chi_1},{\chi_2})\equiv\left(\begin{array}{cc}
{\dd\rho\over\dd{\chi_1}}&~~~{\dd \rho\over\dd{\chi_2}}\\
\\
{\dd z\over\dd{\chi_1}}&~~~ {\dd z\over\dd{\chi_2}}\\
\end{array}
\right)=2\, \left(\begin{array}{cc}
{{\chi_1}}&~~~{-{\chi_2}}\\
\\
{-\zeta{\chi_2}}&~~~ {\zeta(\sqrt{p}-{\chi_1})}\\
\end{array}
\right) \; ,\label{Jaco} \ee is \be  D_2 ({\chi_1},{\chi_2})=
-4\zeta\left[\left({\chi_1}-\frac{\sqrt{p}}{2}\right)^2+{\chi_2}^2-\frac{p}{4}\right]\;
.\label{Det} \ee Caustics occur wherever $D_2$ vanishes, i.e.
wherever the map $(\chi_1, \chi_2)\rightarrow(\rho, z)$ is
singular. Hence, at the caustic, in the limit of zero velocity
dispersion $(\delta v=0)$, the density diverges. The divergence is
cut off, if the $\delta v$ is nonzero, because, the location of
the caustic surface gets smeared over some distance $\d x$ in
space (Eqs. \ref{dxa}-\ref{dxW2}).

Now, to obtain the single valued caustic equations, notice that,
in Eq. \ref{Det}, the curves for which $D_2=0$ are circles (for
$p\neq 0$) centered at $({\chi_1} , {\chi_2})=(\frac{\sqrt{p}}{2}
, 0)$ with radius $\frac{\sqrt{p}}{2}$. (Here we assume that
${\chi_1}$ and ${\chi_2}$ are Cartesian coordinates, if not, the
critical curves are ellipses for $p\neq 0$). The critical circles
degenerate into the singular point $({\chi_1} , {\chi_2})=(0 , 0)$
in the limit $p=0$. The parameter representations of the critical
circles can be given as: \be
{\chi_1}=\frac{\sqrt{p}}{2}\left(1\pm\cos\psi\right)\; ,
\hspace{1cm}{\chi_2}=\frac{\sqrt{p}}{2}\sin\psi\;
,\label{causticpar}\ee where the angular variable $\psi\in [0,
2\pi]$. Since the CDM density diverges where $D_2$ vanishes, by
substituting Eq. \ref{causticpar} into Eqs.
\ref{flownewr}-\ref{flownewz}, we find parametric equations
describing the cross-section of the caustic surface in the $(\rho
, z)$-plane: \be
\rho(\psi)&=&a+\frac{p}{2}\cos\psi(\cos\psi\pm 1)\label{rpsi}\\
z(\psi)&=&\zeta\frac{p}{2}\sin\psi(1\mp\cos\psi)\; . \label{zpsi}
\ee The parameter $a$ is the radius of the caustic ring and $p$ is
the longitudinal dimension of the caustic cross-section. The
parameter $\zeta$, as is shown in the below, is proportional to
the ratio of the latitudinal dimension $q$ to the longitudinal
dimension $p$: $\zeta=\frac{4}{3\sqrt{3}}\frac{q}{p}$. Figure
\ref{fig:fig6} shows a plot of the cross-section, which we call
tricusp.

The tangent vector field on the cross-section \be\vec{{\rm
t}}=\frac{\dd}{\dd\psi}(\rho(\psi)\hat x +
z(\psi)\hat{z})=-\frac{p}{2}\Bigg[\Big(\sin2\psi\pm\sin\psi\Big)\hat
x -\zeta\Big(\cos\psi\mp\cos2\psi\Big)\hat
z\Bigg]\label{tangent}\; ,\ee vanishes at $\psi=0, \frac{2\pi}{3},
\frac{4\pi}{3}$ ($\psi=\frac{\pi}{3}, \pi, \frac{5\pi}{3}$) for
the upper (lower) sign choice of ${\chi_1}$. At these values,
where both $\frac{\dd \rho}{\dd\psi}$ and $\frac{\dd z}{\dd\psi}$
simultaneously vanish, caustic has dual cusps; see also Sect.
\ref{Stability}. Thus, the reason for the existence of all three
of the cusps is the same: simultaneous vanishing of $\frac{\dd
\rho}{\dd\psi}$ and $\frac{\dd z}{\dd\psi}$. This point can not be
seen by the parametrization of Ref. \cite{sing} because, the
tricusp obtained there is made up of two separate pieces on top of
each other. As the parameter $\tau$ ranges in the interval $[0,
\tau_0]$ the functions $\rho(\tau)$ and $z(\tau)$ of \cite{sing}
describing the caustic cross-section plots only the half of the
tricusp. The other half is obtained by the reflection symmetry of
the flow equations, or, i.e. by the double valuedness of the
caustic equation $z(\tau)$. Neither $\rho(\tau)$, nor $z(\tau)$
traverses through the outer cusp. By the new parametrization, on
the other hand, as the angular parameter $\psi$ ranges in the
interval $[0, 2\pi]$, the functions $\rho(\psi)$ and $z(\psi)$
(Eqs. \ref{rpsi}-\ref{zpsi}) plot the whole tricusp as a single
piece, traversing through all the cusps, including the outer one.

Notice also that, the vanishings of both the determinant $D_2$
(Eq. \ref{Det}) and tangent vector $\vec{t}$ (Eq. \ref{tangent})
at three specific values of $\psi$, are independent of $\zeta$.
This means that, the existence of the caustic rings is independent
of the ratio of transverse dimensions $p$ and $q$ (i.e. the
parameter $\zeta$), and the number of cusps is always three. For
$\zeta=1$, the map $\psi\rightarrow \psi+\frac{2\pi}{3}$
transforms the three cusps of the caustic into one another. Thus,
in this case, the tricusp has a $Z_3$ symmetry. In the language of
Catastrophe Theory, the tri-axial tricusp, where the three cusps
make up an equilateral triangle ($\zeta=1$), is a $D_{-4}$
catastrophe. To avoid the clutter of the sign choices, we adopt
the upper sign conventions in Eqs. \ref{rpsi}-\ref{zpsi} and
use\be\rho(\psi)=a+\frac{p}{2}\cos\psi(1+\cos\psi)\; ,
\hspace{1cm}z(\psi)=\zeta\frac{p}{2}\sin\psi(1-\cos\psi)\;
,\label{rhpsi}\ee in the rest of the paper. Since both
$\rho(\psi)$ and $z(\psi)$ are extremized for $\psi=0,
\frac{2\pi}{3}, \frac{4\pi}{3}$, we can deduce the dimensions of
the tricusp by evaluating $\rho(\psi)$ and $z(\psi)$ at those
locations: $(\rho(0), z(0))= (a+p, 0)$, $(\rho(\frac{2\pi}{3}),
z(\frac{2\pi}{3}))= (a-\frac{p}{8}, \zeta\frac{3\sqrt{3}}{8}p)$,
and $(\rho(\frac{4\pi}{3}),z(\frac{4\pi}{3}))= (a-\frac{p}{8},
-\zeta\frac{3\sqrt{3}}{8}p)$. Thus, as is mentioned before, the
dimensions of the cross section in the $\hat{\rho}$ and $\hat{z}$
directions are $p$ and $q= \zeta{3\sqrt{3}\over 4} p$,
respectively. In terms of $p$ and $q$, the locations of the three
cusps in physical $(\rho , z)$-plane are therefore: $(a+p, 0)$,
$(a-\frac{p}{8} , \frac{q}{2})$, and $(a-\frac{p}{8} , -
\frac{q}{2})$.

In physical space, caustics are located where the number of flows
changes by two. Inside (outside) the tricusp, there are four (two)
flows. To count the distinct flows, let us restrict ourselves to
the $z=0$ plane, where, from Eq. \ref{flownewz}, we either have
(i) ${\chi_1} = \sqrt{p}$, or, (ii) ${\chi_2} = 0$. If (i) is the
case, then Eq. \ref{flownewr} gives $\rho =\rho_0
-{\chi_2}^2<\rho_0$. Therefore, at least two distinct flows exist
in the region $\rho<\rho_0$. They are parameterized as $({\chi_1},
{\chi_2})= (\sqrt{p}, \pm\sqrt{\rho_0 -\rho})$. For both of the
flows, the velocities in the $\hat{\rho}$ direction are the same:
\be \frac{\dd \rho}{\dd t}\sim -\frac{\dd
\rho}{\dd\tau}=\sqrt{2u}{\chi_1}=\sqrt{2up}\; .\ee We use $\sim$
because the flow is not stationary, (if one assumes that the flow
is stationary, then $\sim$ is replaced by $=$). The velocities of
the flows in the $\hat{z}$ direction \be \frac{\dd z}{\dd t}\sim
-\frac{\dd z}{\dd\tau}=-\sqrt{2u}\zeta{\chi_2}\; ,\ee are opposite
to each other.
 For the flow $({\chi_1},
{\chi_2})=(\sqrt{p}, \sqrt{\rho_0-\rho})$, $\frac{\dd z}{\dd
t}=-\zeta\sqrt{2u(\rho_0 -\rho)}<0$, thus we call it the
``down-flow.'' For the flow $({\chi_1}, {\chi_2})=(\sqrt{p},
-\sqrt{\rho_0-\rho})$, on the other hand, $\frac{\dd z}{\dd
t}=\zeta\sqrt{2u(\rho_0 -\rho)}>0$, thus we call it the
``up-flow.'' From Eq. \ref{dens}, we know that the contribution of
each of the flows to the density is inversely proportional to the
absolute value of $D_2=4\zeta(\rho-\rho_0)=4\zeta(\rho-a-p)$. Near
$\rho=a$, as $\rho-a\rightarrow 0_\pm$, $D_2\rightarrow -4\zeta
p$. Therefore, near $\rho=a$, each of the contributions of the
down and up flows is finite: $d\propto \frac{1}{4\zeta p}$. If, on
the other hand, one approaches to the outer cusp from the inside
of the tricusp, as $\rho -\rho_0\rightarrow 0_-$, (recall that,
the down and up flows exist only where $\rho<\rho_0$), the
densities of the up and down flows diverge:
$d\sim\frac{1}{\rho_0-\rho}$. If (ii) is the case, then Eq.
\ref{flownewr} gives $\rho =a+ {\chi_1}^2>a$. Therefore, in the
region $\rho>a$, at least two distinct flows exist. They are
parameterized as $({\chi_1}, {\chi_2})= (\pm\sqrt{\rho -a}, 0)$.
For both of the flows, the velocities in the $\hat{z}$ direction
are zero: \be \frac{\dd z}{\dd t}\sim -\frac{\dd
z}{\dd\tau}=-\sqrt{2u}\zeta{\chi_2}=0\; .\ee The velocities of the
flows in the $\hat{\rho}$ direction \be \frac{\dd \rho}{\dd t}\sim
-\frac{\dd \rho}{\dd\tau}=\sqrt{2u}{\chi_1}\; ,\ee are opposite to
each other.
 For the flow $({\chi_1},
{\chi_2})=(\sqrt{\rho-a}, 0)$, $\frac{\dd \rho}{\dd
t}=\sqrt{2u(\rho -a)}>0$, thus, we call it the ``out-flow.'' For
the flow $({\chi_1}, {\chi_2})=(-\sqrt{\rho-a}, 0)$, on the other
hand, $\frac{\dd \rho}{\dd t}=-\sqrt{2u(\rho -a)}<0$, thus, we
call it the ``in-flow.'' For in and out flows, $D_2=-4\zeta(\rho-
a\mp \sqrt{p(\rho-a)})$, respectively. Near $\rho=a$, as
$\rho-a\rightarrow 0_+$ (recall that, the out and in flows exist
only where $\rho>a$), $D_2\rightarrow \pm 4\zeta\sqrt{p(\rho-a)}$.
Therefore, near $\rho=a$, the densities for the out and in flows
diverge: $d\sim\frac{1}{\sqrt{\rho -a}}$.

Next, let us analyze the density properties near the dual cusp,
where the behavior depends upon the direction of approach. For
$z=0$ and $\rho - \rho_0 \rightarrow 0_-$ (inside the caustic
tube), there are four flows: in, out, up and down. For $\rho -
\rho_0 \rightarrow 0_+$ (outside the caustic tube), there are two
flows: in and out. Here, let us use this opportunity to point out
that the density of the out-flow diverges in {\it both} sides of
the cusp, whereas that of the in-flow remains finite. To see this,
let us express the radial coordinate as $\rho=a+p+\epsilon$, where
$\epsilon\rightarrow 0_\pm$, depending on the side we approach to
the cusp. Therefore, we can express the determinant as $D_2\simeq
-4\zeta [p+\epsilon\mp p(1+\frac{\e}{2p}) ]$, for the out and in
flows, respectively. Thus, for the out flow, $D_2\sim -2\zeta\e$,
and hence the density diverges, $d\sim\frac{1}{\mid \e\mid}$, in
both sides of the cusp. For the in flow $D_2\sim -8\zeta p$, thus,
the density of the in flow remains finite in the both sides.
Notice that, near the cusp, unlike the region $\rho-a<0$ where
both of the down and up flows remain finite as one approaches the
fold caustic from the outside, there is a flow (the out flow)
whose density diverges as one approaches the caustic tube from the
outside of the cusp. Hence, the cusps are more divergent then the
folds.

Finally, for the sake of completeness, let us calculate the
density near the cusp, again at $\rho = \rho_0$, but now, as
$z\rightarrow 0_\pm$, in terms of our new parameters. Since $\rho=
a+{\chi_1}^2-{\chi_2}^2$, at $\rho=\rho_0=a+p$, we have \beeq
{\chi_2}^2=({\chi_1}+\sqrt{p})({\chi_1}-\sqrt{p})\;
.\label{sifir}\eneq Using Eqs. \ref{sifir} and \ref{flownewz}, the
determinant $D_2$ (Eq. \ref{Det}) can be written as\beeq
D_2=-4\zeta\left[2({\chi_1}-\sqrt{p})^2+3\sqrt{p}({\chi_1}-\sqrt{p})\right]
=-\frac{2}{\zeta}\left(\frac{z}{{\chi_2}}\right)^2+6\sqrt{p}\frac{z}{{\chi_2}}\;
\label{D2z}. \eneq We also infer from Eqs. \ref{flownewz} and
\ref{sifir} that, near the galactic plane, at $\rho=\rho_0$,
$z\rightarrow-\frac{\zeta{\chi_2}^3}{\sqrt{p}}\rightarrow 0$,
because the parameters $(\chi_1, \chi_2)\rightarrow (\sqrt{p}, 0)$
there. Therefore, in this limit, the second order term
$(\frac{z}{{\chi_2}})^2$ in Eq. \ref{D2z} can be neglected next to
the first order term $\frac{z}{{\chi_2}}$. Hence, we have
$D_2\simeq-6 \left[\zeta p z^2\right]^{1/3}$ for {\it one} of the
flows, and the density $d\sim {1\over\mid z\mid^{2/3}}$ for this
flow. Thus, the density also diverges in this particular limit
near the cusp.

Note also that, the tube caustic collapses to a line caustic in
the limit $p\rightarrow 0$ with $\zeta$ fixed.  In this limit, Eq.
\ref{Det} gives
$D_2=-4\zeta\sqrt{{\chi_1}^4+{\chi_2}^4+2{\chi_1}^2{\chi_2}^2}
=-4\sqrt{\zeta^2(\rho-a)^2+z^2}$ . Hence, the density \beeq
d(\rho, z) = {1\over 8\pi\rho}~{d^2N\over d{\chi_1} d{\chi_2}}
~{1\over\sqrt{\zeta^2(\rho -a)^2 + z^2}}\; ,\label{4.17}\eneq in
the limit $p\rightarrow 0$.
\section{Density at a Caustic Ring}
\label{Density} In the previous section, we have restricted
ourselves to the $\rho$-axis and worked out the density of the
flows near the points $\rho=a$ and $\rho=\rho_0$. They are the
points where the parameters that describe the flow, cross the
degenerate critical circle $\chi_1^2+\chi_2^2-\sqrt{p}\chi_1=0$ on
the $\rho$-axis (see also Sect. \ref{Stability}). These locations
are interesting, because, as we will see in Sect. \ref{Stability},
on the $\rho$-axis at $\rho=a$ a simple fold ($A_2$) catastrophe,
and at $\rho=\rho_0$ a dual cusp ($A_{-3}$) catastrophe,
necessarily occur. In this section, using the equations that
describe the caustic ring in terms of the angular parameter
$\psi$, and following the procedure given in Ref. \cite{lensing},
we derive the density profile near the caustic surface. In Sect.
\ref{subsec:foldcoef}, we derive a density formula valid
everywhere except at the cusps (our mathematical formulation is
invalid at the cusps). In the limit of zero velocity dispersion
($\delta v=0$), the density diverges when one approaches a caustic
surface, on the side which has two extra flows, as the inverse
square root of the distance to the surface. If the velocity
dispersion is small, but  nonzero, this divergence is cut off,
because, the location of the caustic gets smeared out. So the
density at the caustic is no longer infinite, but merely very
large. In Sect. \ref{subsec:densitynearacusp}, we deal with the
density near the cusps. Our results are applied in Sect.
\ref{sec:Lensing}, to derive the gravitational lensing signatures
of the caustic rings.

\subsection{Density everywhere except the cusps}
\label{subsec:foldcoef}

\begin{figure}[ht] \centering
\includegraphics[height=6.5cm,width=13cm]{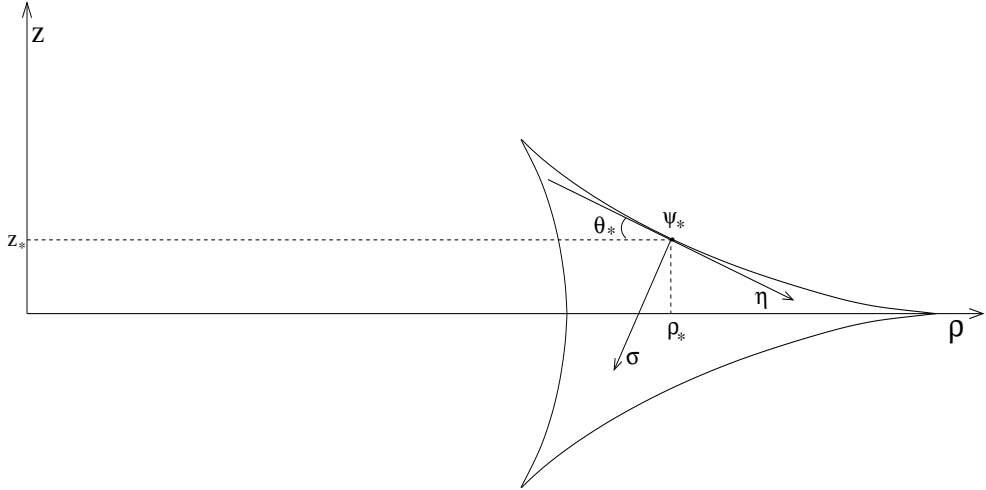}
\caption{An arbitrary point on the tricusp is labeled by $\psi_*$.
Its physical coordinates are $(\rho_*, z_*)$. A new Cartesian
coordinate system ($\sigma , \eta$) is defined there such that
$\hat{\sigma}$ is perpendicular to the caustic surface. It is
rotated relative to the $(\rho, z)$ coordinates by an angle
$\theta(\psi_*)+{{\pi}\over 2}$.} \label{fig:tricusp1}
\end{figure}

We choose an arbitrary point on the surface of the caustic ring
parameterized by $({\chi_1}(\psi_*), {\chi_2}(\psi_*))$ where
${\chi_1}(\psi)$ and ${\chi_2}(\psi)$ are given by Eq.
\ref{causticpar}. We assume that the point is not at one of the
three dual cusps located at $\psi=0, \frac{2\pi}{3}$, and
$\frac{4\pi}{3}$, and use $\psi_*$ to indicate this exclusion.
Whenever $\psi$, without a subscript ``$*$'' is used, the dual
cusps are included. The physical space coordinates $(\rho_*,
z_*)\equiv(\rho(\psi_*),z(\psi_*))$ are given in terms of $\psi_*$
by Eqs. \ref{rhpsi}. The vanishing of $D_2(\psi_*)=\det {\cal
D}(\psi_*)$ implies the existence of a zero eigenvector of the
matrix ${\cal D}(\psi_*)$ and a zero eigenvector of its transpose
\be {\cal D}^T(\psi_*)=2\left(\begin{array}{cc}
{{\chi_1}(\psi_*)}&{-\zeta{\chi_2}(\psi_*)}\\
{-{\chi_2}(\psi_*)}& {\zeta(\sqrt{p}-{\chi_1}(\psi_*))}\\
\end{array}
\right)\; . \label{trans}\ee Let us define
$\theta_*\equiv\theta(\psi_*)$ such that \be {\cal
D}^T(\psi_*)\left(\begin{array}{c}
{\sin{\theta_*}}\\
{\cos{\theta_*}}\\
\end{array}
\right)=0\; . \label{dt1} \ee Hence, we have \beeq
{\chi_1}(\psi_*)\sin{\theta_*}-\zeta{\chi_2}(\psi_*)\cos{\theta_*}=0
\label{bir}\eneq \beeq
-{\chi_2}(\psi_*)\sin{\theta_*}+\zeta(\sqrt{p}-{\chi_1}(\psi_*))\cos{\theta_*}=0
\; .\label{iki}\eneq Equations \ref{bir} and \ref{iki} imply \be
\!\!\!\!\!\!\!\!\!\!\sin{\theta_*}\!\!&=&\!\!\frac{\pm\zeta|{\chi_2}(\psi_*)|}
{\sqrt{{\chi_1}(\psi_*)^2+\zeta^2{\chi_2}(\psi_*)^2}}
=\frac{\pm\zeta(\sqrt{p}-{\chi_1}(\psi_*))}{\sqrt{{\chi_2}(\psi_*)^2
+\zeta^2(\sqrt{p}-{\chi_1}(\psi_*))^2}}
=\frac{\pm\zeta}{\sqrt{\zeta^2 + \cot^2\frac{\psi_*}{2}}}\label{cossin1}\\
\!\!\!\!\!\!\!\!\!\!\cos{\theta_*}\!\!&=&\!\!\frac{\pm{\chi_1}(\psi_*)}{\sqrt{{\chi_1}(\psi_*)^2
+\zeta^2{\chi_2}(\psi_*)^2}}
=\frac{\pm|{\chi_2}(\psi_*)|}{\sqrt{{\chi_2}(\psi_*)^2+\zeta^2(\sqrt{p}-{\chi_1}(\psi_*))^2}}
=\frac{\pm 1}{\sqrt{1+\zeta^2\tan^2\frac{\psi_*}{2}}} \;
.\label{cossin2}\ee Now, let us define a new pair of Cartesian
coordinates $(\eta, \sigma)$ related to $(\rho - \rho_*, z - z_*)$
by a rotation of angle $\theta_*+{\pi\over 2}$: \be
\left(\begin{array}{c}
{\eta}\\
{\sigma}\\
\end{array}
\right)=  \left(\begin{array}{cc}
{\cos{\theta_*}}&{-\sin{\theta_*}}\\
{-\sin{\theta_*}}&{-\cos{\theta_*}}\\
\end{array}
\right)\left(\begin{array}{c}
{\rho-\rho_*}\\
{z -z_*}\\
\end{array}
\right)\; . \label{mateq} \ee We now show that $\sigma$ is the
coordinate in the direction orthogonal to the caustic surface at
$(\rho_*, z_*)$. Consider small deviations about
$({\chi_1}(\psi_*) , {\chi_2}(\psi_*))$ in parameter space:
$({\chi_1},{\chi_2})=({\chi_1}(\psi_*)+\Delta {\chi_1},
{\chi_2}(\psi_*)+\Delta {\chi_2})$. Equations \ref{flownewr},
\ref{flownewz} and \ref{Jaco} imply \be \left(\begin{array}{c}
{{\Delta\rho}}\\
{{\Delta z}}\\
\end{array}
\right)= {\cal D}(\psi_*)\left(\begin{array}{c}
{{\Delta{\chi_1}}}\\
{{\Delta{\chi_2}}}\\
\end{array}
\right)+O(\Delta{\chi_1}^2 , \Delta{\chi_2}^2 ,
\Delta{\chi_1}\Delta{\chi_2})\; , \label{DrDz} \ee where \be
\!\!\!\!\!\!\Delta\rho\!\!&\equiv&\!\!\rho-\rho_*=2({\chi_1}(\psi_*)\D{\chi_1}
-{\chi_2}(\psi_*)\D{\chi_2})+O(\Delta{\chi_1}^2 , \Delta{\chi_2}^2
, \Delta{\chi_1}\Delta{\chi_2})\;
\nonumber\\
\!\!\!\!\!\!\Delta z\!\!&\equiv&\!\!
z-z_*=2\zeta(-{\chi_2}(\psi_*)\D{\chi_1}+(\sqrt{p}-{\chi_1}(\psi_*))\D{\chi_2})+O(\Delta{\chi_1}^2
, \Delta{\chi_2}^2 , \Delta{\chi_1}\Delta{\chi_2})\label{delta}\;
.\ee The expansion of $\sigma$, defined by Eq. \ref{mateq}, in
powers of $\Delta {\chi_1}$ and $\Delta {\chi_2}$ yields \beeq
\sigma=O(\Delta{\chi_1}^2 , \Delta{\chi_2}^2 ,
\Delta{\chi_1}\Delta{\chi_2})\; , \eneq because, due to Eq.
\ref{dt1}, the first order terms vanish. The fact that $\sigma$ is
second order in $\Delta{\chi_1}$ and $\Delta{\chi_2}$ shows that
$\sigma$ is the coordinate in the direction perpendicular to the
caustic surface. This can be seen as follows. The scalar product
of the tangent vector $d{\rm\vec{t}}=\hat\rho d\rho+\hat{z}dz$ and
the normal vector $\hat{n}=\hat{\rho}n_\rho+\hat{z}n_z$ at
$\psi_*$ is zero. Thus, at $\psi_*$, using Eq. \ref{DrDz}, in
first order, we have\be d{\rm\vec{t}}\cdot\hat{n}=(\D \rho\;\;\;\D
z)\left(\begin{array}{c}
{n_\rho}\\
{n_z}\\
\end{array}
\right)=(\D \chi_1\;\;\;\D
\chi_2)D^T(\psi_*)\left(\begin{array}{c} {n_\rho}\\
{n_z}\\
\end{array}
\right)=0\; ,\ee which implies $D^T(\psi_*)\hat{n}=0$. Recall,
however, that we defined the zero eigenvector, and hence
$\theta(\psi_*)$, such that Eq. \ref{dt1} is satisfied. Therefore,
the zero eigenvector must be the normal vector
$\hat{n}=\hat\rho\sin\theta_*+\hat{z}\cos\theta_*$, where
$\theta_*$ is the angle between the $\rho$-axis and the tangent
vector at $(\rho_*, z_*)$; see Fig.
\ref{fig:tricusp1}. Thus, in
first order, $\hat\sigma$ is parallel to the normal. We choose its
direction pointing inward, therefore, at the given $\psi_*$ of
Fig. \ref{fig:tricusp1} we have \beeq
\hat{\sigma}=-\hat{\rho}\sin{\theta_*} -\hat{z}\cos{\theta_*} \;
.\label{s} \eneq To obtain the density profile of the caustic ring
near the point under consideration, we need $D_2(\eta, \sigma)$ to
order $\sqrt{\sigma}$. So, we calculate $\sigma$ to second order
in powers of $\Delta {\chi_1}$ and $\Delta {\chi_2}$, and $D_2$
and $\eta$ to first order.  Using Eq. \ref{mateq} in the
expansion: \beeq \sigma=\frac{1}{2}\left[\frac{\partial^2
\sigma}{\partial{\chi_1}^2}\Big|_{\psi_*}\Delta{\chi_1}^2
+2\frac{\partial^2\sigma}{\partial{\chi_1}\partial{\chi_2}}\Big|_{\psi_*}
\Delta{\chi_1}\Delta{\chi_2}
+\frac{\partial^2\sigma}{\partial{\chi_2}^2}\Big|_{\psi_*}\Delta{\chi_2}^2\right]\;
, \eneq we find \be \sigma=(\Delta{\chi_1}\,\,\,\,\Delta{\chi_2})
\left(\begin{array}{cc}
{-\sin{\theta_*}}&{\zeta\cos{\theta_*}}\\
{\zeta\cos{\theta_*}}&{\sin{\theta_*}}\\
\end{array}\right)\left(\begin{array}{c}
{{\Delta{\chi_1}}}\\
{{\Delta{\chi_2}}}\\
\end{array} \right)\; .
\label{y'} \ee Equation \ref{Det} yields\be D_2&=&\frac{\partial
D_2}{\partial{\chi_1}}\Big |_{\psi_*}\Delta{\chi_1}+\frac{\partial
D_2}{\partial{\chi_2}}\Big |_{\psi_*}\Delta{\chi_2} +
O(\Delta{\chi_1}^2 , \Delta{\chi_2}^2 ,
\Delta{\chi_1}\Delta{\chi_2})\nonumber\\
&=&-8\zeta\left[\left({\chi_1}(\psi_*)-\frac{\sqrt{p}}{2}\right)\D{\chi_1}
+{\chi_2}(\psi_*)\D{\chi_2}\right]+O(\Delta{\chi_1}^2 ,
\Delta{\chi_2}^2 , \Delta{\chi_1}\Delta{\chi_2})\;
.\label{newD}\ee To find $\eta$, we use Eqs. \ref{delta} in Eq.
\ref{mateq}: \be
\!\!\!\!\!\!\!\!\eta\!\!&=&\!\!\frac{\partial\eta}{\partial{\chi_1}}\Big
|_{\psi_*}\!\Delta{\chi_1}+\frac{\partial\eta}{\partial{\chi_2}}\Big
|_{\psi_*}\!\Delta{\chi_2} + O(\Delta{\chi_1}^2 ,
\Delta{\chi_2}^2 , \Delta{\chi_1}\Delta{\chi_2})\nonumber\\
\!\!\!\!\!\!\!\!\!\!&=&\!\!2[{\chi_1}(\psi_*) \cos{\theta_*}\!+\!
\zeta {\chi_2}(\psi_*) \sin{\theta_*}] \Delta {\chi_1}\!-\!2
[{\chi_2}(\psi_*)\cos{\theta_*}\!+\!\zeta(\sqrt{p}-\!{\chi_1}(\psi_*))
\sin{\theta_*}] \Delta {\chi_2}\; . \label{D2e} \ee Equation
\ref{D2e} can be recast using Eqs. \ref{cossin1} and \ref{cossin2}
as \beeq
\eta=2\sqrt{{\chi_1}^2(\psi_*)+\zeta^2{{\chi_2}^2}(\psi_*)}\Delta{\chi_1}
-
2\sqrt{{\chi_2}^2(\psi_*)+\zeta^2(\sqrt{p}-{\chi_1}(\psi_*))^2}\Delta{\chi_2}
\; . \label{eta}\eneq The fact that $D_2(\psi_*)=0$ implies
$\sqrt{p}-{\chi_1}(\psi_*)={\chi_2}^2(\psi_*)/\chi_1(\psi_*)$. We
can, therefore, write Eqs. \ref{newD} and \ref{eta}, at any
$\psi_*$, as a matrix equation: \be \left(\begin{array}{c}
{{-\frac{D_2}{8\zeta}}}\\
\\
{{\frac{\eta{\chi_1}(\psi_*)}{2\sqrt{{\chi_1}^2(\psi_*)+\zeta^2{\chi_2}^2(\psi_*)}}}}\\
\end{array} \right)= \left(\begin{array}{cc}
{{\chi_1}(\psi_*)\!-\!\frac{\sqrt{p}}{2}}&{|{\chi_2}(\psi_*)|}\\
\\
{{\chi_1}(\psi_*)}&{-|{\chi_2}(\psi_*)|}\\
\end{array}\right)\left(\begin{array}{c}
{{\Delta{\chi_1}}}\\
\\
{{\Delta{\chi_2}}}\\
\end{array} \right)\; . \label{matrix}\ee
Equation \ref{matrix} can be inverted to obtain $\Delta{\chi_1}$
and $\Delta{\chi_2}$ as functions of $D_2$ and $\eta$. When the
result is inserted into Eq. \ref{y'}, we obtain \beeq
\sigma(D_2,\eta)=\frac{\zeta{\chi_1}(\psi_*)\left(\frac{D_2^2}{4\zeta^2}
-\frac{\sqrt{p}\,{\chi_1}(\psi_*)\eta^2}{{\chi_1}^2(\psi_*)
+\zeta^2{\chi_2}^2(\psi_*)}\right)}{16\left({\chi_1}(\psi_*)
-\frac{\sqrt{p}}{4}\right)|{\chi_2}(\psi_*)|\sqrt{{\chi_1}^2(\psi_*)
+\zeta^2{\chi_2}^2(\psi_*)}} \;\, . \eneq This gives \be
|D_2(\eta,
\sigma)|\!=\!\!\sqrt{\!8\zeta\Bigg[p\sqrt{(1\!+\!\cos\!\psi_*)^2\!+\!\zeta^2\!\sin^2\!\psi_*}
\left|(1\!+\!2\cos\!\psi_*)\!\tan\!\frac{\psi_*}{2}\right|\sigma
\!+\!\frac{\zeta\eta^2}{1\!+\!\zeta^2\!+\!(1\!-\!\zeta^2)\!\cos\!\psi_*}\Bigg]}
\; . \nonumber\\\ee Along the $\hat{\sigma}$ direction ($\eta=0$),
we have \be |D_2(\sigma)|= 2\sqrt{2\zeta p\, \sigma} C(\psi_*)\; ,
\label{D2sC} \ee where \beeq
C(\psi_*)=\sqrt{\left|(1+2\cos\psi_*)\tan\frac{\psi_*}{2}\right|
\sqrt{(1+\cos\psi_*)^2+\zeta^2\sin^2\psi_*} }\; . \eneq The
derivative of $C(\psi_*)$ implies that, near the caustic surface,
the density decreases as one approaches to the central points
between the cusps where $\psi_*=\frac{\pi}{3},\pi$, and
$\frac{5\pi}{3}$, whereas it increases as one approaches to the
cusps where $\psi=0, \frac{2 \pi}{3}$, and $\frac{4 \pi}{3}$.
Combining Eqs. \ref{dens} and \ref{D2sC}, and minding the factor
of two because two flows contribute, we obtain the density profile
near the surface along the $\hat\sigma$ direction, as \beeq
d(\psi_*, \sigma) = {A(\psi_*) \over \sqrt{\sigma}} \Theta(\sigma)
\eneq with \beeq A(\psi_*)= \frac{d^2M}{d\Omega
dt}\sqrt{\frac{2\zeta}{p}}
\frac{\cos{\alpha(\psi_*)}}{bC(\psi_*)\rho(\psi_*)}\; ,
\label{Apsi}\eneq where we used \beeq \frac{d^2M}{d\chi_1
d\chi_2}=\frac{d^2M}{d\alpha
d\tau}\frac{2}{\sqrt{us}}=\frac{d^2M}{d\alpha
d\tau}\frac{2\zeta}{b}=\frac{d^2M}{d\Omega
dt}\frac{4\pi\zeta}{b}\cos{\alpha(\psi_*)}\; .\label{conversion}
\eneq Here, we define $\frac{d^2M}{d\O dt}=\frac{d^2M}{d\alpha dt
\,2\pi\cos\alpha}$ as the mass falling in, per unit time and unit
solid angle. Note that, $A(\psi)$ diverges at each of the three
dual cusps because $C$ vanishes there. Moreover, one has \beeq
\left.\frac{d^2 M}{d\Omega dt}\right|_n =f_n
v_n\frac{v^2_{\rm{rot}}} {4\pi G} \; , \label{f} \eneq where $v_n$
is the velocity of the particles in the $n$th caustic ring, and
the dimensionless coefficients $f_n$ characterize the density of
the $n$th in and out flows.  In the self-similar model \cite{cr},
we have\beeq \{f_n: n=1, 2,.\; .\; .\}\simeq
(13,~5.5,~3.5,~2.5,~2,.\; .\; .)\cdot 10^{-2} \eneq for
$\epsilon=0.2$. The $\frac{dM}{d\Omega dt}\Big|_n$ does not depend
sharply upon $\epsilon$.  Our estimates are for $\epsilon = 0.2$
because they can most readily be obtained from Ref. \cite{cr} in
that case. Combining Eqs. \ref{Apsi} and \ref{f}, we find \beeq
A_{n}(\psi_*) =\frac{v^2_{ \rm{rot}}}{4\pi
G}\,\frac{f_n}{a_n}\,\frac{v_n}{b_n}\,\sqrt{\frac{2\zeta_n}{p_n}}
\cos{\alpha(\psi_*)}{\mathcal F}_n(\psi_*)\; , \label{Anv2} \eneq
where \beeq {\mathcal
F}_n(\psi_*)=\frac{1}{\frac{\rho_n(\psi_*)}{a_n}C_n(\psi_*)}=
\frac{[1+\frac{p_n}{2a_n}\cos\psi_*(1+\cos\psi_*)]^{-1}}
{\sqrt{|(1+2\cos\psi_*)\tan\frac{\psi_*}{2}|
[(1+\cos\psi_*)^2+\zeta^2_n\sin^2\psi_*]^{\frac{1}{2}}} }\;\;
.\eneq Let us note, as a check, that $C_n(\pi)=\sqrt{2\zeta_n}$,
and hence $A_n(\pi)$ reduce to $A_{0,n}$ of Ref. \cite{lensing},
which are the fold coefficients evaluated at $(\rho, z)=(a_n, 0)$.
(The $A_n$ are minimum at $\psi_*=\frac{\pi}{3}, \pi,
\frac{5\pi}{3}$, they increase as one approaches to the cusps
where $\psi_*=0, \frac{2\pi}{3}, \frac{4\pi}{3})$. It was shown in
Ref. \cite{sing} that $b_n$ and $v_n$ are of the same order of
magnitude. Moreover, the ten rises in the rotation curve of the
Milky Way were interpreted \cite{IRAS} as the effect of caustic
rings. In that case, the widths $p_n$ of caustic rings are
determined from the observed widths of the rises. Typically one
finds $p_n\sim 0.1\, a_n$. Using this, $v_n \sim b_n$ and
$\cos\alpha(\psi_*)\simeq 1$ in Eq. \ref{Anv2},
yields\be\!\!\!\!\{A_n(\psi_*)\!:\! n\!=\!1, 2,.\; .\; .\}\!\sim\!
(5, 6, 6, 7, 8,.\;.\;.)10^{-4}\frac{{\mathcal F}_n(\psi_*)\,{\rm
gr}}{\rm cm^2\,kpc^{\frac{1}{2}}}\!
\left(\frac{0.27}{j_{\rm{max}}}\right)^{\frac{3}{2}}
\!\!\left(\frac{h}{0.7}\right)^{\frac{3}{2}}\!\!
\left(\frac{v_{\rm{rot}}}{220\, {\rm
{km}/{s}}}\right)^{\frac{1}{2}}\; . \label{Anring} \ee Let us now
estimate the fold coefficient more precisely for the fifth caustic
ring of the Milky Way, which is the one closest to us and believed
to be the most constrained by observation. For that ring $a_5=8.31
\,{\rm kpc}$, $b_5=516\, {\rm km/s}$ (if $\tau_0<0$) or 657\,km/s
(if $\tau_0>0$), $p_5=0.134$ kpc, $q_5=0.2$ kpc, $f_5=0.02$,
$v_5=480$ km/s and
$\zeta_5=\frac{4}{3\sqrt{3}}\frac{q_5}{p_5}=1.15$. Therefore, if
$\tau_0<0$ \beeq A_5(\psi_*)=1.76\cdot 10^{-3}\frac{{\mathcal
F}_5(\psi_*)\,{\rm gr}}{\rm cm^2\,kpc^{\frac{1}{2}}}
\left(\frac{v_{\rm{rot}}}{220\, {\rm {km}/{s}}}\right)^2\; ,
\label{A5-1}\eneq or, if $\tau_0>0$ \beeq A_5(\psi_*)=1.38\cdot
10^{-3}\frac{{\mathcal F}_5(\psi_*)\,{\rm gr}}{\rm
cm^2\,kpc^{\frac{1}{2}}} \left(\frac{v_{\rm{rot}}}{220\, {\rm
{km}/{s}}}\right)^2\; , \label{A5-2}\eneq where we have taken
$\cos(\alpha(\psi_*))=1$. If we expand ${\mathcal F}(\psi_*)$ as a
power series around the outer cusp, where $\psi_*=\vartheta$ or
$\psi_*=2\pi-\vartheta$, we find\beeq {\mathcal
F}_n(\vartheta)={\mathcal
F}_n(2\pi-\vartheta)=\frac{\vartheta^{-\frac{1}{2}}}{\sqrt{3}(1+\frac{p_n}{a_n})}
+\frac{[4-\zeta_n^2+(16-\zeta_n^2)\frac{p_n}{a_n}]\vartheta^{\frac{3}{2}}}{16\sqrt{3}
(1+\frac{p_n}{a_n})^2}+O(\vartheta^{\frac{7}{2}})\; .
\label{fnout}\eneq The points $\psi_*=\vartheta$ and
$\psi_*=2\pi-\vartheta$ are located at \be
\rho_n(\vartheta)&=&\rho_n(2\pi-\vartheta)=a_n+p_n(1-\frac{3}{4}\vartheta^2
+\frac{3}{16}\vartheta^4+O(\vartheta^6))\nonumber\\
z_n(\vartheta)&=&-z_n(2\pi-\vartheta)=\zeta_n\frac{p_n}{4}(\vartheta^3
-\frac{\vartheta^5}{4}+O(\vartheta^7))\; . \ee We see that, as
$\psi_*\rightarrow 0$, ${\mathcal F}(\psi_*)$ diverges as
$\psi_*^{-\frac{1}{2}}$ (for ${p_n}= 0.1~ a_n$, ${\mathcal
F}(\vartheta)\simeq 0.525~ \vartheta^{-\frac{1}{2}}$, for ${p_5}=
0.016~ a_5$, ${\mathcal F}_5(\vartheta)\simeq 0.568~
\vartheta^{-\frac{1}{2}}$). However, as is discussed in Sect.
\ref{sec:intro}, the divergence is cut off, because, the location
of the caustic surface gets smeared out over some distance $\delta
x$ (Eqs. \ref{dxa}-\ref{dxW2}) and the cusps are smoothed out.
Recall, however, that these distance scales are estimated taking
only the primordial velocity dispersions (Eqs.
\ref{delvax}-\ref{delvWIMP2}) into account. Therefore, they
provide theoretical lower bounds for $\d x$. CDM flow may have an
effective velocity dispersion $\delta v_{\rm eff}$ which is larger
than its primordial velocity dispersion. Unfortunately, very
little is known about the size of the $\delta v_{\rm eff}$ of CDM
flows in galactic halos. As is noted in Ref. \cite{lensing}, the
sharpness of the edges of the triangular feature in the IRAS map
\cite{IRAS} implies a rough upper bound of 15-20 pc on $\d x$. We
use this upper bound and the lower bounds for $\delta x$s, which
are calculated using the primordial velocity dispersions of axions
and WIMPs, to estimate the upper and lower limits for the density
and magnification at the smoothed dual cusps, in the presence of
finite velocity dispersion. To find an upper bound, we first
assume that the velocity dispersion of the particles are zero, and
then estimate the effects at a location on the surface, whose
transverse distances $(\D\r, \D z)$ to the cusp are greater than
the distance that a caustic surface would smear out, if the
particles had a finite primordial velocity dispersion. The lower
bounds for the effects are estimated at a location whose distance
to the cusp is about the maximum smearing out distance of the
caustic surface implied by the triangular feature in the IRAS map
of the Milky Way. Thus, to estimate the lower limits of the
density and magnification for the smoothed cusps, we choose the
locations that are $\Delta\psi_*=\frac{\pi}{7.5}$ radian away from
the locations that the cusps would occur if the velocity
dispersion were zero. Similarly, the upper bounds for the density
and lensing effects at the smoothed cusps are estimated at a
location on the surface, whose transverse distances $(\D\r, \D z)$
to the cusp are greater than the minimum distance that a caustic
surface would smear out, if the particles had only the primordial
velocity dispersion. We choose the locations that are respectively
$\Delta\psi_*=\frac{\pi}{7500}$ and $\Delta\psi_*=\frac{\pi}{75}$
radian away from the cusps of the axion and WIMP caustic rings, to
estimate the upper limits.

Thus, the lower bounds for the density and magnification at the
smoothed cusps are obtained at $\psi_*=\pm\frac{\pi}{7.5},
\frac{2\pi}{3}\pm \frac{\pi}{7.5}, \frac{4\pi}{3}\pm
\frac{\pi}{7.5}$. The locations of these points with respect to
the nearby cusps, in the limit $\delta v=0$, are: $(\D\rho_n, \D
z_n)(\pm\frac{\pi}{7.5})\simeq (1.26, \pm 1.76\cdot
10^{-1}\zeta_n)\cdot 10^{-1}p_n$, $(\D\rho_n, \D
z_n)(\frac{2\pi}{3}-\frac{\pi}{7.5})=(\D\rho_n, \D
z_n)(\frac{4\pi}{3}+\frac{\pi}{7.5})\simeq (7.82\cdot 10^{-1},
\zeta_n)\cdot 10^{-1}p_n$, and $(\D\rho_n, \D
z_n)(\frac{2\pi}{3}+\frac{\pi}{7.5})=(\D\rho_n, \D
z_n)(\frac{4\pi}{3}-\frac{\pi}{7.5})\simeq (4.78\cdot 10^{-1},
1.18\zeta_n)\cdot 10^{-1}p_n$. We have, for the fifth ring,
$(\D\rho_5, \D z_5)(\pm\frac{\pi}{7.5})\simeq (16.9\,\rm{pc}, \pm
2.7\,\rm{pc})$, $(\D\rho_5, \D
z_5)(\frac{2\pi}{3}-\frac{\pi}{7.5})=(\D\rho_5, \D
z_5)(\frac{4\pi}{3}+\frac{\pi}{7.5})\simeq (10.48\,\rm{pc},
15.44\,\rm{pc})$, $(\D\rho_5, \D
z_5)(\frac{2\pi}{3}+\frac{\pi}{7.5})=(\D\rho_5, \D
z_5)(\frac{4\pi}{3}-\frac{\pi}{7.5})\simeq (6.40\,\rm{pc},
18.15\,\rm{pc})$, which are all about the size of the upper limit
for smearing out of the cusp locations estimated using the
triangular feature in the IRAS map of the Milky Way. To estimate
the upper bounds for the density and magnification near the cusps
of the axion caustic rings, we choose $\psi_*=\pm\frac{\pi}{7500},
\frac{2\pi}{3}\pm\frac{\pi}{7500},
\frac{4\pi}{3}\pm\frac{\pi}{7500}$. The locations of these points,
with respect to the nearby cusps, in the limit $\delta v=0$, are:
$(\D\rho_n, \D z_n)(\pm\frac{\pi}{7500})\simeq (1.32, \pm
1.84\cdot 10^{-4}\zeta_n)\cdot 10^{-7}p_n$, $(\D\rho_n, \D
z_n)(\frac{2\pi}{3}\pm\frac{\pi}{7500})\simeq (6.58\cdot 10^{-1},
1.14\zeta_n)\cdot 10^{-7}p_n$. For the fifth ring $(\Delta\rho_5,
\Delta z_5)(\pm\frac{\pi}{7500})\simeq(1.76\cdot 10^{-5}\,
\rm{pc}, \pm 2.83\cdot 10^{-9}\,\rm{pc})$, $(\D\rho_5, \D
z_5)(\frac{2\pi}{3}\pm\frac{\pi}{7500})=(\D\rho_5, \D
z_5)(\frac{4\pi}{3}\mp\frac{\pi}{7500})\simeq (8.82\cdot
10^{-6}\,\rm{pc}, 1.75\cdot 10^{-5}\,\rm{pc})$. These distances
are all grater than the lower bound for smearing out $\delta x_a$
of the caustic surface location, estimated using the primordial
velocity dispersion of the axions. For the WIMP caustic rings,
considering Eq. \ref{dxW2} (i.e. assuming the particles were
affected by the reheating), the upper bounds for the density and
magnification are estimated at $\psi_*=\pm\frac{\pi}{75},
\frac{2\pi}{3}\pm\frac{\pi}{75}, \frac{4\pi}{3}\pm\frac{\pi}{75}$.
The locations of these points with respect to the nearby cusps
are: $(\Delta\rho_n, \Delta z_n)(\pm\frac{\pi}{75})\simeq(1.32,
\pm 1.84\cdot 10^{-2}\zeta_n)\cdot 10^{-3}p_n$, $(\Delta\rho_n,
\Delta z_n)(\frac{2\pi}{3}+\frac{\pi}{75})=(\Delta\rho_n, \Delta
z_n)(\frac{4\pi}{3}-\frac{\pi}{75})\simeq(6.42\cdot 10^{-1},
1.15\zeta_n)\cdot 10^{-3}p_n$, $(\Delta\rho_n, \Delta
z_n)(\frac{2\pi}{3}-\frac{\pi}{75})=(\Delta\rho_n, \Delta
z_n)(\frac{4\pi}{3}+\frac{\pi}{75})\simeq(6.74\cdot 10^{-1},
1.13\zeta_n)\cdot 10^{-3}p_n$. For the fifth ring, we have
$(\Delta\rho_5, \Delta z_5)(\pm\frac{\pi}{75})\simeq(1.76\cdot
10^{-1}\, \rm{pc}, \pm 2.83\cdot 10^{-3}\,\rm{pc})$,
$(\Delta\rho_5, \Delta
z_5)(\frac{2\pi}{3}+\frac{\pi}{75})=(\Delta\rho_5, \Delta
z_5)(\frac{4\pi}{3}-\frac{\pi}{75})\simeq(8.60\cdot
10^{-2}\,\rm{pc}, 1.77\cdot 10^{-1}\,\rm{pc})$, $(\Delta\rho_5,
\Delta z_5)(\frac{2\pi}{3}-\frac{\pi}{75})=(\Delta\rho_5, \Delta
z_5)(\frac{4\pi}{3}+\frac{\pi}{75})\simeq(9.03\cdot
10^{-2}\,\rm{pc}, 1.74\cdot 10^{-1}\,\rm{pc})$. These distances
are all grater than the lower bound $\delta x_\chi$ for smearing
out of the caustic surface location, estimated using the
primordial velocity dispersion of the WIMPs.

We, therefore, estimate the lower bound of the ${\mathcal F}$ at
the outer cusps of the caustic rings, assuming ${p_n}=0.1~a_n$,
$\zeta_n=1$, as ${\mathcal F}(\pm\frac{\pi}{7.5})\simeq 0.85$ (for
the fifth ring of the Milky Way, ${\mathcal
F}_5(\pm\frac{\pi}{7.5})\simeq 0.9$). To predict the upper bounds
for the ${\mathcal F}$ near the cusps of the caustic rings, we
distinguish between the axion and WIMP caustic rings. At the outer
cusp of a caustic ring, assuming ${a_n}=0.1~a_n$ and $\zeta_n=1$,
we find an upper bound as ${\mathcal F}(\frac{\pm\pi}{7500})\simeq
25.6$ for the axion rings. If the fifth caustic ring of the Milky
Way (where ${p_5}=0.016~a_5, \zeta_5\simeq 1.15$) is axionic, then
the upper bound is ${\mathcal F}_5(\frac{\pm\pi}{7500})\simeq
27.8$. Thus, the function ${\mathcal F}_n$ in Eq. \ref{Anring} may
enhance the density at the outer cusp of an axion caustic ring in
the range between 0.9 and 26 times, depending on the size of the
effective velocity dispersion. Taking a sample point at
$\psi_*=\vartheta=\frac{\pi}{100}$ and assuming ${p_n}=0.1~a_n$,
$\zeta_n=1$, we find ${\mathcal F}(\frac{\pi}{100})={\mathcal
F}(\frac{199\pi}{100})\simeq 3$, (for the fifth ring of the Milky
Way ${\mathcal F}_5(\frac{\pi}{100})\simeq 3.2$). Therefore, at
the point $\psi_*=\frac{\pi}{100}$, whose location with respect to
the outer cusp is $(\Delta\rho_n, \Delta z_n)=(7.4\cdot 10^{-4},
7.75\cdot 10^{-6}\zeta_n)p_n$, the factor $\mathcal{F}$ triples
the fold coefficient. For the fifth caustic ring of the Milky Way,
the location given by $\psi_*=\frac{\pi}{100}$ is $9.92\cdot
10^{-2}$pc (about $3\cdot 10^{12}$km) away from the location of
the cusp that would occur in the limit $\delta v=0$. The fold
coefficient at this location $A_5(\frac{\pi}{100})\simeq 5\cdot
10^{-3}{{\rm gr}}/{\rm cm^2\,kpc^{\frac{1}{2}}}$, where we
averaged Eqs. \ref{A5-1} and \ref{A5-2}. For the WIMP caustic
rings, the upper bound is $\mathcal{F}(\pm\frac{\pi}{75})\simeq
2.6$ (if the fifth ring of the Milky Way is a WIMP caustic, then
the upper bound is $\mathcal{F}_5(\pm\frac{\pi}{75})\simeq 2.8$).
Therefore, the function ${\mathcal F}$ in Eq. \ref{Anring} may
enhance the density at the outer cusp of a WIMP caustic ring in
the range between 0.9 and 2.6 times, depending on the size of the
effective velocity dispersion.

Next, to estimate the fold coefficient at points around the other
two non-planer cusps, which are located at\be
\rho_n(\frac{2\pi}{3}\pm\vartheta)&=&\rho_n(\frac{4\pi}{3}\mp\vartheta)
=a_n-\frac{p_n}{8}\left(1-3\vartheta^2\pm\sqrt{3}\vartheta^3
+\frac{3\vartheta^4}{4}+O(\vartheta^5)\right)\\
z_n(\frac{2\pi}{3}\pm\vartheta)&=&-z_n(\frac{4\pi}{3}\mp\vartheta)
=\zeta_np_n\frac{\sqrt{27}}{8}
\left(1-\vartheta^2\mp\frac{\vartheta^3}{\sqrt{27}}+\frac{\vartheta^4}{4}
+O(\vartheta^5)\right) \; ,\ee where we expand ${\mathcal F}_n$
as: \beeq {\mathcal F}_n(\frac{2\pi}{3}\pm\vartheta)={\mathcal
F}_n(\frac{4\pi}{3}\mp\vartheta)
=\frac{8\sqrt{2}\vartheta^{-\frac{1}{2}}}
{\sqrt{3}(8-\frac{p_n}{a_n})(1+3\zeta_n^2)^\frac{1}{4}}
\pm\frac{2\sqrt{2}(1-\zeta_n^2)\vartheta^{\frac{1}{2}}}{
(8-\frac{p_n}{a_n})(1+3\zeta_n^2)^\frac{5}{4}}+O(\vartheta^{\frac{3}{2}})
\; .\label{fnnon}\eneq For ${p_n}= 0.1~a_n$, $\zeta_n= 1$, we find
${\mathcal F}(\frac{2\pi}{3}\pm\vartheta)={\mathcal
F}(\frac{4\pi}{3}\mp\vartheta)\simeq 0.585~
\vartheta^{-\frac{1}{2}}$, which diverges as one approaches to the
cusps. Near the non-planer cusps of the rings (taking ${p_n}=
0.1~a_n$, $\zeta_n= 1$) we find the upper bound of ${\mathcal F}$
for the axion caustic rings as ${\mathcal F}(\frac{2\pi}{3}\pm
\frac{\pi}{7500})={\mathcal F}(\frac{4\pi}{3}\mp
\frac{\pi}{7500})\simeq 28.6$ (if the fifth ring of the Milky Way
is an axion caustic, then ${\mathcal F}_5(\frac{2\pi}{3}\pm
\frac{\pi}{7500})={\mathcal F}_5(\frac{4\pi}{3}\mp
\frac{\pi}{7500})\simeq 26.8$). Thus, depending on the size of the
velocity dispersion, the factor ${\mathcal F}$ in Eq. \ref{Anring}
may enhance the density in the range between 0.9 and 28.6 times,
near the non-planer cusps of the axion caustics. For the WIMP
caustic rings, on the other hand, the upper bound of ${\mathcal
F}$ near the non-planer cusps, is ${\mathcal F}(\frac{2\pi}{3}\pm
\frac{\pi}{75})={\mathcal F}(\frac{4\pi}{3}\mp
\frac{\pi}{75})\simeq 2.9$ (if the fifth ring of the Milky Way is
a WIMP caustic, then ${\mathcal F}_5(\frac{2\pi}{3}\pm
\frac{\pi}{75})={\mathcal F}_5(\frac{4\pi}{3}\mp
\frac{\pi}{75})\simeq 2.7$ is the upper limit of ${\mathcal F}$ at
the non-planer cusps). Therefore, the factor ${\mathcal F}$ in Eq.
\ref{Anring} may enhance the density in the range between 0.9 and
2.9 times, near the non-planer cusps of the WIMP caustics. The
locations that differ by $\pm\frac{\pi}{75}$ from the non-planer
cusps of the fifth ring of the Milky Way, are about $0.18$ pc away
from the locations where the cusps would occur if the velocity
dispersion of the flow were zero.

\subsection{Density at a dual cusp}
\label{subsec:densitynearacusp}

In this section, following the procedure given in \cite{lensing}
and assuming that the velocity dispersion $\delta v=0$, we derive
the dark matter density profile near a dual cusp, in terms of the
new parameters. We choose the outer cusp which lies in the $z=0$
plane at $\rho = a + p \equiv \rho_0$. Close to the $z=0$-plane,
from Eq. \ref{flownewz}, we have either ${\chi_1}\simeq\sqrt{p}$,
or ${\chi_2}\simeq 0$, or both. Near $\rho=\rho_0$, however, from
Eq. \ref{flownewr}, we have ${\chi_1}^2-{\chi_2}^2\simeq p$.
Therefore, near the outer cusp, we must have both ${\chi_1}\simeq
\sqrt{p}$ and ${\chi_2}\simeq 0$. Defining new dimensionless
quantities\beeq A\equiv\frac{{\chi_2}}{\sqrt{p}}\, ,\hskip 0.5 cm
T\equiv 1-\frac{{\chi_1}}{\sqrt{p}}\, ,\hskip 0.5 cm
X\equiv{{\rho-\rho_0}\over p}\, ,\hskip 0.5 cm
Z\equiv\frac{z}{\zeta p}\; , \label{capvar}  \eneq we recast Eqs.
\ref{flownewr}-\ref{flownewz} as: \beeq X=-2T+T^2-A^2 ,\hskip 0.5
cm Z=2AT ,\hskip 0.5 cm D_2(A,T)= 4\zeta
p\left(T-T^2-A^2\right)\label{kem21}\; . \eneq Close to the cusp,
where ${\chi_1}\simeq \sqrt{p}$, we may neglect the terms of order
$T^2$ in Eqs. \ref{kem21}. The term of order $A^2$ cannot be
neglected. Therefore, near the cusp, we have \be X=-2T-A^2\,
,\hskip 0.5 cm Z=2AT\, ,\hskip 0.5 cm D_2(A,T)= 4\zeta
p\left(T-A^2\right)\; . \label{dat} \ee Inserting $T=\frac{Z}{2A}$
into the equation for $X$, one obtains the third order polynomial
equation: \beeq A^3+X A+ Z=0\; . \label{pol} \eneq The
discriminant is:
$\delta=\left(\frac{Z}{2}\right)^2+\left(\frac{X}{3}\right)^3$. If
$\delta>0$, the cubic equation has one real root, and two complex
roots which are complex conjugate of each other. If $\delta < 0$,
all the roots are real and unequal.  For $\delta=0$, all the roots
are real and at least two are equal. The number of real roots is
the number of flows at a given location. The tricusp has two flows
outside and four inside.  In the neighborhood of a cusp, however,
one of the flows of the tricusp is nonsingular and does not
participate in the cusp caustic.  To include the root
corresponding to the nonsingular flow near $(\rho, z) = ( \rho_0,
0)$, one must keep the terms of order $T^2$ in Eq. \ref{kem21}.

The equation for the caustic surface in physical space is
$\delta=0$. Indeed, Eqs. \ref{dat} imply that at the dual cusp
caustic, where $D_2$ vanishes, $T=A^2$ (therefore,
$Z=2A^3=2T^{3/2}$ and $X=-3A^2=-3T$ at the cusp). The
discriminant, using Eqs. \ref{dat}, can also be written as
$\delta=\frac{D_2}{108\,\zeta p}(A^4+7A^2T-8T^2)$. Thus $D_2=0$
implies $\delta=0$, however, the converse is not true: $\delta=0$
does not necessarily imply $D_2=0$, because not all flows at the
location of the caustic surface are singular.

Inserting the $D_2$ obtained in Eq. \ref{dat} into Eq. \ref{dens},
yields \beeq d = {1 \over 8\pi\zeta} {1\over \rho_0\, p}
{{d^2M}\over{d{\chi_1} d{\chi_2}}} \sum_{j=1}^n {1 \over |T -
A^2|_j}= {1 \over 4\pi\zeta} {1\over \rho_0\, p}
{{d^2M}\over{d{\chi_1} d{\chi_2}}} \sum_{j=1}^n {1 \over |X + 3
A^2|_j}\; , \label{dTA} \eneq where the sum is over the flows
(i.e., the real roots of the cubic polynomial Eq. \ref{pol}). In
the last equality we used the equation for $X$ (Eq. \ref{dat}), to
rewrite $T-A^2=-\frac{1}{2}(X+3A^2)$.  If the discriminant
$\delta=(Z/2)^2+(X/3)^3>0$, the one real root is \beeq A=(-{Z\over
2}+\sqrt{\delta})^{1/3}-({Z\over 2}+\sqrt{\delta})^{1/3}.
\label{A1} \eneq This describes the one flow outside the cusp.
Inserting Eq. \ref{A1} in Eq. \ref{dTA}, we obtain \beeq
d=\frac{1}{4\pi\zeta}\frac{1}{\rho_0\, p}\frac{d^2M}{d{\chi_1}
d{\chi_2}} \frac{1}{|X-3(-\frac{Z}{2}+\sqrt{\delta})^{2/3}
-3(\frac{Z}{2}+\sqrt{\delta})^{2/3}|} \; . \label{d1rbp}\eneq

Next, we calculate the density inside the cusp, where $\delta <0$.
The three real roots of the polynomial Eq. \ref{pol} are
\cite{lensing}\be A_1=2\sqrt{\frac{-X}{3}}\cos{\theta} ,\;\;\;
A_2=2\sqrt{\frac{-X}{3}}\cos{(\theta +\frac{2\pi}{3})} ,\;\;\;
A_3=2\sqrt{\frac{-X}{3}}\cos{(\theta +\frac{4\pi}{3})}\; , \ee
where $\cos{3\theta}\equiv
-\frac{Z}{2}\left(-\frac{3}{X}\right)^{3/2}$ and
$0\leq\theta\leq{\pi\over 3}$.  Inserting them into Eq. \ref{dTA},
and adding the individual flow densities, we find\beeq
d=\frac{1}{2\pi\zeta}\frac{1}{\rho_0\, p}\frac{d^2M}{d\chi_1
d\chi_2}\frac{1}{|X|}\frac{1}{(\sqrt{3}-\tan{\theta})
\sin{2\theta}}\; . \label{dtan}\eneq Note that using Eq.
\ref{conversion} and $\cos\alpha\simeq 1$ reproduces the results
obtained in Ref. \cite{lensing}.
\section{Differential Geometry of Caustic Rings} \label{difgeo}

In this section, differential geometry of the axially symmetric
caustic rings is studied. We exclude the cusps from the discussion
and apply the method of moving frames through out the section. We
calculate the local quantities, such as metric, Gaussian, mean,
and the principal curvatures of the caustic surface. The principle
curvature radii, which are the inverses of the principal
curvatures (see the Appendix), are used in the lensing
applications in Sect. \ref{sec:Lensing}.

The caustic surface is completely described by the three vector
\be \vec{X}(\psi,
\phi)=\rho(\psi)\cos{\phi}\;\hat{x}+\rho(\psi)\sin{\phi}\;\hat{y}+z(\psi)\;\hat{z}\;
,\ee where $\psi$ and $\phi$ are angular variables ranging in the
interval $[0, 2\pi]$, and $(\hat{x}, \hat{y}, \hat{z})$ are the
Cartesian unit vectors. The functions $\rho(\psi)$ and $z(\psi)$
are given in Eq. \ref{rhpsi} which we copy here \be
\rho(\psi)=a+\frac{p}{2}\cos\psi(1+\cos\psi)\;,
\hspace{0.5cm}z(\psi)=\zeta\frac{p}{2}\sin\psi(1-\cos\psi)\;
.\nonumber \ee

At each point $P$ of the caustic surface, excluding the dual cusps
where $\psi=0, \frac{2\pi}{3}$, and $\frac{4\pi}{3}$ (denoting
this exclusion by $*$ as in the previous section), we choose three
differentiable orthonormal vectors $\hat{e}_i$, $i=\psi_*, \phi,
n$, with the inner product \beeq
\hat{e}_i\cdot\hat{e}_j=\delta_{ij}\; ,\label{on}\eneq such that
$\hat{e}_n$ is the unit normal $\hat n$, and hence
$\hat{e}_{\psi_*}$ and $\hat{e}_\phi$ span the tangent plane: \be
\hat{e}_{\psi_*}&\equiv&\frac{\vec X_{,{\psi_*}}}{\|\vec
X_{,{\psi_*}}\|}=
\frac{\rho'(\cos\phi\;\hat{x}+\sin\phi\;\hat{y})+z'\hat{z}}{\sqrt{\rho'^2+z'^2}}\\
\hat{e}_\phi&\equiv&\frac{\vec X_{,\phi}}{\|\vec
X_{,\phi}\|}=-\sin\phi\;\hat x+\cos\phi\;\hat y\\
\hat{e}_n&\equiv&\hat n=\hat{e}_{\psi_*} \times \hat{e}_\phi=
\frac{-z'(\cos\phi\;\hat{x}+\sin\phi\;\hat{y})+\rho'\hat{z}}{\sqrt{\rho'^2+z'^2}}\;
. \ee The subscript comma ``$,$'' indicates partial
differentiation with respect to the index that follows it. The
prime denotes derivative with respect to $\psi_*$. The reason for
the exclusion of the cusps from our considerations can be
understood here. Although both $\frac{\dd \rho}{\dd\psi}$ and
$\frac{\dd z}{\dd\psi}$ are well defined (zero) at the cusps, the
left and right limits of $\frac{\rho'}{\sqrt{\rho'^2+z'^2}}$ and
$\frac{z'}{\sqrt{\rho'^2+z'^2}}$ differ by an overall sign at the
cusps (only $\frac{z'}{\sqrt{\rho'^2+z'^2}}$ is continuous at
$\psi=0$ where it vanishes). Hence, $\hat{e}_n$ and
$\hat{e}_{\psi_*}$ are not well defined at the cusps.

Because $\vec{X}$ is constrained to move in the surface, the
differential $d\vec{X}$, which is a vector with one-form
coefficients, must lie in the tangent plane: \be
d\vec{X}&=&\vec{X}_{,{\psi_*}} d\psi_* + \vec{X}_{,\phi}
d\phi=\|\vec{X}_{,{\psi_*}}\|d\psi_*\hat{e}_{\psi_*}+
\|\vec{X}_{,\phi}\|d\phi\hat{e}_\phi\nonumber\\
&\equiv&\theta_{\psi_*}\hat{e}_{\psi_*}+\theta_\phi\hat{e}_\phi\;\label{Xt}
.\ee The differential one-forms on the caustic surface are
therefore \be \theta_{\psi_*}\!
&=&\!\sqrt{\rho'^2+z'^2}d\psi_*\!=\!p\sqrt{\frac{1+\zeta^2}{2}
+\frac{(1-\zeta^2)}{2}\cos\psi_*}\,\,\left|\sin\!\frac{3\psi_*}{2}\right| d\psi_*\, \\
\theta_\phi\!&=&\!\rho\, d\phi
=\left[a+\frac{p}{2}\cos\psi_*(1+\cos\psi_*)\right]
d\phi\\\theta_n\!&=&\!0\; .\ee The one-form $\theta_\psi$ vanishes
only at the cusps. The line element, or the metric,
$ds^2=d\vec{X}\cdot
d\vec{X}=\theta_{\psi_*}\otimes\theta_{\psi_*}+\theta_\phi\otimes\theta_\phi$
for the ring surface is\be
ds^2\!\!&=&\!\!g_{{\psi_*}{\psi_*}}d^2\psi_*+g_{\phi\phi}d^2\phi=(\rho'^2+z'^2)d^2\psi_*+\rho^2d^2\phi\nonumber\\
\!\!&=&\!\!p^2\left[{\frac{1+\zeta^2}{2}
+\frac{(1-\zeta^2)}{2}\cos\psi_*}\right]\sin^2\!\frac{3\psi_*}{2}d^2\psi_*
\!+\!\left[a+\frac{p}{2}\cos\psi_*(1+\cos\psi_*)\right]^2\!\!
d^2\phi\; . \ee The area ${{\mathcal{A}}}$ of the caustic surface
can be calculated by integrating the area element, proportional to
the square root of the determinant of the above metric, over the
whole surface. For the triaxial caustic rings ($\zeta=1$), we find
\beeq
{{\mathcal{A}}}=\int_0^{2\pi}d\phi\int_0^{2\pi}d\psi_*\sqrt{\det{g_{ij}}}=8\pi
p\left(a+\frac{p}{4}\right)\; .\eneq The length ${\ell}$ of the
tricusp cross-section can be calculated by integrating the square
root of the line element with a constant $\phi$, along the
cross-section. For the triaxial caustic rings, we find \beeq
\ell=\int_0^{2\pi}d\psi_*\sqrt{ds^2}=4p\; . \eneq

Each vector field $\hat{e}_i(\psi_*)$ is a differentiable map, and
the differential $d\hat{e}_i$ at point $P$ is a linear map. Thus,
for each $P$, we can write \be
d\hat{e}_i\equiv\omega_{ij}\hat{e}_j \label{est1}\; .\ee Since
$\hat{e}_i$ is a differentiable vector field, the $\omega_{ij}$
are differential one-forms. They are called the connection forms
in the moving frame $\hat{e}_i$. For the caustic ring, we have \be
d\vec{e}_{\psi_*}&=&\frac{(\rho'z''-z'\rho'')d\psi_*}{\rho'^2+z'^2}\vec{e}_n
+\frac{\rho'd\phi}{\sqrt{\rho'^2+z'^2}}\vec{e}_\phi
=\omega_{{\psi_*}\phi}\vec{e}_\phi+\omega_{{\psi_*} n}\vec{e}_n\\
d\vec{e}_\phi&=&-\frac{\rho'
d\phi}{\sqrt{\rho'^2+z'^2}}\vec{e}_{\psi_*}+\frac{z'd\phi}{\sqrt{\rho'^2+z'^2}}\vec{e}_n
=\omega_{\phi{\psi_*}}\vec{e}_{\psi_*}
+\omega_{\phi n}\vec{e}_n\\
d\vec{e}_n&=&-\frac{(\rho'z''-z'\rho'')d\psi_*}{\rho'^2+z'^2}\vec{e}_n
-\frac{z'd\phi}{\sqrt{\rho'^2+z'^2}}\vec{e}_\phi=\omega_{n{\psi_*}}\vec{e}_{\psi_*}
+\omega_{n\phi}\vec{e}_\phi\; .
 \ee
Not all of the connection forms are independent. Differentiating
the orthonormality condition \ref{on}, and using Eq. \ref{est1},
one finds that the connection forms are antisymmetric. Therefore,
\be \omega_{\psi_*\phi}\!\!\!&=&\!\!\!-\omega_{\phi\psi_*}=
\frac{\rho' d\phi}{\sqrt{\rho'^2+z'^2}}=\frac{\rho'
\theta_\phi}{\rho\sqrt{\rho'^2+z'^2}}= -\frac{(\sin\psi_*+\sin
2\psi_*)d\phi}{\sqrt{2[1+\zeta^2+(1-\zeta^2)\cos{\psi_*}]}
\left|\sin{\frac{3\psi_*}{2}}\right|}\\
\omega_{\psi_* n}\!\!\!&=&\!\!\!-\omega_{n\psi_*}= \frac{(\rho'
z''-z'\rho'') d\psi_*}{\rho'^2+z'^2}=\frac{(\rho' z''-z'\rho'')
\theta_{\psi_*}}{(\rho'^2+z'^2)^\frac{3}{2}}=-\frac{\zeta
d\psi_*}{1+\zeta^2+(1-\zeta^2)\cos{\psi_*}}\label{npsi}\\
\omega_{\phi n}\!\!\!&=&\!\!\!-\omega_{n \phi}= \frac{z'
d\phi}{\sqrt{\rho'^2+z'^2}}= \frac{z'
\theta_\phi}{\rho\sqrt{\rho'^2+z'^2}}=\frac{\zeta(\cos\psi_*-\cos
2\psi_*)d\phi}{\sqrt{2[1+\zeta^2+(1-\zeta^2)\cos{\psi_*}]}
\left|\sin{\frac{3\psi_*}{2}}\right|} \; .\label{nphi}\ee The
connection one-forms $\omega_{\psi\phi}$ and $\omega_{\phi n}$ are
not well defined at the cusps --- except for $\omega_{\phi n}$ at
$\psi=0$, where it vanishes. Although finite, their left and right
limits differ by an overall sign at the cusps. $\omega_{\psi n}$,
on the other hand, is continuous everywhere.

Equations \ref{Xt} and \ref{est1}, that define $\theta_i$ and
$\omega_{ij}$ via $d\vec{X}$ and $d\hat{e}_i$, describe how the
moving frame varies as we move along a curve on the surface. The
identities $d(d\vec{X})=0$ and $d(d\hat{e}_i)=0$ (Poincare Lemma)
imply \be 0&=&d(d\vec{X})=d(\theta_i \hat{e}_i)=d\theta_i
\hat{e}_i-\theta_i\wedge
d\hat{e}_i=(d\theta_i-\omega_{ij}\wedge\theta_j)\hat{e}_i\label{C1}\\
0&=&d(d\hat{e}_i)=d(\omega_{ij} \hat{e}_j)=d\omega_{ij}
\hat{e}_j-\omega_{ij}\wedge
d\hat{e}_j=(d\omega_{ij}-\omega_{ik}\wedge\omega_{kj})\hat{e}_j\label{C2}\;
.\ee The last equations in \ref{C1} and \ref{C2} that the forms
and connection coefficients satisfy, \beeq
d\theta_i=\omega_{ij}\wedge\theta_j\;, \hspace{0.6cm}
d\omega_{ij}=\omega_{ik}\wedge\omega_{kj}\; ,\eneq are called the
Cartan's first and second structure equations, respectively. Since
there is only one linearly independent two-form on a two
dimensional surface, we have \be d\omega_{\psi\phi}= \omega_{\psi
n}\wedge\omega_{n\phi}=-\kappa_G \theta_\psi\wedge\theta_\phi
\label{kap}\; ,\ee where the scalar $\kappa_G$ is the Gaussian
curvature, and is independent of the choice of $\hat e_i$.
Similarly, the two form\beeq \theta_\psi\wedge\omega_{\phi
n}-\theta_\phi\wedge\omega_{\psi
n}=-2H\theta_\psi\wedge\theta_\phi \; ,\label{H}\eneq defines the
scalar $H$, called the mean curvature of the surface. For the
caustic ring  Eqs. \ref{kap}-\ref{H} yield\beeq
\kappa_G(\psi_*)=\frac{z'(\rho' z'' -z' \rho'')}{\rho
(\rho'^2+z'^2)^2}=-\frac{2\zeta^2[1+\zeta^2+(1-\zeta^2)\cos{\psi_*}]^{-2}
}{p(1+2\cos\psi_*)[a+\frac{p}{2}\cos\psi_*(1+\cos\psi_*)]}\;
,\eneq and \be
\!\!\!\!\!\!\!\!\!\!\!\!&&\!\!\!\!\!\!H(\psi_*)=-\frac{1}{2}
\left(\frac{\rho' z''-z'\rho''}{(\rho'^2+
z'^2)^{\frac{3}{2}}}+\frac{z'}{\rho\sqrt{\rho'^2+z'^2}}\right)\nonumber\\
\!\!\!\!\!\!\!\!\!\!\!\!&&\!\!\!\!\!\!=\!\frac{\zeta}{p}\!
\left[\frac{1}{\sqrt{2}[1+\zeta^2+(1-\zeta^2)\cos\psi_*]^\frac{3}{2}
\left|\sin\frac{3\psi_*}{2}\right|}-\frac{\rm{Sign(\cos\psi_*-\cos
2\psi_*)}}{[\frac{2a}{p}
+\cos\psi_*(1+\cos\psi_*)]\sqrt{\zeta^2+\cot^2(\frac{\psi_*}{2})}}\right]\ee
respectively. The Gaussian curvature at the outer cusp $\kappa_G
(0)=-\frac{\zeta^2}{6p\rho_0}$. However, $\kappa_G$ is not well
defined at the two non-planer cusps. It diverges with differing
signs in the left and right limits at each of them. The mean
curvature $H$ is also divergent at the cusps.

The connection one-forms $\omega_{\phi n}$ and $\omega_{\psi n}$
are linear combinations of $\theta_\psi$ and $\theta_\phi$, for
two dimensional surfaces, in general. Because of the relation
$d\theta_n=0=\omega_{n\psi}\wedge\theta_\psi +
\omega_{n\phi}\wedge\theta_\phi$ we have symmetry in the expansion
coefficients, hence \be \left(\begin{array}{c}
{{\omega_{n \psi}}}\\
{\omega_{n \phi}}\\
\end{array} \right)= \left(\begin{array}{cc}
{h_{\psi\psi}}&{h_{\psi\phi}}\\
{h_{\psi\phi}}&{h_{\phi\phi}}\\
\end{array}\right)\left(\begin{array}{c}
{{\theta_\psi}}\\
{{\theta_\phi}}\\
\end{array} \right)\; . \label{oht}\ee
Using Eqs. \ref{kap}, \ref{H} and \ref{oht}, we find \be
\kappa_G&=&h_{\psi\psi} h_{\phi\phi}- h^2_{\psi\phi}=\det(h_{ij})\\
H&=&\frac{1}{2}(h_{\psi\psi}+h_{\phi\phi})= \frac{1}{2}{\rm
trace}\, (h_{ij})\; .\ee Characteristic roots of the symmetric
matrix $(h_{ij})$ are called the principal curvatures $\kappa_i$
of the surface. (Geometrically, they are the inverse ``radii''
along the principal directions of the surface, see the Appendix).
We consequently have \beeq \kappa_i^2-2H\kappa_i+\kappa_G=0 \;
,\eneq where \beeq \kappa_G=\kappa_\psi\, \kappa_\phi\; ,\;
H=\frac{1}{2}(\kappa_\psi +\kappa_\phi)\; .\label{meangauss}\eneq
Hence \beeq \kappa_i=H\pm \sqrt{H^2-\kappa_G}\; .\label{kij}\eneq
The matrix $(h_{ij})$ is diagonal for the caustic rings, thus \be
\kappa_{\psi_*}(\psi_*)&=&h_{\psi_*\psi_*}=\frac{z'\rho''-\rho'z''}{(\rho'^2+z'^2)^\frac{3}{2}}
=\frac{\sqrt{2}\zeta}{p[1+\zeta^2+(1-\zeta^2)\cos{\psi_*}]^{\frac{3}{2}}
\left|\sin{\frac{3\psi_*}{2}}\right|}\label{kappapsi}\\
h_{\psi_*\phi}&=&h_{\phi\psi_*}=0\\
\kappa_\phi(\psi_*)&=&h_{\phi\phi}=-\frac{z'}{\rho\sqrt{\rho'^2+z'^2}}=-\frac{\zeta\,
\rm{Sign}(\cos\psi_*-\cos2\psi_*)}{[a+\frac{p}{2}\cos\psi_*(1+\cos\psi_*)]
\sqrt{\zeta^2+\cot^2\frac{\psi_*}{2}}}\; . \label{kappaphi}\ee The
principal curvature $\kappa_{\psi_*}$ is always positive. Its
power series expansions \beeq
\kappa_{\psi_*}(\vartheta)=\kappa_{\psi_*}(2\pi-\vartheta)=\frac{\zeta\vartheta^{-1}}{3p}
+\frac{\zeta(2-\zeta^2)\vartheta}{8p}+O(\vartheta^3)\; , \eneq and
\beeq
\kappa_{\psi_*}(\frac{2\pi}{3}\mp\vartheta)\!=\!\kappa_{\psi_*}(\frac{4\pi}{3}\pm\vartheta)
\!=\!\frac{8\zeta\vartheta^{-1}}{3p(1\!+\!3\zeta^2)^{\frac{3}{2}}}
\mp\frac{4\sqrt{3}\zeta(1\!-\!\zeta^2)}{p(1\!+\!3\zeta^2)^{\frac{5}{2}}}+
\frac{2\zeta(7\!-\!14\zeta^2\!+\!15\zeta^4)\vartheta}{p(1\!+\!3\zeta^2)^{\frac{7}{2}}}
+O(\vartheta^2) \eneq show that $\kappa_\psi$ diverges as
$\frac{1}{\vartheta}$ at the cusps. $\kappa_\phi$, on the other
hand, is negative in the regions $0<\psi_*<\frac{2\pi}{3}$ and
$\frac{4\pi}{3}<\psi_*<2\pi$, and positive in the region
$\frac{2\pi}{3}<\psi_*<\frac{4\pi}{3}$. As can be seen from the
expansion \beeq
\kappa_\phi(\vartheta)=\kappa_\phi(2\pi-\vartheta)=-\frac{\zeta\vartheta}{2(a+p)}+O(\vartheta^3)
\; ,\eneq $\kappa_\phi$ vanishes at the outer cusp. Although
discontinuous, it remains finite at the two non-planer cusps:
\beeq
\kappa_\phi(\frac{2\pi}{3}\mp\vartheta)=\kappa_\phi(\frac{4\pi}{3}\pm\vartheta)
=\mp\frac{8\sqrt{3}\zeta}{(8a-p)\sqrt{1+3\zeta^2}}
+\frac{16\zeta\vartheta}{(8a-p)(1+3\zeta^2)^{\frac{3}{2}}}+O(\vartheta^2)\;
. \eneq Finite left and right limits at the cusps, differ by a
sign.

The (normal) curvature $\kappa_n$ at a point on a surface in the
direction of a tangent line $L$ is calculated by using the Euler's
Theorem \beeq \kappa_n(\psi_*, \omega)=
\kappa_{\psi_*}(\psi_*)\cos^2\omega
+\kappa_\phi(\psi_*)\sin^2\omega \; ,\label{Eulers}\eneq where
$\omega$ is the angle between $L$ and a tangent line in the
principal direction associated with $\kappa_{\psi_*}$. As is shown
in the Appendix, the inverses of $\kappa_{\psi_*}$ and
$\kappa_\phi$ are respectively the curvature radii $R_{\psi_*}$
and $R_\phi$ along the principal directions on the surface. In the
next section, we study the gravitational lensing effects of the
dark matter caustic rings. Equations \ref{kappaphi} and
\ref{Eulers} are used to find the curvature radii along the
directions of the line of sights.
\section{Gravitational Lensing by Caustic Rings}
\label{sec:Lensing}

The Caustic Ring Model of galactic halos precisely predict the
density (Sect. \ref{Density}) and the geometry (Sect.
\ref{difgeo}) of the CDM distribution in the caustic neighborhood.
Gravitational lensing may be a useful tool to test these
predictions. Dark matter caustics have calculable lensing
signatures \cite{Hogan,lensing,Thesis,probing}. In Ref.
\cite{lensing}, we derived the lensing equations for the outer and
ring caustics, and estimated the image magnification at a number
of sample locations of the line of sight on the caustic surface.
The gravitational lensing effects of a caustic surface are largest
when the line of sight is near tangent to the surface, because the
contrast in column density, defined in Eq. \ref{S}, is largest
there \cite{lensing}. The effects, as can be seen in Eq. \ref{ep},
are proportional to (i) the fold coefficient $A$ at the point
where the line of sight is near tangent to the caustic surface,
(ii) the square root of the curvature radius $R$ of the surface
along the direction associated with the line of sight, and
inversely proportional to (iii) the critical surface density
$\Sigma_c$, defined in Eq. \ref{scri}. Our focus is upon the
caustic rings in this paper. Because, the fold coefficient
increases proportional to $\vartheta^{-\frac{1}{2}}$ (Eq.
\ref{fnout} and \ref{fnnon}) as one approaches to the cusps of the
rings (where $\psi_*=\pm\vartheta, \frac{2\pi}{3}\pm\vartheta,
\frac{4\pi}{3}\pm\vartheta$), the effects are larger near the
cusps. Because the curvature radius is largest near the outer cusp
in particular, where $R_\phi\sim \vartheta^{-1}$ (Eq. \ref{rphi}),
the effects are largest there. Near the non-planer cusps, on the
other hand, $R_\phi\sim$const (Eqs. \ref{rphinonplaner}), hence
the effects are not as large there. In this section, to obtain the
magnification for a line of sight tangential at an arbitrary point
--- except at the cusps where the lensing effects are infinite
 if $\d v=0$, and our formulation fails --- near the caustic surface, we use
the equations for the fold coefficient and the curvature radius
along the line of sight, obtained in Sects. \ref{Density} and
\ref{difgeo} respectively, in the lensing equations of Ref.
\cite{lensing}. We estimate the magnification by the rings at
cosmological distances, where the critical surface density
$\Sigma_c$ is minimized (hence the lensing effects are maximized),
and also by the nearby fifth caustic ring of our own galaxy, at
several sample locations of the line of sight in the caustic
neighborhood, near the cusps. Like Ref. \cite{lensing}, we
consider only the line of sights that are parallel to the galactic
plane of the caustic ring and near tangent to the surface at a
given $\psi_*$. Thus, all the line of sights lie in the plane
$z=z(\psi_*)$ in this paper. Unlike Ref. \cite{lensing}, we pick
$\psi_*$ near the cusps, to estimate the lensing effects.

\begin{figure}[ht] \centering
\includegraphics[height=3cm,width=12cm]{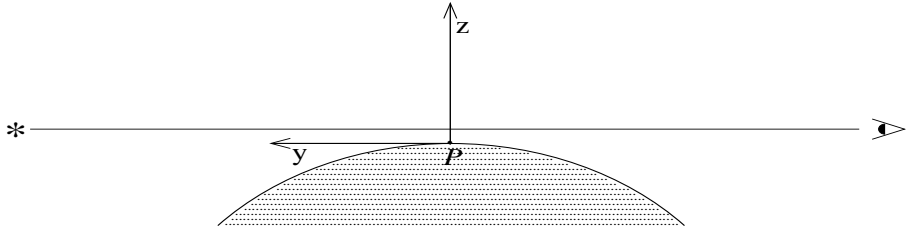}
\caption{Lensing by a concave fold. The arc is the intersection of
the caustic surface with the plane containing the normal
($\hat{z}$) to the surface and the line of sight ($\hat{y}$).  The
shaded area indicates the side with the two extra flows.}
\label{fig:concave}
\end{figure}

We consider three cases of gravitational lensing by a caustic ring
surface. In all the cases considered the line of sight is near
tangent to the surface where a simple fold catastrophe is located.
The three cases are distinguished by the curvature of the surface
at the tangent point in the direction of the line of sight. In the
first case, the line of sight is near tangent to the caustic
surface that curves toward the side with two extra flows; see Fig.
\ref{fig:concave}. We call such a surface ``concave.''

\begin{figure}[ht] \centering
\includegraphics[height=3cm,width=12cm]{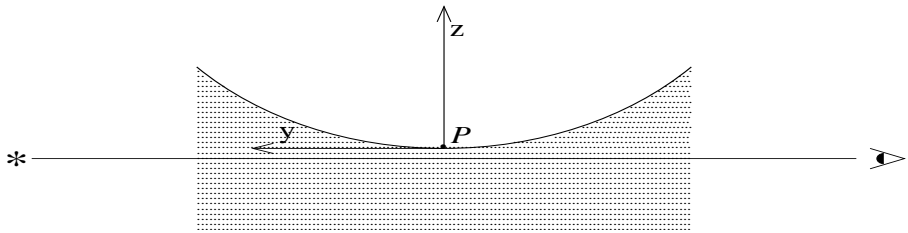}
\caption{Lensing by a convex fold.  Same as Fig. \ref{fig:concave}
except that now the caustic surface curves away from the side with
two extra flows.} \label{fig:convex}
\end{figure}

In the second case, the surface is ``convex,'' (i.e., it curves
away from the side with two extra flows); see Fig.
\ref{fig:convex}.

In the third case, the caustic surface has zero curvature at the
tangent point (the radius of curvature is infinite), but the
tangent line is entirely outside the side with two extra flows.

Although we exclude the cusp locations (where the lensing effects
diverge if $\d v=0$) from our formulation, we pick sample points
that are near (but sufficiently far from) the cusps. In the
presence of finite velocity dispersion, we constrain the expected
effects at the smoothed cusps, considering the smearing out of the
caustic surfaces of the axion and WIMP flows. To find an upper
bound, as explained in Sect. \ref{subsec:foldcoef}, we first set
the velocity dispersion of the CDM flow to zero, and then estimate
the effects at the points on the surface that are respectively
$\Delta\psi_*=\frac{\pi}{7500}$ and $\Delta\psi_*=\frac{\pi}{75}$
radian away from the cusps of the axion and WIMP caustic rings.
Similarly, we estimate the lower limits at the points that are
$\Delta\psi_*=\frac{\pi}{7.5}$ radian away from the locations that
cups would occur if velocity dispersion were zero.

Let us summarize the gravitational lensing phenomenon briefly. In
linear approximation, the deflection angle $\vec\theta$ of a light
ray due to a gravitational field is given by \beeq \vec\theta
=\vec{\nabla} {2\over{c^2}}\int V~dy \label{v} \; ,\eneq where
$V(x, y, z)$ is the Newtonian potential. We choose the $y$-axis in
the direction of propagation of light. Geometrically,
$\vec{\theta}$ is related to the angular shift
$~\vec{\xi_I}-\vec{\xi_S}~$ on the sky of the apparent direction
of a source: \beeq \vec\theta (\vec{\xi_{
I}})={{D_S}\over{D_{LS}}}(\vec{\xi_I}-\vec{\xi_S})\; ,
\label{fund} \eneq where $D_S$ and $D_{LS}$ are the distances of
the source to the observer and to the lens respectively.
${\vec{\xi}}_{ S}$ is the angular position of the source in the
absence of the lens while ${\vec{\xi}}_{ I}$ is the angular
position of the image with the lens present. The angles carry
components in the $x$ and $z$ directions: $\vec{\theta}= (\theta_x
, \theta_z)$, $\vec{\xi} = (\xi_x,\xi_z)$, etc.  Unless otherwise
stated, we mean by a vector, a quantity with components in the $x$
and $z$ directions. We have $\vec{x}=(x, z)=D_L{\vec{\xi}}_I$. It
is convenient to introduce a 2D potential $\psi({\vec{\xi}}_{ I})$
so that $ \vec\theta={D_{ S} \over D_{ LS}} \vec{\nabla}_{\xi_{
I}}\psi({\vec{\xi}}_{ I})$ where $\vec{\nabla}_{{\xi}_{ I}} =
D_L\vec{\nabla}$, and $D_L$ is the distance of the observer to the
lens. Then, Eq. \ref{fund} becomes \beeq \vec{\xi}_{
I}=\vec{\xi}_{ S}+\vec{\nabla}_{\xi_{ I}}\psi(\vec{\xi}_{ I})\; .
\label{lens} \eneq It gives the map $\vec{\xi}_{ S}(\vec{\xi_{
I}})$ from the image plane to the source plane. The inverse map
may be one to one, or one to many. In the latter case, there are
multiple images and infinite magnification when a pair of images
merge. The potential $\psi$ obeys the Poisson equation: \beeq
\nabla^2_{\xi_{ I}}\psi={{8\pi G}\over{c^2}}{{D_L
D_{LS}}\over{D_{S}}}\Sigma= 2 {{\Sigma}\over{\Sigma_c}}\; ,
\label{Poisson} \eneq where $\Sigma(\xi_{ Ix}, \xi_{ Iz})$ is the
column density (i.e., the integral of the volume density along the
line of sight): \beeq \Sigma(\vec{\xi}_{ I})=\int dy\, d(D_L\xi_{
Ix},y,D_L\xi_{ Iz})\; , \label{S} \eneq and $\Sigma_c$ is the
critical surface density \beeq \Sigma_c={{c^2D_{S}} \over{4\pi G
D_L D_{LS}}}=0.347 \, {\rm{g/cm^2}} \left({{D_{S}}\over{D_L
D_{LS}}}\,{\rm{Gpc}} \right)\, . \label{scri} \eneq A uniform
sheet of density $\Sigma_c$ focuses radiation from the source to
the observer. In general, the shift in image position,
$\D\xi\equiv\xi_I-\xi_S$, is the gradient of a potential whose 2D
Laplacian is the column density. For an arbitrary mass
distribution, the procedure for calculating the shift involves two
steps.  First, the matter density is integrated along the line of
sight to obtain the column density. Second, the potential is
obtained by convoluting the column density with the 2D Green's
function.  In our applications, however, the caustic has contrast
in only one of the dimensions transverse to the line of sight. The
procedure can be simplified by expressing the shift directly as an
integral over the parameter space of the dark matter flow forming
the caustic. In this paper, we use the equations we derived in
\cite{lensing} to estimate the lensing effects in Sects.
\ref{sub:concave}-C.

Gravitational lensing produces the map $\vec\xi_S(\vec\xi_I)$,
given in Eq. \ref{lens}, of an object surface onto an image
surface. The inverse map may be one to one, or one to many. In the
latter case, there are multiple images and infinite magnification
when a pair of images merge. The image structure, distortion, and
magnification are given by the Jacobian matrix of the map from
image to source: \beeq K_{ij}\equiv {{\partial \xi_{
Si}}\over{\partial \xi_{ Ij}}} =\delta_{ij}-\psi_{ij} \label{K} \;
 , \eneq where $\psi_{ij}\equiv\frac{\partial^2\psi} {\partial
\xi_{ Ii}\partial\xi_{ Ij}}$. Because gravitational lensing does
not change surface brightness, the magnification ${\mathcal{M}}$
is the ratio of image area to source area. Therefore, \beeq
{\mathcal{M}}={1\over{|\det{(K_{ij})}|}}\;  . \eneq To first
order, for $\psi_{ij} \ll 1$, \beeq
{\mathcal{M}}=1+\nabla_{\xi_I}^2\psi = 1+2{\Sigma\over\Sigma_c}\;
, \label{mag} \eneq where we used Eq. \ref{Poisson} in the last
equality. In the convex (Sect. \ref{sub:convex}) and zero
curvature (Sect. \ref{sub:zero}) cases considered, a point source
can have multiple images \cite{lensing}. In those cases, when two
images merge, the Jacobian of the map vanishes and the
magnification diverges. As we will see in Sects.
\ref{sub:concave}-C, the magnification is larger near the cusps.
We find that, for the line of sights considered, the magnification
is proportional to $\vartheta^{-\frac{1}{2}}$ near the non-planer
cusps where the case is either concave or convex, on the other
hand, it is proportional to $\vartheta^{-1}$ near the outer cusp,
where the case is concave (recall that $\vartheta\rightarrow 0$ at
the cusps). Unfortunately, the gravitational lensing due to the
fifth caustic ring of the Milky Way, which is only 55 pc away from
us, turns out to be too weak to be observed --- even near the
cusps
--- with current instruments (Sects. \ref{sub:concave}-C). To
obtain the largest lensing effects, we wish to minimize
$\Sigma_c$, given in Eq. \ref{scri}. For fixed $D_S$, the minimum
occurs when the lens is situated half-way between the source and
the observer. Also, $D_S$ should be as large as possible. In our
estimates, to get large effects, we assume that the source is at
cosmological distances (e.g., $2D_L=2D_{LS}=D_S = 1 {\rm Gpc}$, in
which case $\Sigma_c = 1.39$ g/cm$^2$). We find in Sects.
\ref{sub:concave}-B that, near the cusps of the caustic rings at
cosmological distances, the effects for point-like sources are
quite promising. The images of extended sources may also show
distortions that can be unambiguously attributed to the dark
matter caustics \cite{Hogan,lensing}. Observation of the
calculated lensing signatures would give strong evidence for
caustics and CDM in galactic halos.

As is mentioned earlier, the lensing by a caustic ring surface
also depends on the square root of the curvature radius $R$ along
the line of sight $L$. The principal curvature radii $R_{\psi}$
and $R_\phi$ are the inverses of the principal curvatures
$\kappa_{\psi}$ and $\kappa_\phi$, respectively. As is shown in
the Appendix, $R_\psi$ is the radius along the cross-sectional
plane of the caustic ring, and, $R_\phi$ is the radius along the
direction perpendicular to the cross-sectional plane. Therefore,
$R$ can be obtained from Eq. \ref{Eulers} as \beeq {1 \over R} =
{\cos^2{\omega} \over R_\psi} + {\sin^2{\omega} \over R_\phi} \,
,\label{Rad} \eneq where $\omega$ is the angle between the line of
sight and the direction associated with $R_\psi$.  In this paper,
we adopt the convention that $R$ is negative (positive) if, along
the line of sight, the surface curves toward (away from) the side
with two extra flows. If $R$ is negative, the surface is called
``concave.'' If $R$ is positive, the surface is called ``convex.''
$R_\psi>0$ everywhere, except at the cusps where it vanishes.
$R_\phi$, on the other hand, changes sign at the non-planer cusps.
In the regions where $0<\psi_*<\frac{2\pi}{3}$ and
$\frac{4\pi}{3}<\psi_*<2\pi$, we have $R_\phi<0$. Between these
regions, where $\frac{2\pi}{3}<\psi_*<\frac{4\pi}{3}$, we have
$R_\phi>0$. The behavior of $R$ near the cusps, where we expect
large lensing effects to occur, can be obtained by expanding
$R_\psi$ and $R_\phi$ as power series in the neighborhood of the
cusps. Expansions around $\psi=0, \frac{2\pi}{3}$, and
$\frac{4\pi}{3}$ show that $R_\psi$ vanishes at all the cusps
linearly: \beeq
R_\psi(\vartheta)=R_\psi(2\pi-\vartheta)=\frac{3p\vartheta}{\zeta}
-\frac{9p(2-\zeta^2)\vartheta^3}{8\zeta}+O(\vartheta^5) \eneq
\beeq
R_\psi(\frac{2\pi}{3}\mp\vartheta)=R_\psi(\frac{4\pi}{3}\pm\vartheta)
=\frac{3p(1+3\zeta^2)^{\frac{3}{2}}\vartheta}{8\zeta}
\pm\frac{9\sqrt{3}p\sqrt{1+3\zeta^2}(1-\zeta^2)\vartheta^2}{16\zeta}+O(\vartheta^3)\;
. \eneq $R_\phi$, on the other hand, diverges as $\vartheta^{-1}$
at the outer cusp: \beeq
R_\phi(\vartheta)=R_\phi(2\pi-\vartheta)=-\frac{2(a+p)\vartheta^{-1}}{\zeta}
+\frac{[(2-3\zeta^2)(a+p)+18p]\vartheta}{12\zeta}+O(\vartheta^3)\label{rphi}\;
,\eneq whereas it remains finite (although discontinuous) at the
other two cusps: \beeq
R_\phi(\frac{2\pi}{3}\mp\vartheta)=R_\phi(\frac{4\pi}{3}\pm\vartheta)
=\mp\frac{(8a-p)\sqrt{1+3\zeta^2}}{8\sqrt{3}\zeta}
-\frac{(8a-p)\vartheta}{12\zeta\sqrt{1+3\zeta^2}}
+O(\vartheta^2)\; . \label{rphinonplaner}\eneq Thus, for any
$\psi_*$ in the regions where $0<\psi_*< {2 \over 3}\pi$, or
$\frac{4}{3}\pi<\psi_*< 2\pi$, there is a pair of lines of sight
for which the curvature vanishes.  They are at angles: \beeq
\omega = \pm \arctan{\sqrt{- {R_\phi \over R_\psi}}}
\label{zerodirection}\eneq relative to the cross-sectional plane.
Gravitational lensing by a fold of zero curvature is discussed in
Section \ref{sub:zero}.
\subsection{Lensing by a Concave Fold of a Caustic Ring}
\label{sub:concave} 
\begin{figure}[ht] \centering
\includegraphics[height=8.5cm,width=6cm]{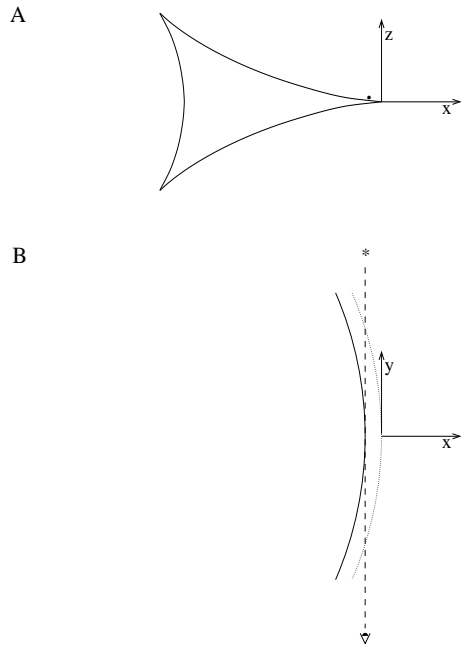}
\caption{Lensing by a concave fold of the caustic ring for a line
of sight near tangent to the surface at a point $\psi_*$ close to
the outer cusp, where $(\rho, z)=(\rho_0, 0)$. The line of sight
lies in the $z=z(\psi_*)$-plane. We define $x\equiv \rho-\rho_0$.
A) Side view in the direction of the line of sight. The latter is
represented by the dot near $x=z=0$. B) Top view. The solid curve
is the location of the concave fold in the $z=z(\psi_*)$-plane,
and the dotted curve is the location of the outer cusps of the
ring.} \label{fig:lensingbycusp}
\end{figure}

When the line of sight is near tangent to a
caustic surface which curves towards the side with two extra
flows, we define the fold surface as concave (Fig.
\ref{fig:concave}). To estimate the lensing by concave folds, in
Ref. \cite{lensing}, we considered only the outer caustics which
are topological spheres surrounding the galaxies. The outer
caustics are concave everywhere on the surface. Here, in this
paper, we study the gravitational lensing by the concave folds of
the caustic rings. If the line of sight is chosen along the
principal direction associated with $R_\phi$, two thirds of a
caustic ring surface, where $0<\psi_*<\frac{2\pi}{3}$ and
$\frac{4\pi}{3}<\psi_*<2\pi$ provides a cross-section where the
fold is concave; see Fig. \ref{fig:lensingbycusp}. The convex and
zero curvature cases are discussed in Sects. \ref{sub:convex}-C.

Image shift and magnification for a concave fold are given
\cite{lensing} as \be \Delta\xi&=&\xi_{ I}-\xi_{ S} = \eta\,\xi_{
I}\,\Theta{(-\xi_{ I})} \label{Dks}\\{\mathcal{M}}&=&{{d\xi_{
I}}\over{d\xi_{S}}} = 1 + \eta\;\Theta{(-\xi_I)} + 0(\eta^2)\; ,
\label{Mm} \ee where \beeq \eta = {{2\pi
A\sqrt{2|R|}}\over{\Sigma_c}} \; .\label{ep} \eneq When the line
of sight of a moving source crosses the surface of a simple
concave fold, the component of its apparent velocity perpendicular
to the fold changes abruptly. Also, a discontinuity occurs in the
magnification of the image. Both effects are of order $\eta$.
Here, using Eqs. \ref{fnout}, \ref{fnnon}, \ref{rphi} and
\ref{rphinonplaner}, we see that, for the concave case, the
maximum effects should be expected near the cusps. Near the outer
cusp $\eta\sim \vartheta^{-1}$ where $\psi_*=\vartheta$ or $2\pi -
\vartheta$. Near the non-planer cusps $\eta\sim
\vartheta^{-\frac{1}{2}}$ where $\psi_*=\frac{2\pi}{3}-\vartheta$
and $\psi_*=\frac{4\pi}{3}+\vartheta$. Inserting the fold
coefficient $A$ for the rings, given in Eq. \ref{Anv2}, and
$R=R_\phi(\psi_*)=\kappa^{-1}_\phi(\psi_*)$ given in Eq.
\ref{kappaphi}, into Eq. \ref{ep}, we obtain \beeq
\eta=\frac{4\pi}{\Sigma_c}\frac{d^2M}{d\Omega
d\tau}\frac{\cos{\alpha(\psi_*)}}{b\sqrt{ap}} {\mathcal
G}(\psi_*)\; ,\label{ETA}\eneq where \beeq {\mathcal G}(\psi_*)=
\sqrt{\frac{|\csc\psi_*|}
{[1+\frac{p}{2a}\cos\psi_*(1+\cos\psi_*)]|(1+2\cos\psi_*)\tan\frac{\psi_*}{2}|}}\;\;
. \eneq Notice that, $\zeta$ dependencies of $A$, and $R_\phi$
cancel each other in Eq. \ref{ep}, hence $\eta$ in Eq. \ref{ETA}
is independent of $\zeta$. Combining Eqs. \ref{f} and \ref{ETA}
gives $\eta_n$, for the line of sights that are parallel to the
galactic plane of the caustic ring under consideration and near
tangent to its surface at any $\psi_*$, as \beeq
\eta_n(\psi_*)=\frac{v^2_{ \rm{rot}}}{\Sigma_c
G}f_n\frac{v_n}{b_n}\frac{\cos{\alpha(\psi_*)}}{\sqrt{a_np_n}}
{\mathcal G}_n(\psi_*)\; .\label{esteta}\eneq It was shown in
reference \cite{sing} that $b_n$ and $v_n$ are of the same order
of magnitude. Moreover, Sikivie \cite{IRAS} interpreted the ten
rises in the rotation curve of the Milky Way as the effect of
caustic rings. In that case, the widths $p_n$ of caustic rings are
determined from the observed widths of the rises.  Typically, one
finds $p_n\sim 0.1\, a_n$. Using this, $v_n \sim b_n$, and
$\cos{\alpha(\psi_*)}\simeq 1$, Eq. \ref{esteta} yields \beeq
\{\eta_n: n=1,2,. . .\}\sim (7, 6, 6, 5, 5,. . .)\cdot 10^{-2}
{\mathcal G}_n(\psi_*)\frac{D_L\,D_{LS}}{D_S\,{\rm Gpc}}
\left(\frac{0.27}{j_{\rm{max}}}\right)\!\!
\left(\frac{h}{0.7}\right)\!\! \left(\frac{v_{{\rm
rot}}}{220\,{\rm km/s}}\right)\; .\label{etan}\eneq Comparing the
estimates Eq. \ref{etan} with the estimates of $\eta_n$ for the
outer caustics considered in \cite{lensing}, where the simple fold
is always concave, we see that the lensing effects of the caustic
rings are about $10\, {\mathcal G}_n(\psi_*)$ times larger than
the lensing effects of the outer caustics. We may expand the
function ${\mathcal G}_n(\psi_*)$ in Eq. \ref{esteta}, around
$\psi_*=0, \frac{2\pi}{3}$, and $\frac{4\pi}{3}$. Near the outer
cusp \beeq {\mathcal G}_n(\vartheta)\!=\!{\mathcal
G}_n(2\pi-\vartheta)\!=\!\frac{\sqrt{2}
\vartheta^{-1}}{\sqrt{3}(1\!+\!\frac{p_n}{a_n})^{\frac{1}{2}}}
\!+\!\frac{(5\!+\!14\frac{p_n}{a_n})\vartheta}{12\sqrt{6}(1\!+\!\frac{p_n}{a_n})^{\frac{3}{2}}}
\!+\!\frac{[23+\!4\frac{p_n}{a_n}(9\!+\!37\frac{p_n}{a_n})]\vartheta^3}{320\sqrt{6}
(1\!+\!\frac{p_n}{a_n})^{\frac{5}{2}}}\!+\!O(\vartheta^5) \;
.\eneq For $\vartheta\ll 1$, the first term dominates, hence
${\mathcal G}_n(\vartheta)$ increases proportional to
$\vartheta^{-1}$, as $\vartheta$ decreases. For ${p}_n= 0.1~a_n$,
${\mathcal G}(\vartheta)\simeq 0.778$ $\vartheta^{-1}$. Near the
non-planer cusps (where the cases are concave) we have\beeq
{\mathcal G}_n(\frac{2\pi}{3}-\vartheta)={\mathcal
G}_n(\frac{4\pi}{3}+\vartheta)=\frac{4\vartheta^{-\frac{1}{2}}}{3^{\frac{3}{4}}
(8-\frac{p_n}{a_n})^{\frac{1}{2}}}+
\frac{\vartheta^{\frac{1}{2}}}{3^{\frac{5}{4}}(8-\frac{p_n}{a_n})^{\frac{1}{2}}}
+\frac{(3-\frac{19}{8}\frac{p_n}{a_n})\vartheta^{\frac{3}{2}}}{3^{-\frac{1}{4}}
(8-\frac{p_n}{a_n})^{\frac{3}{2}}}+O(\vartheta^\frac{5}{2}) \;
.\label{expandGnonpl}\eneq For small $\vartheta$, the first term
dominates. Taking ${p}_n=0.1~a_n$, we find ${\mathcal
G}(\frac{2\pi}{3}-\vartheta)\simeq
0.624\cdot\vartheta^{-\frac{1}{2}}$. The expansions for ${\mathcal
G}_n(\frac{2\pi}{3}+\vartheta)$ and ${\mathcal
G}_n(\frac{4\pi}{3}-\vartheta)$ will be used in Sect.
\ref{sub:convex}, where the cases are convex. We investigate the
gravitational lensing properties near the cusps of axion and WIMP
caustic rings in the next two sections. As in Sect.
\ref{subsec:foldcoef}, to estimate the upper bounds at the
smoothed cusps (when $\d v\neq 0$), we pick
$\vartheta=\frac{\pi}{7500}$ and $\vartheta=\frac{\pi}{75}$
(implied by the minimum primordial $\d v$ of the particles) for
the axion and WIMP caustic rings, respectively. The lower bounds
are estimated choosing $\vartheta=\frac{\pi}{7.5}$ (implied by the
triangular feature \cite{IRAS} in the IRAS map of the Milky Way
galactic plane).

\subsubsection{Lensing by a concave fold near the cusps of axion
caustic rings}

At the outer cusp of an axion caustic ring, taking $p_n=0.1~ a_n$,
we constrain the range of function $\mathcal{G}_n$ between
$\mathcal{G}(\pm\frac{\pi}{7.5})\simeq 1.9$ and
$\mathcal{G}(\pm\frac{\pi}{7500})\simeq 1858.5$ (for the fifth
ring of the Milky Way, we have
$\mathcal{G}_5(\pm\frac{\pi}{7.5})\simeq 2$, and
$\mathcal{G}_5(\pm\frac{\pi}{7500})\simeq 1933.7$). Assuming that
the line of sights are parallel to the galactic plane of the
caustic ring, and near tangent to the surface at the points
$\psi_*=\pm\frac{\pi}{7500}$ and $\psi_*=\pm\frac{\pi}{7.5}$
respectively; see Fig. \ref{fig:lensingbycusp}, we find
$\eta(\pm\frac{\pi}{7.5})\simeq 2.91\cdot 10^{-2}$ and
$\eta(\pm\frac{\pi}{7500})\simeq 27.88$ for the sources at
cosmological distances (e.g. $2D_L=2D_{LS}=D_S=1$ Gpc). Thus,
depending on the magnitude of the effective velocity dispersion of
the axion flow in a galactic halo, $\eta$ may range between $0.03$
and $28$ at the outer cusp of an axion caustic ring. This means
that the magnification of a point-like source can be between $3\%$
and $2800\%$. The upper bound, however, is estimated at a point
whose distance to the location that the cusp would occur if
$\delta v$ were zero, is about the minimum smearing out distance
$\d x_a$ due to the primordial velocity dispersion of the flow
(see Sect. \ref{subsec:foldcoef}). Because the effective velocity
dispersion of the flow is expected to be larger than its
primordial velocity dispersion, caustics are likely to be smeared
out more than the size $\d x_a$ estimated in Eq. \ref{dxa}. To
make another estimation at a point further than
$\psi_*=\frac{\pi}{7500}$ from the outer cusp, let us choose a
sample point at $\psi_*=\frac{\pi}{100}$, where ${\mathcal
G}(\frac{\pi}{100})\simeq 24.79$. For the sources at cosmological
distances, we obtain $\eta(\frac{\pi}{100})\simeq 0.37$, which
implies about $37\%$ magnification. This is about $250$ times
larger than the effect estimated (of order $0.15\%$) for the
concave folds \cite{lensing} considering the outer caustics at
cosmological distances. The $37\%$ magnification is about 37 times
larger than the largest effect we estimated (of order $1\%$) in
Ref. \cite{lensing}, considering a line of sight near tangent to a
cosmological caustic ring at $(\rho,z)=(a,0)$, where the fold is
convex. At this location, however, the lensing effect of the
convex fold is minimum, because, the fold coefficient $A$ and the
curvature radius $R_\phi$ are minimum there. Both the $A$ and
$R_\phi$ increase monotonically, as the line of sight approaches
to the cusps. At the non-planer cusps of a cosmological axion
caustic ring, we have
$\eta(\frac{2\pi}{3}-\frac{\pi}{7.5})=\eta(\frac{4\pi}{3}+\frac{\pi}{7.5})\simeq
0.016$, and $ \eta(\frac{2\pi}{3}-\frac{\pi}{7500})=
\eta(\frac{4\pi}{3}+\frac{\pi}{7500})\simeq 0.46$. This implies
that the magnification at the non-planer cusps of the axion
caustic rings may range between $\% 2$ and $\% 46$. At sample
points that are $\frac{\pi}{100}$ radian away from the non-planer
cusps (i.e. choosing $\vartheta=\frac{\pi}{100}$ in Eq.
\ref{expandGnonpl}), we find ${\mathcal
G}(\frac{197\pi}{300})={\mathcal G}(\frac{403\pi}{300})\simeq
3.54$. Therefore
$\eta(\frac{197\pi}{300})=\eta(\frac{403\pi}{300})\simeq 0.053$.
Hence, the magnification is about 5\%, at the sample points near
the non-planer cusps of the cosmological axion caustic rings, when
the case is concave.

Let us estimate the $\eta$, for the fifth caustic ring of the
Milky Way, which is the one closest to us and believed to be the
most constrained by observation. We have $a_5=8.31$ kpc, $b_5=657$
km/s (if $\tau_0>0$) or $516$ km/s (if $\tau_0<0$), $p_5=0.134$
kpc, $q_5=0.2$ kpc, $f_5=0.02$ and $v_5=480$ km/s. The closest
distance to the fifth ring is $55$ pc. The tangential distance to
the ring, $D_L$, is therefore $965$ pc. Thus\beeq\eta_5=1.26\cdot
10^{-7}\frac{v_5}{b_5} \cos\alpha(\psi_*){\mathcal
G}_5(\psi_*)\frac{D_{LS}}{D_S}(\frac{v_{\rm{rot}}}{220\, {\rm
{km}/{s}}})^2\; ,\eneq where ${v_5}/{b_5}$ is about 1.37 if
$\tau_0>0$, or about 1.27 if $\tau_0<0$. Considering the
cosmological sources for which  ${D_{LS}}/{D_S}\simeq 1$, we find
$\eta_5(\psi_*)\sim 1.7\cdot 10^{-7} {\mathcal G}_5(\psi_*)$. Near
the outer cusp ${\mathcal G}_5(\vartheta)\simeq 0.81\cdot
\vartheta^{-1}$. Unfortunately, even at $\psi_*=\frac{\pi}{7500}$
(assuming the ring is axionic and the velocity dispersion of the
flow is about the primordial value), we find that $\eta_5\simeq
3.33\cdot 10^{-4}$ if $\tau_0>0$, or $\eta_5\simeq 3.10\cdot
10^{-4}$ if $\tau_0<0$. Thus, the magnification and image
distortion are negligible (of order $\% 0.03$). However, as noted
in Refs. \cite{Hogan,lensing}, the images of extended sources may
be modified in recognizable ways. In particular, straight jets
would be seen with an abrupt bend where their line of sight
crosses a fold. If a jet makes an angle $\alpha$ with the normal,
it appears bent \cite{lensing} by an angle $\d\equiv
\frac{1}{2}\eta\sin{2\alpha}$.

\subsubsection{Lensing by a concave fold near the cusps of WIMP caustic rings}

At the outer cusp of a WIMP caustic ring, taking $p_n=0.1~a_n$, we
constrain the range of function $\mathcal{G}_n$ between
$\mathcal{G}(\pm\frac{\pi}{7.5})\simeq 1.9$ and
$\mathcal{G}(\pm\frac{\pi}{75})\simeq 18.6$ (for the fifth ring of
the Milky Way, we have $\mathcal{G}_5(\pm\frac{\pi}{7.5})\simeq 2$
and $\mathcal{G}_5(\pm\frac{\pi}{75})\simeq 19.3$). Assuming that
the line of sights are parallel to the galactic plane of the
caustic ring, and tangent to the surface at the points
$\psi_*=\pm\frac{\pi}{7.5}$ and $\psi_*=\pm\frac{\pi}{75}$
respectively, we find $\eta(\pm\frac{\pi}{7.5})\simeq 0.029$ and
$\eta(\pm\frac{\pi}{75})\simeq 0.28$ for the sources at
cosmological distances. Thus, depending on the magnitude of the
effective velocity dispersion of the axion flow in a galactic
halo, $\eta$ can be in the range between $0.03$ and $0.28$ at the
outer cusp of a WIMP caustic ring. This implies between $3\%$ and
$28\%$ magnification near the outer cusp. At the non-planer cusps
of a cosmological WIMP caustic ring, we have
$\eta(\frac{2\pi}{3}-\frac{\pi}{7.5})=\eta(\frac{4\pi}{3}+\frac{\pi}{7.5})\simeq
0.016$ and
$\eta(\frac{2\pi}{3}-\frac{\pi}{75})=\eta(\frac{4\pi}{3}+\frac{\pi}{75})\simeq
0.046$. This implies that the magnification at the non-planer
cusps of WIMP caustic rings may range between $\% 2$ and $\% 5$
when the fold is concave. If the nearby ring is a WIMP caustic,
its lensing effects near the cusps are negligible (at most, we
find $\eta_5(\pm\frac{\pi}{75})\simeq 3.3\cdot 10^{-6}$, and
$\eta_5(\frac{2\pi}{3}-\frac{\pi}{75})=\eta_5(\frac{4\pi}{3}+\frac{\pi}{75})\simeq
5.2\cdot 10^{-7}$). In the next section, we consider the
gravitational lensing effects of caustics rings when the line of
sight is tangent to the surface where a convex fold catastrophe is
located.
\subsection{Lensing by a Convex Fold of a Caustic Ring}
\label{sub:convex} By definition, a convex fold is curved in the
direction opposite to the side with two extra flows (Fig.
\ref{fig:convex}) and radius of curvature of the surface along the
line of sight is positive. For a caustic ring, in the region where
$\frac{2\pi}{3}<\psi_*<\frac{4\pi}{3}$, $R_\phi>0$, hence the
simple folds associated with $R_\phi$ in this region are convex;
see Fig. \ref{fig:convex}. We assume that the line of sight is
parallel to the galactic plane of the caustic ring, and passes by
the surface near tangent as in Fig. \ref{fig:convex} (hence the
line of sight lies in the $z=z(\psi_*)$-plane). The
$\xi_{I}$-dependent shift is given \cite{lensing} as \beeq
\Delta\xi=\xi_{ I} -\xi_{ S}
=-\frac{\eta}{\pi}\left[\,\ln{\left({{R_\phi D_L|\xi_{ I}|}
\over{2 L^2}}\right)}-1\right]\xi_{I}\; , \label{shifc} \eneq
where the length scale $L$ is a cutoff beyond which our
description of the flow is invalid. The function $\eta$ is given
in Eq. \ref{esteta}. The $\xi_{I}$-independent part of the shift,
$-\frac{L^2 \eta}{\pi D_L R_\phi}$, is uninteresting, and
therefore subtracted out in Eq. \ref{shifc}. The magnification is
given \cite{lensing} as\beeq {\mathcal{M}}=\frac{d\xi_{I}}{d\xi_{
S}}=\left|\frac{d\xi_{S}}{d\xi_{I}}\right|^{-1} =\left|1 +
\frac{\eta}{\pi}~ \ln{\left({{R_\phi\,D_L\,|\xi_{ I}|} \over{2
L^2}}\right)}\right|^{-1} \; . \label{magc} \eneq The cutoff $L$
has an effect on the magnification and elongation of the image in
the direction normal to the caustic surface, but that effect is
$\xi_I$ independent. $L$ has a global effect on the image, as
opposed to an effect localized near $\xi_I = 0$. In particular,
when the source is exactly behind the caustic ($\xi_S = 0$), the
images are at $\xi_I = -\xi_c$, 0, and $+\xi_c$, with \beeq \xi_c
= {2 L^2 \over R_\phi D_L} \exp{(-{\pi \over \eta} + 1)}\; .
\label{xic} \eneq A point-like source has a single image
sufficiently far from the caustic, say at $\xi_{ I1}>0$. When the
line of sight approaches the caustic surface tangent point,  two
new images appear on top of each other at $\xi_{ I2}=\xi_{ I3} =
-\xi_c/e$. At that moment, the magnification at $\xi_{I2}$ is
infinite, and $\xi_S = \eta \xi_c/e \pi$. As the source crosses
the caustic, $\xi_{ I2}$ moves toward $\xi_{ I1}$ and finally
merges with it. When $\xi_{ I1} = \xi_{ I2} = +\xi_c/e$, the
magnification diverges again. After that, only the image at $\xi_{
I3}$ remains. The image of an extended object is modified
\cite{lensing} in the direction perpendicular to the convex fold
by the relative amount \beeq {\mathcal{M}} - 1 = - {\eta \over
\pi} \ln{\left({|\xi_I| \over \xi_d}\right)}\; ,\label{Str} \eneq
where\beeq \xi_d = {2 L^2 \over R_\phi D_L}\; . \label{xid} \eneq
The image is stretched for $\xi_I < \xi_d$, and compressed for
$\xi_I > \xi_d$. The line of sights we consider in this section
are near tangent to the caustic surface at sample points
$\psi_*=\frac{2\pi}{3}+\vartheta$
($\psi_*=\frac{4\pi}{3}-\vartheta$), where $\vartheta$ are small,
and lie in the $z=z(\frac{2\pi}{3}+\vartheta)$-plane
($z=z(\frac{4\pi}{3}-\vartheta)$-plane). For such a line of sight,
$L^2$ is of order ${\sqrt{3}}(8a-p)p\vartheta^3/16$. Therefore,
using Eqs. \ref{rphinonplaner} and \ref{xid}, we obtain $\xi_d
\sim {3\zeta p\vartheta^3}/{\sqrt{1+3\zeta^2}D_L}$. Taking
$D_L=0.5$ Gpc, $p=100$ pc and $\zeta=1$, we estimate $\xi_d\sim
3\cdot 10^{-7}\vartheta^3$ at the sample locations of cosmological
caustics. For the fifth caustic ring of the Milky Way, we find
$\xi_{d5}\sim 0.2\cdot \vartheta^3$. Because $p/D_L$ is the
transverse angular size of the caustic ring, our description
certainly fails for $\xi_I > p/D_L$. We have $p/D_L\sim 2\cdot
10^{-7}$ for the rings at cosmological distances, whereas
$p_5/D_{L5}\simeq 0.14$ for the fifth ring of the Milky Way.

The estimates for $\eta$ (Eq. \ref{ep}) in Eqs. \ref{shifc} and
\ref{magc} are the same as in Eq. \ref{etan}, except that, now,
${\mathcal G}(\psi_*)$ has to be evaluated in the region where
$\frac{2\pi}{3}<\psi_*<\frac{4\pi}{3}$. Note that at $\psi_*=\pi$,
i.e at $(\rho, z)=(a, 0)$, for $p_n=0.1~a_n$ we have ${\mathcal
G}_n(\pi)=\frac{1}{\sqrt{2}}$, and hence, $\eta_n(\pi)$ are equal
to the estimates obtained (of order $10^{-2}$) for the convex case
in Ref. \cite{lensing} (denoted by $\eta'_n$). However, near the
non-planer cusps, we have \beeq {\mathcal
G}_n(\frac{2\pi}{3}+\vartheta)={\mathcal
G}_n(\frac{4\pi}{3}-\vartheta)
=\frac{4\vartheta^{-\frac{1}{2}}}{3^{\frac{3}{4}}
\sqrt{8-\frac{p_n}{a_n}}}-
\frac{\vartheta^{\frac{1}{2}}}{3^{\frac{5}{4}}\sqrt{8-\frac{p_n}{a_n}}}
+\frac{(3-\frac{19}{8}\frac{p_n}{a_n})\vartheta^{\frac{3}{2}}}{3^{-\frac{1}{4}}
(8-\frac{p_n}{a_n})^{\frac{3}{2}}}+O(\vartheta^\frac{5}{2}) \;
\label{Gnonpl}.\eneq Because ${\mathcal G}_n$ diverge as
$\vartheta^{-\frac{1}{2}}$ near the non-planer cusps,
$\eta_n(\psi_*)$ are larger than $\eta_n(\pi)$ near the cusps.
Depending on the magnitude of the effective velocity dispersion of
the CDM flow in a galactic halo, the range of $\eta$ can be
constrained at the smoothed non-planer cusps. For the axion
caustic rings, taking $p_n=0.1~ a_n$, we find
$\eta(\frac{2\pi}{3}+\frac{\pi}{7500})=0.46$ and
$\eta(\frac{2\pi}{3}+\frac{\pi}{7.5})=0.03$, for the rings at
cosmological distances. (Due to the reflection symmetry of the
caustic rings, the same results hold at
$\psi_*=\frac{4\pi}{3}-\frac{\pi}{7500}$ and
$\psi_*=\frac{4\pi}{3}-\frac{\pi}{7.5}$ respectively). For the
WIMP caustic rings, taking $p_n=0.1~ a_n$, we constrain the range
of $\eta$ at the smoothed non-planer cusps between
$\eta(\frac{2\pi}{3}+\frac{\pi}{75})\simeq 0.05$ and
$\eta(\frac{2\pi}{3}+\frac{\pi}{7.5})\simeq 0.03$, for the sources
at cosmological distances. (Due to the reflection symmetry of the
caustic rings, the same results hold for
$\psi_*=\frac{4\pi}{3}-\frac{\pi}{75}$ and
$\psi_*=\frac{4\pi}{3}-\frac{\pi}{7.5}$). Unfortunately, although
$\eta$ increases as the point at which the line of sight is near
tangent to the convex fold approaches to the non-planer cusps, the
magnification (Eq. \ref{Str}) is not. This is because, the line of
sight stays close over shorter depths, as it approaches to the
non-planer cusps. For an angular distance $\xi_I\sim 10^{-9}$, we
find about $1\%$ magnification at the sample point
$\psi_*=\frac{2\pi}{3}+\frac{\pi}{10}$ for the cosmological
caustic rings. Even at angular distances as small as $\xi_I\sim
10^{-10}$, the magnification is about $3\%$ at this point. The
lensing effects by the convex fold of the nearby fifth ring are
negligible (of order $10^{-6}$). Finally, in the next section, we
consider the gravitational lensing by a fold with zero curvature.

\subsection{Lensing by a Zero Curvature Fold of a Caustic Ring}
\label{sub:zero}

We found in Eq. \ref{zerodirection} that, in the regions
$0<\psi_*<\frac{2\pi}{3}$ and $\frac{4\pi}{3}<\psi_*<2\pi$ a
caustic ring has a pair of tangent lines along which the curvature
vanishes. For such a line of sight, the image shift is given
\cite{lensing} as \beeq \Delta\xi=-\Theta(-\xi_I)
\left(-\xi_0\,\xi_I^3 \right)^{1/4}\; , \label{dxio} \eneq where
\beeq \xi_0 \equiv (9.89~\frac{A}{\Sigma_c})^4 \frac{U}{D_L}\; .
\label{xi0} \eneq Here, $U$ is a positive constant and has
dimensions of (length)$^3$ \cite{lensing}. The fold coefficient
$A$ is estimated in Eq. \ref{Anring}. The magnification is given
\cite{lensing} as: \beeq {\mathcal{M}} = \left|1 - {3 \over
4}\Theta(-\xi_I)
\left(-\frac{\xi_0}{\xi_I}\right)^{1/4}\right|^{-1} \; . \eneq

Triple images occur when $|\xi_I| \leq \xi_0$. Unfortunately, for
the zero curvature tangents of caustic rings, $\xi_0$ is very
small. To have some order of estimates for $\xi_0$, let us
consider the line of sights along the zero curvature tangents at
$\psi_*=\frac{\pi}{75}$ and $\psi_*=\frac{\pi}{7.5}$ of a
cosmological ring surface ($2D_L=2D_{LS}=D_S= {\rm Gpc}$). From
Eq. \ref{Anring}, we find $A(\frac{\pi}{75})\sim 1.5 \cdot 10^{-3}
{{\rm gr} \over {\rm cm}^2 {\rm kpc}^{1 \over 2}}~$ and
$A(\frac{\pi}{7.5})\sim 5.1 \cdot 10^{-4} {{\rm gr} \over {\rm
cm}^2 {\rm kpc}^{1 \over 2}}~$. Taking $U = ({\rm kpc})^3$, one
finds $\xi_0(\frac{\pi}{75}) \simeq 2.9 \cdot 10^{-14}$ and
$\xi_0(\frac{\pi}{7.5}) \simeq 3.5 \cdot 10^{-16}$. Hence, the
triple images cannot be resolved. At an angular distance as small
as $\xi_I \sim 10^{-9}$ ($10^{-10}$) the magnification and image
distortion are of order 2\% (3\%) for the line of sight near
tangent at $\psi_*=\frac{\pi}{7.5}$.   For the line of sight near
tangent at $\psi_*=\frac{\pi}{75}$, the effects are of order $6\%$
for $\xi_I\sim 10^{-9}$ and $11\%$ for $\xi_I\sim 10^{-10}$. For
the fifth caustic ring of the Milky Way the effects are negligible
(of order $10^{-5}$ for $\xi_I\sim 10^{-9}$ and
$\psi_*=\frac{\pi}{75}$).

In the next section, we study the structural stability of the
caustic rings, using the tools of the Catastrophe Theory.

\section{Structural Stability of Caustic Rings}
\label{Stability}

In this section, we analyze the dark matter caustic rings in the
Catastrophe Theory \cite{Gilmore} point of view. Catastrophe
Theory studies phenomena distinguished by sudden changes in
behavior, arising from smooth alterations in circumstances. It
analyzes how the qualitative nature of equation {\it solutions}
that describe the {\it states} of the phenomena, depends on the
parameters that appear in the equations. In general, the problem
of determining the solutions (states), let alone analyzing how
these solutions change, as the parameters that control the
circumstances change, is a formidable task. Considerable amount of
simplifications occur, however, if the equations can be derived as
the gradient (with respect to the state variables) of a
catastrophe function (or, potential function) $\Phi$ of both the
state variables and control parameters. In particular, Catastrophe
Theory studies how the equilibria (or the critical points),
$\nabla\Phi=0$, of the gradient systems change, as the control
parameters change.

The catastrophe function of the triaxial caustic rings
($\zeta=1$), whose equilibrium points are given by the flow
equations (Eqs. \ref{flownewr}-\ref{flownewz}) near the caustic,
can be given as\be  \Phi({\chi_1},{\chi_2}, p,
\rho-a,z)=\frac{1}{3}\,{\chi_1}^3-{\chi_1}{\chi_2}^2
+\sqrt{p}\,{\chi_2}^2-(\rho-a){\chi_1}-z{\chi_2} \;
.\label{pot}\ee In the Catastrophe Theory terminology, the
$\chi_i$ are called the {\it state variables}, $p$, the
``longitude'' of the caustic cross-section, and $(\rho-a, z)$, the
physical space coordinates on the transverse cross-sectional plane
(Fig. \ref{fig:fig6}), are called the {\it control parameters} of
the $\Phi$. The term $\frac{1}{3}\,{\chi_1}^3-{\chi_1}{\chi_2}^2$,
which is purely composed of the state variables, is called the
{\it germ}, and the terms proportional to the control parameters
are called the {\it perturbations}. We have seen that neither the
existence of the caustic rings (Eq. \ref{Det}), nor their lensing
effects (Eq. \ref{ETA}) depend on $\zeta$. (Both the function
$\eta$ and the vanishing of determinant $D_2$ are independent of
$\zeta$). The parameter $\zeta$ determines the ratio of the
latitudinal dimension $q$, and the longitudinal dimension $p$.
Recall, however, that the flow equations near the caustic rings
are derived \cite{sing} assuming that the caustics are tight, i.e.
$p\ll a$ and $q\ll a$. The fifth caustic ring of the Milky Way
appears tight in the IRAS map \cite{IRAS}, and
$\zeta_5=\frac{4}{3\sqrt{3}}\frac{q_5}{p_5}=1.15$ in that case.
When $\zeta=1$ ($\zeta\neq 1$), the three dual cusps of the
caustic cross-section make up an equilateral (arbitrary) triangle.

The potential given in Eq. \ref{pot} is symmetric under the
discrete transformations $(\chi_1, \chi_2)\rightarrow(\chi_1,
-\chi_2)$ and $(p, \rho-a, z)\rightarrow (p, \rho-a, -z)$. The
equilibrium, $\nabla\Phi=0$, of the caustic ring potential indeed
implies the flow equations (Eq. \ref{flownewr}-\ref{flownewz} with
$\zeta=1$) \be
\frac{\dd\Phi}{\dd\chi_1}&=&0\Rightarrow\rho=a+\chi_1^2-\chi_2^2\label{catr2}\\
\frac{\dd\Phi}{\dd\chi_2}&=&0\Rightarrow
z=2(\sqrt{p}-{\chi_1})\chi_2\label{catz2}\; . \ee If the system
described by $\Phi$ is in equilibrium (stable or unstable) at a
particular point in the state space, the stability properties of
the equilibrium can be determined from the Stability (Hessian)
Matrix \be \Phi_{ij}\equiv \frac{\dd \Phi}{\dd \chi_i
\dd\chi_j}\equiv\left(\begin{array}{cc}
{\dd\rho\over\dd{\chi_1}}&~~~{\dd \rho\over\dd{\chi_2}}\\
\\
{\dd z\over\dd{\chi_1}}&~~~ {\dd z\over\dd{\chi_2}}\\
\end{array}
\right)=2\, \left(\begin{array}{cc}
{{\chi_1}}&~~~{-{\chi_2}}\\
\\
{-{\chi_2}}&~~~ {\sqrt{p}-{\chi_1}}\\
\end{array}
\right) \; . \label{stab}\ee This is, in fact, nothing but the
Jacobian matrix we have obtained in Eq. \ref{Jaco}. The
equilibrium points, or the critical points, at which
$\nabla\Phi=0$ are called nonisolated, degenerate, or non-Morse
critical points. The critical points at which $\det\Phi_{ij}\neq
0$ are called isolated, nondegenerate, or Morse critical points.
The fourfold degenerate critical point of $\Phi$ occurs when the
stability matrix $\Phi_{ij}$ vanishes identically. This implies
$\chi_1=\chi_2=p=0$. Because all the critical points a priori
satisfy Eqs. \ref{catr2}-\ref{catz2}, we also have $\rho=a$, and
$z=0$ at the fourfold degenerate critical point. Thus, the point
at $(p, \rho, z)=(0, a, 0)\in\mathrm{R}^3$ in control parameter
space, parameterizes the function \be\Phi(\chi_1, \chi_2, 0, 0,
0)=\frac{1}{3} \chi_1^3-\chi_1\chi_2^2\; ,\ee which has fourfold
degenerate critical point at $(\chi_1,
\chi_2)=(0,0)\in\mathrm{R}^2$, in state variable space.

\begin{figure}[ht] \centering
\includegraphics[height=7.5cm,width=15.5cm]{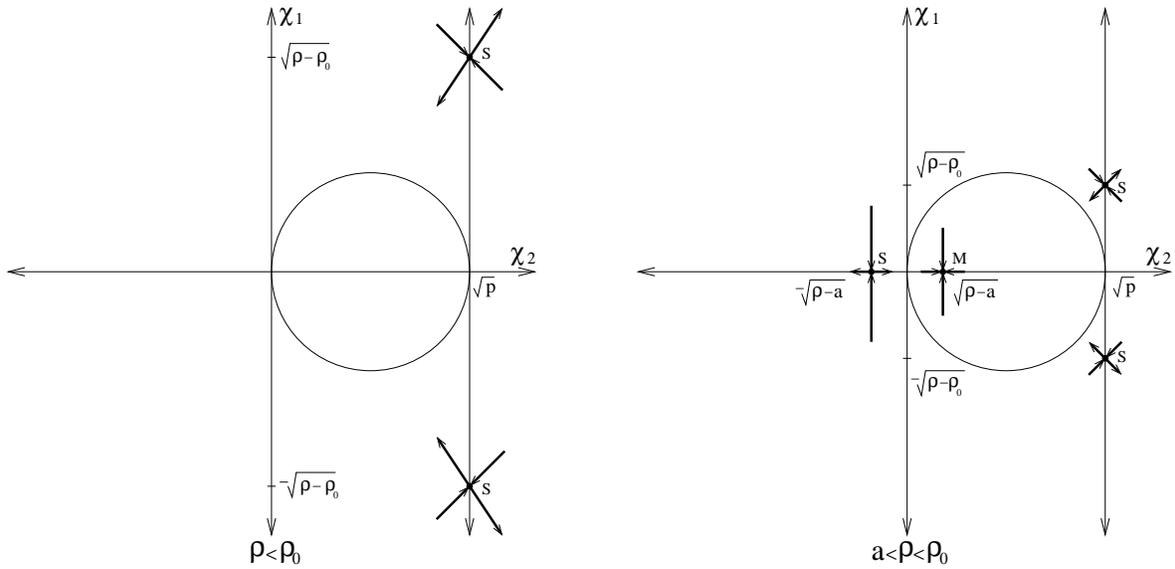}
\caption{The critical points of the catastrophe function for the
caustic rings are shown in the $(\chi_1$-$\chi_2)$-plane as a
function of the control parameter $\rho$ for given ring radius $a$
and caustic longitude $p$. In the left panel $\rho<\rho_0$,
whereas on the right panel $a<\rho<\rho_0$. The circular sets of
degenerate critical points are located where the determinant of
the stability matrix vanishes. The S and M denote the saddle and
local minimum respectively. The signs of the eigenvalues along the
principal directions at the isolated critical points are indicated
by the thick arrows.} \label{fig:catastrophe}
\end{figure}

\begin{figure}[ht] \centering
\includegraphics[height=8.5cm,width=12cm]{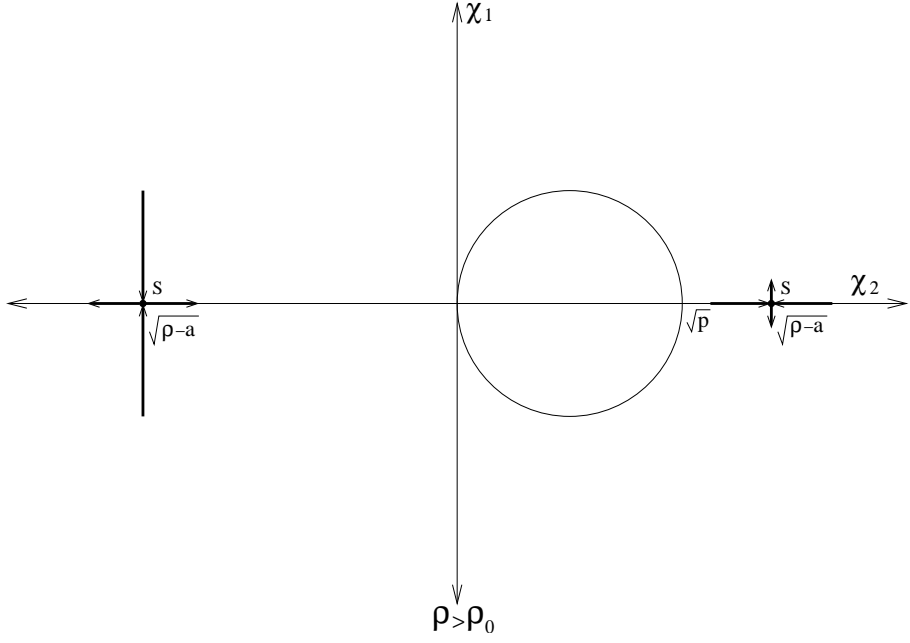}
\caption{ The critical points of the catastrophe function for the
caustic rings. Same as Fig. \ref{fig:catastrophe} except that now
the control parameter $\rho>\rho_0$.} \label{fig:catas2}
\end{figure}

The twofold and threefold degenerate critical points are found by
requiring one of the eigenvalues of the stability matrix to
vanish. When this occurs, the determinant of $\Phi_{ij}$ is zero:
\beeq D_2 ({\chi_1},{\chi_2}) =
-4\left[\left({\chi_1}-\frac{\sqrt{p}}{2}\right)^2+{\chi_2}^2-\frac{p}{4}\right]=0\;
. \eneq So, the set of degenerate (Non-Morse) critical points of
$\Phi$ is the circle \beeq
\left({\chi_1}-\frac{\sqrt{p}}{2}\right)^2+{\chi_2}^2=\frac{p}{4}\;
. \label{ci}\eneq This means that, whenever a critical point lies
on this circle, it is doubly or triply degenerate (fourfold if
$p=0$ also). As we have seen in Sect. \ref{sec:FlowEq}, points of
this circle in the state variable space, produce the caustic
cross-section with longitude $p$ in physical $(\rho-a, z)$-plane.
Now, to study the stability properties of the caustics (degenerate
critical points), we can use the two state variables $\chi_1$ and
$\chi_2$, to give a parametric representation for the three
control parameters in $\mathrm{R}^3$. The parametric
representation for $p(\chi_1, \chi_2)$ that can be obtained from
Eq. \ref{ci} is\beeq
p=\left(\frac{\chi_1^2+\chi_2^2}{\chi_1}\right)^2 \; .\eneq The
parametric representation for $\rho$ and $z$ are given in Eqs.
\ref{catr2}-\ref{catz2}. Instead of working with these parametric
equations directly, to study the stability properties easily, we
may take advantage of the scaling relations among the state and
control parameters. As can be seen through Eqs.
\ref{pot}-\ref{stab}, if the state variables $\chi_1$ and $\chi_2$
both scale like $\lambda$, then all the control parameters $p$,
$\rho-a$, and $z$ scale like $\l^2$:\beeq (\chi_1,
\chi_2)\rightarrow\lambda(\chi_1, \chi_2)\Rightarrow (p,\rho-a,
z)\rightarrow\l^2(p, \rho-a, z) \; .\label{scale}\eneq Moreover,
because the longitudinal size of the caustic rings $p\geq 0$, it
is sufficient to consider the two dimensional $(\rho-a, z)$
cross-sections of the three dimensional control parameter space,
for $p=0$, and for any $p>0$ (recall that we study the tight
caustics for which $0<p\ll a$, in this paper). Any other case can
be considered via the scaling relations Eq. \ref{scale}. For
$p>0$, we determine how the circular set (Eq. \ref{ci}) of
degenerate critical points parameterizes the $(\rho-a, z)$
cross-section by substituting a parameter representation (Eq.
\ref{causticpar}) of this circle:\be
{\chi_1}=\frac{\sqrt{p}}{2}\left(1\pm\cos\psi\right)\; ,\;\;\;
{\chi_2}=\frac{\sqrt{p}}{2}\sin\psi\; ,\nonumber\ee where
$\psi\in[0, 2\pi]$ is an angular variable, into Eqs.
\ref{catr2}-\ref{catz2}. We find\beeq
\rho(\psi)=a+\frac{p}{2}\cos\psi(\cos\psi\pm 1)\; ,\;\;\;
z(\psi)=\frac{p}{2}\sin\psi(1\mp\cos\psi)\; . \label{causz}\eneq
These are the caustic equations (Eqs. \ref{rpsi}-\ref{zpsi}) with
$\zeta=1$. Thus, by substituting the parametric equations (Eq.
\ref{causticpar}) for $\chi_1$ and $\chi_2$ satisfying the two and
three fold degeneracy condition ($\det\Phi_{ij}=0$), into the
equations for critical points (flow equations), we have found, in
Eq. \ref{causz}, the doubly and triply degenerate critical points
(folds and cusps respectively) of the catastrophe function $\Phi$.
The symmetry $(\chi_1, \chi_2)\rightarrow(\chi_1, -\chi_2)$ and
$(p, \rho-a, z)\rightarrow(p, \rho-a, -z)$, which was noted
earlier, is realized as $\psi\rightarrow -\psi$ in Eqs. \ref{ci}
and \ref{causz}. When $p=0$, Eqs. \ref{catr2}-\ref{catz2} and
\ref{ci} imply that the caustic cross-section reduces to the point
$(0,0)$ in the $(\rho-a,z)$-plane. In this case, the caustic tube
becomes an isolated caustic circle with $(\rho, z)=(a, 0)$. (It
becomes a topological circle if the axial symmetry is not
assumed). The plot of Eqs. \ref{causz}, for a given $p>0$, in
$(\rho-a, z)$-plane is called the $D_{-4}$ catastrophe which we
call tricusp; see Fig. \ref{fig:fig6}. For any $p>0$, the tricusp
on the $(\rho-a, z)$-plane of the control parameter space, divides
the plane into three open regions which describe catastrophe
functions of three qualitatively different types. The qualitative
properties of functions parameterized by points within any one
region, however, are the same. These properties change, as we pass
through the tricusp on the plane. Therefore, to determine the
qualitative properties of the catastrophe functions, it is
sufficient to consider convenient points in each of these regions.
We restrict ourselves to the $\rho-a$ axis. On this axis, one
passes through the tricusp at $\rho-a=0$ and $\rho=\rho_0=a+p$.
Hence, the three open regions are: $\rho<a$, $a<\rho<\rho_0$, and
$\rho>\rho_0$; see Fig. \ref{fig:fig6}. The equations
\ref{catr2}-\ref{catz2}, which determine the critical points of
$\Phi$, on this axis are: \be
\rho-a-\chi_1^2+\chi_2^2=0\label{r0}\\
2(\sqrt{p}-\chi_1)\chi_2=0\label{z0}\; .\ee We infer, from Eq.
\ref{z0} that the critical points must have either (i) $\chi_2=0$,
or (ii) $\chi_1=\sqrt{p}$. If (i) is the case, Eq. \ref{r0} yields
$\chi_1=\pm\sqrt{\rho-a}$ (out and in flows) for $\rho>a$. If (ii)
is the case, Eq. \ref{r0} yields $\chi_2=\pm\sqrt{\rho_0-\rho}$
(down and up flows) for $\rho<\rho_0$. Thus, for any given $p>0$,
along the $\rho-a$ axis of the control parameter space of the
catastrophe function $\Phi$, there are two sets of critical points
for $\rho<a$: $(\chi_1,\chi_2)=(\sqrt{p},
\pm\sqrt{\rho_0-\rho})$(down and up flows), four sets of critical
points for $a<\rho<\rho_0$: $(\chi_1, \chi_2)=(\sqrt{p},
\pm\sqrt{\rho_0-\rho})$ and $(\chi_1,
\chi_2)=(\pm\sqrt{\rho_0-\rho}, 0)$ (down, up, in and out flows);
see Fig. \ref{fig:catastrophe},
and two sets of critical points for $\rho>\rho_0$: $(\chi_1,
\chi_2)=(\pm\sqrt{\rho_0-\rho}, 0)$ (in and out flows); see Fig.
\ref{fig:catas2}.

Once the locations of the critical points are known, their
stability properties can be determined from the Stability Matrix
Eq. \ref{stab}. In case (i), where $(\chi_1, \chi_2)=(\pm
\sqrt{\rho-a}, 0)$, (for $\rho>a$), we have\be\Phi_{ij}=2\,
\left(\begin{array}{cc}
{{\pm\sqrt{\rho-a}}}&~{{0}}\\
{0}&~ {\sqrt{p}\mp{\sqrt{\rho-a}}}\\
\end{array}
\right) \; .\ee Thus, in this case the $\chi_1$ and the $\chi_2$
axes are the principal directions with the eigenvalues
$\l_{\chi_1}=\pm 2\sqrt{\rho-a}$ and
$\l_{\chi_2}=2(\sqrt{p}\mp\sqrt{\rho-a})$ respectively. The
$\chi_1$ eigenvalue of the upper (lower) critical point
$\l^u_{\chi_1}$ ($\l^l_{{\chi_1}}$) is always positive (negative),
except at $\rho=a$, where it vanishes. The $\chi_2$ eigenvalue of
the upper critical point, $\l^u_{{\chi_2}}=
2(\sqrt{p}-{\sqrt{\rho-a}})$ changes sign when $\rho$ goes through
$\rho_0=a+p$: $\l^u_{{\chi_2}}$ is positive for $a<\rho<\rho_0$,
whereas it is negative for $\rho>\rho_0$. Thus, the upper critical
point $(\chi_1, \chi_2)=(\sqrt{\rho-a}, 0)$ with the eigenvalues
$(\lambda^u_{{\chi_1}}, \lambda^u_{{\chi_2}})=2(\sqrt{\rho-a},
\sqrt{p}-\sqrt{\rho-a})$ is a local minimum for $a<\rho<\rho_0$,
whereas it is a saddle for $\rho>\rho_0$ ($\lambda^u_{{\chi_2}}$
vanishes when $\rho=\rho_0$); see Figs.
\ref{fig:catastrophe}-\ref{fig:catas2}. The lower critical point
$(\chi_1, \chi_2)=(-\sqrt{\rho-a}, 0)$, on the other hand, has the
$\chi_1$ eigenvalue $\lambda^l_{{\chi_1}}=-\sqrt{\rho-a}\leq0$
(vanishes when $\rho=a$), and $\chi_2$ eigenvalue
$\lambda^l_{{\chi_2}}=\sqrt{p}+\sqrt{\rho-a}>0$. Thus, the lower
critical point is always a saddle for any $\rho>a$. Note that, the
points $\rho=a$ and $\rho=\rho_0$ are the locations at which the
critical points $(\chi_1,\chi_2)=(\pm\sqrt{\rho-a},0)$ cross the
circular degenerate critical set
$\chi_1^2+\chi_2^2-\sqrt{p}\chi_1=0$ (Eq. \ref{ci}) on the state
variable plane. We, therefore, expect these locations to be
special.

The stability properties and principal axes of the critical points
of case (ii), $(\chi_1, \chi_2)=(\sqrt{p},
\pm\sqrt{\rho_0-\rho})$, where $\rho<\rho_0$, are determined
similarly. The stability matrix Eq. \ref{stab}, for this case is
\be\Phi_{ij}=2\, \left(\begin{array}{cc}
{\sqrt{p}}&~{{\mp\sqrt{\rho_0-\rho}}}\\
{\mp\sqrt{\rho_0-\rho}}&~ {0}\\
\end{array}
\right) \; .\ee The principal directions are no longer along the
$\chi_1$ and $\chi_2$ axes. So, they must be determined by matrix
diagonalization. We find that, the eigenvalues $\l_1$ and $\l_2$
associated with the principal directions $\vec{v}_1$ and
$\vec{v}_2$ are the same for the upper and lower critical points:
\beeq(\l^u_{{v}_1}, \l^u_{{v_2}})=(\l^l_{{v_1}},
\l^l_{{v_2}})=(\sqrt{p}+\sqrt{p+4(\rho_0-\rho)},
\sqrt{p}-\sqrt{p+4(\rho_0-\rho)})\; .\eneq The eigendirections at
the upper and lower critical points are \be \vec{v}^u_{1}=
\left(\begin{array}{c}
{1}\\
{\frac{\sqrt{p}-\sqrt{p+4(\rho_0-\rho)}}{2\sqrt{\rho_0-\rho}}}\\
\end{array}\right) ,\;\; {{\vec{v}_2}^u}= \left(\begin{array}{c}
{1}\\
{\frac{\sqrt{p}+\sqrt{p+4(\rho_0-\rho)}}{2\sqrt{\rho_0-\rho}}}\\
\end{array}\right)\; ,\ee
and \be \vec{v}^l_1= \left(\begin{array}{c}
{1}\\
{\frac{-\sqrt{p}+\sqrt{p+4(\rho_0-\rho)}}{2\sqrt{\rho_0-\rho}}}\\
\end{array}\right) ,\;\; \vec{v}^l_2= \left(\begin{array}{c}
{1}\\
{\frac{-\sqrt{p}-\sqrt{p+4(\rho_0-\rho)}}{2\sqrt{\rho_0-\rho}}}\\
\end{array}\right)\; ,\ee
respectively. The eigenvalues $\lambda^u_{v_1}=\lambda^l_{v_1}=
\sqrt{p}+\sqrt{p+4(\rho_0-\rho)}>0$, whereas
$\lambda^u_{v_2}=\lambda^l_{v_2}=
\sqrt{p}-\sqrt{p+4(\rho_0-\rho)}\leq 0$ (vanishes at
$\rho=\rho_0$). Thus, both of the critical points $(\chi_1,
\chi_2)=(\sqrt{p}, \pm\sqrt{\rho_0-\rho})$, where $\rho<\rho_0$,
are saddle points; see Fig. \ref{fig:catastrophe}. Where
$\rho=\rho_0$, the critical points cross the circular degenerate
critical set in the state variable plane, and, $\lambda^u_{v_2}$
and $\lambda^l_{v_2}$ vanish. We expect this point to be special
too. Before investigating the special locations at $\rho=a$, and
at $\rho=\rho_0$, let us summarize what we have learned. In each
of the regions divided by the triangular shape of the $(\rho-a,
z)$ cross-section of the control parameter space, different types
of catastrophe functions $\Phi$ are parameterized. The interior of
the triangular region (the region where $a<\rho<\rho_0$ on the
$\rho$-axis) parameterizes the functions which have three saddles,
and one local minimum. The saddles exist outside the circular set
of degenerate (Non-Morse) critical points, and the local minimum
exists inside the circular set (Figs.
\ref{fig:catastrophe}-\ref{fig:catas2}). The region outside the
triangular domain (where $\rho<a$ and $\rho>\rho_0$ on the
$\rho$-axis) parameterizes the functions with two saddles, both of
which lie outside the circular set of degenerate critical points
of the state variable plane. Let us, now, investigate what happens
at the special locations $\rho=a$ and $\rho=\rho_0$, as $\rho$
increases along the line $z=0$ for a given $p>0$. For $\rho<a$,
there is a pair of saddle points at $(\chi_1, \chi_2)=(\sqrt{p},
\pm\sqrt{\rho_0-\rho})$, each of which has the eigenvalues
$\lambda_{{v_1},{v_2}}=\sqrt{p}\pm\sqrt{p+4(\rho_0-\rho)}$ in the
corresponding principal directions. As $\rho$ increases through
$a$, a fold catastrophe occurs at $(\chi_1,\chi_2)=(0,0)$ (when
$\rho=a$). This is because, once $\rho>a$, a saddle comes into
existence outside the circular set, at
$(\chi_1,\chi_2)=(-\sqrt{\rho-a}, 0)$ with $(\l_{\chi_1},
\l_{\chi_2})=2(-\sqrt{\rho-a}, \sqrt{p}+\sqrt{\rho-a})$, and a
local minimum comes into existence inside the circular set, at
$(\chi_1, \chi_2)=(\sqrt{\rho-a}, 0)$ with $(\lambda_{\chi_1},
\lambda_{\chi_2})=2(\sqrt{\rho-a}, \sqrt{p}-\sqrt{\rho-a})$. The
new critical points at $(\chi_1, \chi_2)=(\pm\sqrt{\rho-a}, 0)$
approach each other as $\rho$ approaches $a$ from above and become
degenerate at $(\chi_1, \chi_2)=(0,0)$ when $\rho=a$, and
disappear for $\rho<a$. Meanwhile, the eigenvalues in the $\chi_2$
direction approach to $\sqrt{p}$, and the eigenvalues in the
$\chi_1$ direction decrease monotonically, and vanish at $\rho=a$.
This can also be seen from the fact that, at the doubly degenerate
critical point $(\chi_1, \chi_2)=(0,0)$ (where $\rho=a$ on the
$\rho$-axis), the catastrophe function becomes\beeq
\Phi=\frac{1}{3}\chi_1^3-\chi_1\chi_2^2+\sqrt{p}\chi_2^2\label{PHI}\;
,\eneq and the stability matrix $\Phi_{ij}$ takes the form\be
\Phi_{ij}\Big|_{(0,0)}=\left(\begin{array}{cc}
{0}&~{{0}}\\
{0}&~ {2\sqrt{p}}\\
\end{array}
\right)\; .\ee Thus, the eigenvalue in the $\chi_1$ direction is
zero, whereas the eigenvalue in the $\chi_2$ direction is nonzero
and positive. There exists a catastrophe at $\rho=a$, because, the
number of isolated critical points in the $(\chi_1, \chi_2)$-plane
changes as $\rho$ passes through $a$, where $\det{\Phi}_{ij}=0$.
Moreover, the point $\rho=a$ on the $\rho$-axis is expected to be
a location for a fold $(A_2)$ catastrophe, because two critical
points are involved there. The rigorous way to prove this, is to
check if the form of the catastrophe function Eq. \ref{PHI} can be
reduced to the canonical form for an $A_2$ catastrophe, by making
a smooth change of variables at $\rho=a$. Since we expect the
catastrophe germ is of type $A_2$, which is
$c_2\chi_2'^2+c_1\chi_1'^3$, and from Eq. \ref{pot} we have
$\sqrt{p}$ as the coefficient of $\chi_2^2$, the canonical form at
$\rho=a$ that we seek is $\sqrt{p}\chi_2'^2+c_1\chi_1'^3$. We
introduce the following non-linear transformation\be
\chi_1'\!&=&\!\chi_1+\Delta_1=\chi_1+A_{20}\chi_1^2+A_{11}\chi_1\chi_2
+A_{02}\chi_2^2\nonumber\\
\chi_2'\!&=&\!\chi_2+\Delta_2=\chi_2+B_{20}\chi_1^2+B_{11}\chi_1\chi_2+B_{02}\chi_2^2
\; , \label{axpr}\ee and see if the coefficients $A_{ij}$, and
$B_{ij}$ can be chosen to satisfy \be
\!\!\!\!\!\!\!\!\sqrt{p}\chi_2^2-\chi_1\chi_2^2+\frac{1}{3}\chi_1^3\!\!&=&\!\!
\sqrt{p}\chi_2'^2+c_1\chi_1'^3\nonumber\\\!\!&=&\!\!\sqrt{p}(\chi_2^2+2\chi_2\D_2+\D_2^2)
+c_1(\chi_1^3+3\chi_1^2\D_1+3\chi_1\D_1^2+\D_1^3)\;
\label{lhs}.\ee The equality is already satisfied for the terms of
degree two. For the terms of degree three, we have \beeq
-\chi_1\chi_2^2+\frac{1}{3}\chi_1^3=2\sqrt{p}\chi_2\D_2+c_1\chi_1^3=2\sqrt{p}(B_{20}\chi_1^2\chi_2
+B_{11}\chi_1\chi_2^2+B_{02}\chi_2^3)+c_1\chi_1^3\; .\eneq The
solution is \beeq B_{20}=0\; ,\;\; B_{11}=\frac{-1}{2\sqrt{p}}\;
,\;\; B_{02}=0\; ,\;\; c_1=\frac{1}{3}\; . \eneq On the left hand
side of Eq. \ref{lhs}, the coefficients for the terms of degree
four and higher are all zero. The corresponding terms on the right
hand side of Eq. \ref{lhs}, however, depend on the disposable
coefficients $A_{ij}$ and $B_{ij}$ (for $i+j\geq 2$). Because, for
a given degree, there are more disposable coefficients on the
right hand side of Eq. \ref{lhs}, than the zero coefficients of
the left hand side, it is possible to choose $A_{ij}$ and $B_{ij}$
so that all the monomials of degree greater than or equal to four,
have vanishing coefficients. Thus, at $\rho=a$, the catastrophe
function can be put into the form\beeq
\Phi=\sqrt{p}\chi'^2_2+\frac{1}{3}\chi_1'^3\; . \eneq By rescaling
the control parameters $p^\frac{1}{4}\chi_2'\equiv y$ and
$3^{-\frac{1}{3}}\chi_1'\equiv x$, we obtain the canonical form of
$A_2$ catastrophe: $\Phi=y^2+x^3$. This proves that $\rho=a$ is a
fold location.

We now turn our attention to the other interesting location on the
$\rho$-axis, that is, to the point $\rho=\rho_0$. As $\rho$
increases further toward $\rho_0$, the new saddle on the $\chi_1$
axis at $(\chi_1, \chi_2)=(-\sqrt{\rho-a}, 0)$ with
$(\lambda_{\chi_1},
\l_{\chi_2})=2(-\sqrt{\rho-a},\sqrt{p}-\sqrt{\rho-a})$ moves away
to the west. The local minimum $(\chi_1, \chi_2)=(\sqrt{\rho-a},
0)$ with $(\lambda_{\chi_1},
\l_{\chi_2})=2(\sqrt{\rho-a},\sqrt{p}-\sqrt{\rho-a})$ moves across
the degenerate critical circle, toward the point
$(\chi_1,\chi_2)=(\sqrt{p}, 0)$. The two original saddles at
$(\chi_1, \chi_2)=(\sqrt{p}, \pm\sqrt{\rho_0-\rho})$, with
$\lambda_{{v_1},{v_2}}=\sqrt{p}\pm\sqrt{p+4(\rho_0-\rho)}$, move
toward the same point $(\sqrt{p}, 0)$ along the line
$\chi_1=\sqrt{p}$. As the critical points approach to $(\sqrt{p},
0)$, the two principal directions of the two saddles rotate toward
the $\chi_1$ and $\chi_2$ axes, and the eigenvalues all approach
to $(2\sqrt{p}, 0)$. As $\rho$ passes through $\rho_0$, the two
original saddles and the local minimum collide at $(\chi_1,
\chi_2)=(\sqrt{p}, 0)$, and form a triply degenerate critical
point. By expanding the catastrophe function Eq. \ref{pot} for
$z=0$ and $\rho=\rho_0$ about the three fold degenerate critical
point $(\chi_1,\chi_2)=(\sqrt{p}, 0)$ using
$\chi_1=(\chi_1-\sqrt{p})+\sqrt{p}\equiv X_1+\sqrt{p}$ and
$\chi_2\equiv X_2$, and dropping the unimportant constant term, we
find \beeq\Phi(X_1, X_2)=\frac{1}{3}X_1^3+\sqrt{p}X_1^2-X_1X_2^2\;
.\label{cuspcat}\eneq The stability matrix $\Phi(X_1, X_2)$ at the
critical point $(X_1, X_2)=(0,0)$ is\be
\Phi_{ij}\Big|_{(0,0)}=\left(\begin{array}{cc}
{2\sqrt{p}}&~{{0}}\\
{0}&~ {0}\\
\end{array}
\right)\; .\ee Thus, at $(\chi_1, \chi_2)=(\sqrt{p}, 0)$, as is
mentioned above, the eigenvalue in the $\chi_1$ direction
$\l_{\chi_1}=2\sqrt{p}$ and the eigenvalue in the $\chi_2$
direction $\l_{\chi_2}=0$. For $\rho>\rho_0$, the two saddles and
the local minimum are combined to form a single saddle at
$(\chi_1,\chi_2)=(\sqrt{\rho-a}, 0)$, with
$(\l_{\chi_1},\l_{\chi_2})=2(\sqrt{\rho-a},\sqrt{p}-\sqrt{\rho-a}
)$. The $\chi_2$ eigenvalue of the critical point $(\chi_1,
\chi_2)=(\sqrt{\rho-a}, 0)$ changes sign as $\rho$ passes through
$\rho_0$. This implies a dual cusp catastrophe $A_{-3}$ at
$(\chi_1, \chi_2)=(\sqrt{p}, 0)$, where $(\rho, z)=(\rho_0, 0)$.
To prove the existence of the cusp at $(\chi_1, \chi_2)=(\sqrt{p},
0)$ rigorously, we need to reduce the catastrophe function Eq.
\ref{cuspcat} into the canonical form of the dual cusp $A_{-3}$:
$\sqrt{p}X'^2_1+C_2X'^4_2$. Making a smooth change of variables,
the same as Eq. \ref{axpr}, but, with $\chi_i'$ replaced by
$X_i'$, and $\chi_i$ replaced by $X_i$, we obtain the equation
that the catastrophe function at $(X_1, X_2)=(0, 0)$ has to
satisfy \be &&\sqrt{p}X_1^2-X_1X_2^2+\frac{1}{3}X_1^3=
\sqrt{p}X_1'^2+C_2X_2'^4=\sqrt{p}(X_1^2+2X_1\D_1+\D_1^2)\nonumber\\
&&\hspace{5cm}+C_2(X_2^4+4X_2^3\D_2+6X^2_2\D_2^2+4X_2\D_2^3+\D_2^4)\;
.\label{0eq}\ee The equality is already satisfied for the terms of
degree two. For the terms of degree three, we have \be
-X_1X_2^2+\frac{1}{3}X_1^3=2\sqrt{p}X_1\D_1
=2\sqrt{p}(A_{20}X_1^3+A_{11}X_1^2X_2+A_{02}X_1X_2^2)\; ,\ee whose
solution is \beeq A_{20}=\frac{1}{6\sqrt{p}}\; ,\;\; A_{11}=0\;
,\;\; A_{02}=\frac{-1}{2\sqrt{p}}\; .\eneq For all the terms of
degree four, we must have\be
0=2\sqrt{p}X_1\D_1+\sqrt{p}\D_1^2+C_2X^4_2
&=&2\sqrt{p}(A_{30}X_1^4+A_{21}X_1^3X_2+A_{12}X_1^2X_2^2+A_{03}X_1X_2^3)\nonumber\\
&&+\frac{1}{36\sqrt{p}}X_1^4-\frac{1}{6\sqrt{p}}X_1^2 X_2^2
+\frac{1}{4\sqrt{p}}X_2^4+C_2X_2^4\; , \ee which implies \beeq
A_{30}=-\frac{1}{72p}\; ,\;\; A_{21}=0\; ,\;\; A_{12}=\frac{1}{12
p}\; ,\;\; A_{03}=0\; ,\;\; C_2=-\frac{1}{4\sqrt{p}}\; . \eneq
Similarly, it can be shown that, Eq. \ref{0eq} can be satisfied
for the terms of all degrees greater than four. Thus, we have\beeq
\Phi=\sqrt{p}X_1'^2-\frac{1}{4\sqrt{p}}X_2'^4\; .\eneq By
rescaling the control parameters $p^\frac{1}{4}X_1'\equiv x'$ and
$2^{-\frac{1}{2}}p^{-\frac{1}{8}}X'_2\equiv y'$, we obtain the
canonical form of $A_{-3}$ dual cusp catastrophe:
$\Phi=x'^2-y'^4$. This proves the existence of the dual cusp at
$(X_1, X_2)=(0, 0)$, or i.e. at $(\chi_1, \chi_2)=(\sqrt{p}, 0)$,
and hence, at $\rho=\rho_0$ on the $\rho$-axis. For $\rho>\rho_0$,
a second saddle also exists along $\chi_1<0$ axis, at $(\chi_1,
\chi_2)=(-\sqrt{\rho-a}, 0)$ with $(\lambda_{\chi_1},
\lambda_{\chi_2})=2(-\sqrt{\rho-a}, \sqrt{p}+\sqrt{\rho-a})$; see
Fig. \ref{fig:catas2}.

Although the study has been carried out on the symmetry axis $z=0$
to make the calculation simple, the procedure can be generalized
to an arbitrary direction and the results are generic. When the
point in control parameter space moves through the boundaries of
the triangular region in an arbitrary direction, a fold or a dual
cusp catastrophe occurs depending on whether the point moves
through one of the three arc like sides or one of the three
corners (dual cusps). Upon moving through a side line (fold) a
minimum collides and annihilates one of the three saddles. At a
fold, one of the three saddles and the local minimum are on top of
each other. The saddle that is annihilated depends on which
boundary side line the control point moves through. Upon moving
through a dual cusp, two of the saddles and the local minimum
collide, one of the saddles and the minimum annihilate each other,
and one single saddle remains. At a dual cusp, two of the three
saddles and the local minimum are on top of each other. Hence, the
results are (qualitatively) independent of the restriction we made
by focusing the $\rho$-axis only. The CDM flow described by the
$\Phi$ (Eq. \ref{pot}) necessarily generates the tricusp caustic
in physical space. In this sense the caustic rings are
structurally stable.

\section{Conclusions}
\label{Conc} The infall of Cold Dark Matter (CDM) onto isolated
galaxies such as our own Milky Way produces discrete number of
flows and caustics in the halo CDM distribution. Caustics are
locations in space where the density diverges in the limit of zero
velocity dispersion. The generic caustic is a surface at the
boundary between two regions, one of which has $n$ flows and the
other $n+2$ flows. In the limit of zero velocity dispersion, the
density diverges as $\frac{1}{\sqrt{\sigma}}$, where $\sigma$ is
the distance to the surface, on the side with $n+2$ flows. If the
velocity dispersion is small, but nonzero, this divergence is cut
off, because, the location of the caustic gets smeared out. So the
density at the caustic is no longer infinite, but merely very
large. There are two types of caustics in the galactic halos: {\it
outer} and {\it inner}. An outer caustic is a simple fold ($A_2$)
catastrophe, located on a topological sphere enveloping the
galaxy. An inner caustic is a ring whose cross-section is an
elliptic umbilic ($D_{-4}$) catastrophe with three dual cusps
($A_{-3}$ catastrophes); see Fig. \ref{fig:fig6}. We call this
cross-section a {\it tricusp}. Our focus was upon the caustic
rings in this paper. In Sect. \ref{sec:FlowEq}, we reparameterized
the CDM flow near a caustic ring, and obtained the ring equations
in space, as single valued functions $\rho(\psi)$ and $z(\psi)$ of
an angular variable $\psi$ parameterizing the tricusp. The density
profile near the surface of a caustic ring is $d(\psi,
\sigma)=\frac{A(\psi)}{\sqrt{\sigma}}\Theta(\s)$, where $A(\psi)$
is called the fold coefficient. In Sect. \ref{Density}, we derived
the fold coefficient, everywhere on the caustic ring. $A(\psi)$ is
minimum at the middle locations between the cusps. It increases
monotonically as one approaches to the cusps, where it diverges in
the limit of zero velocity dispersion. A caustic ring also has a
specific geometry. Section \ref{difgeo} is devoted to the
differential geometry of the caustic ring surface. We obtained
Gaussian, mean and principal curvatures as functions of $\psi$. In
Sect. \ref{sec:Lensing}, we derived the gravitational lensing
effects of the caustic rings for the line of sights that are
parallel to the galactic plane of the ring and near tangent to the
surface at any given $\psi$. We estimated the image magnifications
due to the caustic rings at cosmological distances and due to the
nearby fifth ring of our own galaxy. For a given critical surface
density, the effects are proportional to the fold coefficient
$A(\psi)$ at the point where the line of sight is near tangent to
the surface, and inversely proportional to the square root of the
curvature of the surface along the direction associated with the
line of sight. The magnification increases as the line of sight
approaches to the cusps where it diverges in the limit of zero
velocity dispersion. If the velocity dispersion is small but
nonzero the cusps are smoothed out, hence the lensing effects are
no longer infinite but merely large at the cusps. We used the
lower and upper bounds of the effective velocity dispersions of
the axion and WIMP flows in galactic halos to constrain the
lensing effects at the cusps. For a cosmological axion caustic
ring, we found that the magnification may range between $3\%$ and
$2800\%$ at the outer cusp, and between $2\%$ and $46\%$ at the
non-planer cusps. For a cosmological WIMP caustic ring, on the
other hand, we constrained the magnification between $3\%$ and
$28\%$ at the outer cusp, and between $2\%$ and $5\%$ at the
non-planer cusps. Because the upper bounds for the magnification
at the cusps are obtained considering the minimum primordial value
of the velocity dispersions of the CDM candidates in space, and
because the observer's line of sight may not exactly be parallel
to the galactic plane of the caustic ring in general, they should
be regarded with precaution. In the limit of zero velocity
dispersion, at a sample point near the outer cusps of cosmological
CDM caustic rings, we found about $37\%$ magnification. This
effect is about 250 times greater than the effect obtained for the
concave case considering the outer caustics, and about $37$ times
greater than the largest effect predicted considering the caustic
rings where the case was convex, before. The nearby caustic ring,
the fifth ring of the Milky Way, is only 1 kpc away (in the
direction of observation) from us. We also estimated the
magnifications of this ring of our own galaxy. Unfortunately, even
near the outer cusp, the lensing effects are too weak to be
observed with present instruments. In Sec. \ref{Stability}, we
presented the correspondence of our formulation with the
Catastrophe Theory. We derived the Catastrophe Function of the
triaxial caustic rings and obtained the flow equations as the
equilibrium points of this Catastrophe Function. The analysis of
the Stability (Hessian) Matrix showed that the caustic rings are
structurally stable; see also the recent simulations
\cite{simulations} confirming that the caustic ring is stable
under perturbations.

\section{Appendix}
\begin{figure}[ht]
\centering
\includegraphics[height=6.5cm,width=11.5cm]{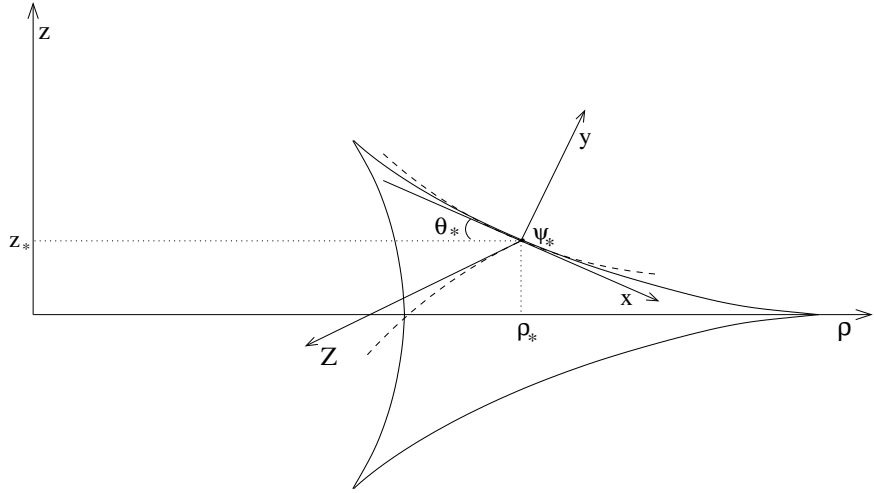}
\caption{Same as Fig. \ref{fig:tricusp1} except that, now, we
define a new Cartesian coordinate system $(x, y, Z)$ at the
arbitrary point $\psi_*$, such that $(\hat{x}, \hat{y})=(\hat\eta,
-\hat\sigma)$. In the neighborhood of $\psi_*$, the caustic
cross-sections in the $(x,y)$ and $(y, Z)$ planes can be
approximated by the circles indicated by the dashed arcs. The
circle in the $(x, y)$-plane has radius $R_\psi$, whereas, the one
in the $(x, Z)$-plane has radius $R_\phi$. }
\label{fig:geoappendix}
\end{figure}

In this appendix, we show geometrically that the principal
curvature radii on the caustic ring surface are the inverses of
the corresponding principal curvatures. On the $(x,y)$ and $(y,
Z)$-planes of Fig. \ref{fig:geoappendix}, the caustic
cross-sections around an arbitrary point $\psi_*$ on the surface,
can be approximated locally by the circles, as indicated by the
dashed arcs of the figure. On the $(x, y)$-plane, the circle that
is tangent to the ring surface at the point $(\rho(\psi_*),
z(\psi_*))$, satisfies the equation\beeq
(\rho-\rho_0)^2+(z-z_0)^2=R_\psi^2 \; ,\label{circ}\eneq where
$(\rho_0, z_0)$ is the center of the circle. By taking the first and second derivatives of both sides of Eq.
\ref{circ} with respect to $\rho$, we obtain\be
\rho-\rho_0&=&-\left(\frac{dz}{d\rho}\right)\;(z-z_0)\label{r1}\\
z-z_0&=&-\left(\frac{d^2z}{d\rho^2}\right)^{-1}\left[1
+\left(\frac{dz}{d\rho}\right)^2\right]\; ,\label{z1}
 \ee
respectively. Using Eqs. \ref{r1}-\ref{z1} in Eq. \ref{circ}, we
find \be R_\psi=\pm\left|\frac{d^2z}{d\rho^2}\right|^{-1}
\left[1+\left(\frac{dz}{d\rho}\right)^2\right]^\frac{3}{2}\Bigg|_{(\rho(\psi_*),
z(\psi_*))} \; .\label{Rpsi}\ee We may express the first and
second derivatives of $z$ with respect to $\rho$ as \beeq
\frac{dz}{d\rho}=\frac{z'}{\rho'} ,\;
\frac{d^2z}{d\rho^2}=\frac{\rho' z'' - z' \rho''}{\rho'^3}
\label{surf}\; ,\eneq where prime denotes derivative with respect
to $\psi$. Inserting, Eq. \ref{surf} into Eq. \ref{Rpsi} and using
the fact that $\rho'z''-z'\rho''<0$ on the whole surface, we
obtain \beeq R_\psi(\psi_*)=
\frac{(\rho'^2(\psi_*)+z'^2(\psi_*))^\frac{3}{2}}
{z'(\psi_*)\rho''(\psi_*)-\rho'(\psi_*)z''(\psi_*)}=\frac{1}{\kappa_\psi(\psi_*)}\;
.\eneq In writing the last equality we compared $R_\psi$ with Eq.
\ref{kappapsi}.

We, next, want to show that $R_\phi$, the principal curvature
radius in the plane defined by $\hat{y}$ and
$\hat{\textsc{z}}\equiv \hat{x}\times\hat{y}$ (Fig.
\ref{fig:geoappendix}), is the inverse of $\kappa_\phi$. Around
the origin of the $(x, y, \textsc{z})$-coordinate system of Fig.
\ref{fig:geoappendix}, which is chosen to be the point $\psi_*$,
the caustic surface equation can be expressed as: \beeq
y=\frac{x^2}{2R_\psi}+\frac{\textsc{z}^2}{2R_\phi}\;
,\label{y}\eneq where $R_\psi>0$ for $0<\psi_*<2\pi$, whereas,
$R_\phi>0$ only in the region
$\frac{2\pi}{3}<\psi_*<\frac{4\pi}{3}$. $R_\phi<0$ in the regions
$0<\psi_*<\frac{2\pi}{3}$ and $\frac{4\pi}{3}<\psi_*<2\pi$; see
Fig. \ref{fig:geoappendix}. The intersection of the caustic
surface with $z=z_*$-plane is a circle of radius $\rho_*$;
therefore, near $(\rho, z)=(\rho_*, z_*)$, the equation for this
circle can be written as\beeq
\rho-\rho_*=-\frac{\textsc{z}^2}{2\rho_*}\; .\label{1}\eneq On the
other hand, from Eq. \ref{mateq}, in the $z=z_*$-plane, we have
\be x=(\rho -\rho_*)\cos\theta_*, \;\; y=(\rho
-\rho_*)\sin\theta_* \label{xy}\; ,\ee where $x=\eta$, and
$y=-\sigma$. Inserting Eqs. \ref{xy} into Eq. \ref{y}, in the
neighborhood of $(\rho_*, z_*)$ we find that the intersection of
the caustic surface with the $z=z_*$-plane satisfies\beeq
y=(\rho-\rho_*)\sin\theta_*=\frac{\textsc{z}^2}{2R_\phi}\label{2}\;
,\eneq where we neglected the term proportional to
$(\rho-\rho_*)^2$. Combining Eqs. \ref{1} and \ref{2} we obtain,
 \beeq R_\phi=-\frac{\rho_*}{\sin\theta_*}\label{R2}\; .
\eneq By the definition of the angle $\theta_*$, \beeq
(\rho-\rho_*)\tan\theta_*=-(z-z_*) \; ,\eneq see Fig.
\ref{fig:geoappendix}. Hence, we can express
$\tan\theta_*=-\frac{z'}{\rho'}$, thus \beeq
\sin\theta_*=\pm\left[1+\left(\frac{d\rho}{dz}\right)^2\right]^{-\frac{1}{2}}\Bigg|_{(\rho(\psi_*),
z(\psi_*))}
=\frac{z'(\psi_*)}{\sqrt{\rho'^2(\psi_*)+z'^2(\psi_*)}}
\label{st}\; .\eneq Insertion of Eq. \ref{st} into Eq. \ref{R2}
yields \beeq
R_\phi(\psi_*)=-\frac{\rho(\psi_*)\sqrt{\rho'^2(\psi_*)+z'^2(\psi_*)}}{z'(\psi_*)}=\frac{1}{\kappa_\phi(\psi_*)}\;
. \eneq In writing the last equality we compared $R_\phi$ with Eq.
\ref{kappaphi}.
\begin{acknowledgements}
The author would like to thank P. Sikivie for reading the
manuscript and providing useful comments. This research was
supported by European Union grant FP-6-012679.
\end{acknowledgements}


\begin{thebibliography}{99}
\bibitem{Bennett}
C.~L.~Bennett et al., Astrophys. J., Suppl. {\bf 148}, 1 (2003).
\bibitem{Ipser}
P.~Sikivie and J.~Ipser, Phys. Lett. B {\bf 291}, 288 (1992);
A.~Natarajan and P.~Sikivie, ``Robostness of Discrete Flows and
Caustics in Cold Dark Matter Cosmology,'' astro-ph/0508049.
\bibitem{Zel}
Y.~B.~Zel'dovich, Astron. Astrophys. {\bf 5}, 84 (1970).
\bibitem{Huchra}
V.~de~Lapparent, M.~J.~Geller, and J.~P.~Huchra,  Astrophys. J.,
Lett. Ed. {\bf 302} L1 (1986).
\bibitem{cr}
P.~Sikivie, Phys. Lett. B {\bf 432}, 139 (1998).
\bibitem{sing}
P.~Sikivie, Phys. Rev. D {\bf 60}, 063501 (1999).
\bibitem{Big}
P.~Sikivie, ``The Big Flow,'' astro-ph/0112072.
\bibitem{Hogan}
C.~Hogan,  Astrophys. J. {\bf 527}, 42 (1999).
\bibitem{lensing}
C.~Charmousis, V.~Onemli, Z.~Qiu, and P.~Sikivie, Phys. Rev. D
{\bf 67}, 103502 (2003).
\bibitem{Thesis}
Vakif~K.~Onemli, ``Gravitational Lensing by Dark Matter
Caustics,'' astro-ph/0401162,\\ Ph. D. dissertation, to be
published as a book, by the Nova Science Publishers, Inc., New
York.
\bibitem{probing}
R.~Gavazzi, R.~Mohayaee, and B.~Fort, ``Probing Dark Matter
Caustics with Weak Lensing,'' astro-ph/0506061; R.~Mohayaee,
S.~Colombi, B.~Fort, R.~Gavazzi, S.~Shandarin, and J.~Touma,
``Caustics in Dark Matter Halos,'' astro-ph/0510575.
\bibitem{Micro}
L. Bergstrom, J. Edsjo, and C. Gunnarsson, Phys. Rev. D {\bf 63},
083515 (2001); C.~Hogan, Phys. Rev. D {\bf 64}, 063515 (2001);
L.~Pieri and E.~Branchini, J. Cosmol. Astropart. Phys. {\bf 0505},
007 (2005); R.~Mohayaee and S.~F.~Shandarin, ``Gravitational
Cooling and Density Profile near Caustics in Collisionless Dark
Matter Haloes,'' astro-ph/0503163.
\bibitem{IRAS}
P.~Sikivie, Phys. Lett. B {\bf 567}, 1 (2003).
\bibitem{KS}
W.~Kinney and P.~Sikivie, Phys. Rev. D {\bf 61}, 087305 (2000).
\bibitem{STW1}
P.~Sikivie, I.~Tkachev, and Y.~Wang, Phys. Rev. Lett {\bf 75},
2911 (1995).
\bibitem{STW2}
P.~Sikivie, I.~Tkachev, and Y.~Wang, Phys. Rev. D {\bf 56}, 1863
(1997).
\bibitem{FG}
J.~A.~Fillmore and P.~Goldreich, Astrophys. J. {\bf 281}, 1
(1984); E.~Bertschinger, Astrophys. J. Suppl. {\bf 58}, 39 (1985).
\bibitem{Mahdavi}
A.~Mahdavi, N.~Trentham, and R.~B.~Tully, ``The NGC 5846 Group:
Dynamics and the Luminosity Function to $M_R=-12$,''
astro-ph/0506737; R.~B.~Tully, ``Observations of Infall and
Caustics,'' astro-ph/0509482.
\bibitem{Gilmore}
R.~Gilmore, {\it Catastrophe Theory for Scientists and Engineers}
(Dover, New York, 1993); V.~I.~Arnold, {\it Singularities of
Caustics and Wave Fronts} (Kluwer Academic Publisers, Dordrecht,
1990).
\bibitem{simulations}
A.~Natarajan and P.~Sikivie, ``The Inner Caustics of Cold Dark
Matter Halos,'' astro-ph/0510743.
\end{thebibliography}
\end{document}